\documentclass[10pt,twocolumn,aps,showpacs,preprintnumbers,nofootinbib,prd,superscriptaddress,groupedaddress]{revtex4-1}
\usepackage{beramono}
\usepackage[T1]{fontenc}
\usepackage[latin9]{inputenc}
\usepackage{color}
\usepackage{verbatim}
\usepackage{mathtools}
\usepackage{amsmath}
\usepackage{amssymb}
\usepackage{graphicx}
\usepackage{wasysym}
\usepackage{esint}
\usepackage[unicode=true,
 bookmarks=false,
 breaklinks=false,pdfborder={0 0 1},colorlinks=true]
 {hyperref}
\hypersetup{
 linktocpage,citecolor=darkblue,linkcolor=darkblue,urlcolor=darkblue}

\makeatletter
\usepackage{amsthm}
\usepackage{amsfonts}
\usepackage{epsfig}
\usepackage{times}

\usepackage{epstopdf}
\usepackage{aas_macros}
\usepackage{tensor}
\usepackage{bm}
\usepackage{url}\usepackage{wasysym}

\definecolor{oxfordblue}{rgb}{0.0, 0.13, 0.28}
\definecolor{burgundy}{rgb}{0.5, 0.0, 0.13}
\definecolor{darkolivegreen}{rgb}{0.33, 0.42, 0.18}
\definecolor{darkblue}{rgb}{0,0,0.5}
\definecolor{richcarmine}{rgb}{0.84, 0.0, 0.25}
\definecolor{darkblue}{rgb}{0,0,0.5}
\definecolor{venetianred}{rgb}{0.78, 0.03, 0.08}
\definecolor{skobeloff}{rgb}{0.0, 0.48, 0.45}

\renewcommand{\vec}[1]{\boldsymbol{#1}}

\newcommand{\ben}{\begin{enumerate}}
\newcommand{\een}{\end{enumerate}}

\def\be{\begin{equation}}
\def\ee{\end{equation}}
\def\bea{\begin{eqnarray}}
\def\eea{\end{eqnarray}}

\newcommand{\beq}{\begin{eqnarray}}
\newcommand{\eeq}{\end{eqnarray}} 
\newcommand{\ba}{\begin{align}}
\newcommand{\ea}{\end{align}}

\def\O#1#2{O({#1})}
\def\Os#1{O({#1})}
\def\keeping{O(\epsilon^4)\equiv}
\def\Pmass{m}

\def\be{\begin{equation}}
\def\ee{\end{equation}}
\def\best{\begin{equation*}}
\def\eest{\end{equation*}}
\def\beqn{\begin{eqnarray}}
\def\eeqn{\end{eqnarray}}

\makeatother

\begin{document}

\title{Gravitational Magnus effect}

\author{L.~Filipe O.~Costa$^{1,\,*}$, Rita Franco$^{2}$, Vitor Cardoso$^{2,3}$ }

\affiliation{${^{1}}$ GAMGSD, Departamento de Matemática, Instituto Superior
Técnico, Universidade de Lisboa, 1049-001 Lisboa, Portugal}

\email{lfpocosta@math.ist.utl.pt}

\affiliation{${^{2}}$ CENTRA, Departamento de F\'{i}sica, Instituto Superior
Técnico, Universidade de Lisboa, Avenida Rovisco Pais 1, 1049 Lisboa,
Portugal}

\affiliation{${^{3}}$ Perimeter Institute for Theoretical Physics, 31 Caroline
Street North Waterloo, Ontario N2L 2Y5, Canada}
\begin{abstract}
It is well known that a spinning body moving in a fluid suffers a
force orthogonal to its velocity and rotation axis --- it is called
the Magnus effect. Recent simulations of spinning black holes and
(indirect) theoretical predictions, suggest that a somewhat analogous
effect may occur for purely gravitational phenomena. The magnitude
and precise direction of this ``gravitational Magnus effect'' is
still the subject of debate. Starting from the rigorous equations
of motion for spinning bodies in general relativity (Mathisson-Papapetrou
equations), we show that indeed such an effect takes place and is
a fundamental part of the spin-curvature force. The effect arises
whenever there is a current of mass/energy, nonparallel to a body's
spin. We compute the effect explicitly for some astrophysical systems
of interest: a galactic dark matter halo, a black hole accretion disk,
and the Friedmann-Lemaître-Robertson-Walker (FLRW) spacetime. It is
seen to lead to secular orbital precessions potentially observable
by future astrometric experiments and gravitational-wave detectors.
Finally, we consider also the reciprocal problem: the ``force''
exerted by the body on the surrounding matter, and show that (from
this perspective) the effect is due to the body's gravitomagnetic
field. We compute it rigorously, showing the matching with its reciprocal,
and clarifying common misconceptions in the literature regarding the
action-reaction law in post-Newtonian gravity. 
\end{abstract}
\maketitle
\tableofcontents{}

\section{Introduction}

\begin{figure}
\includegraphics[width=1\columnwidth]{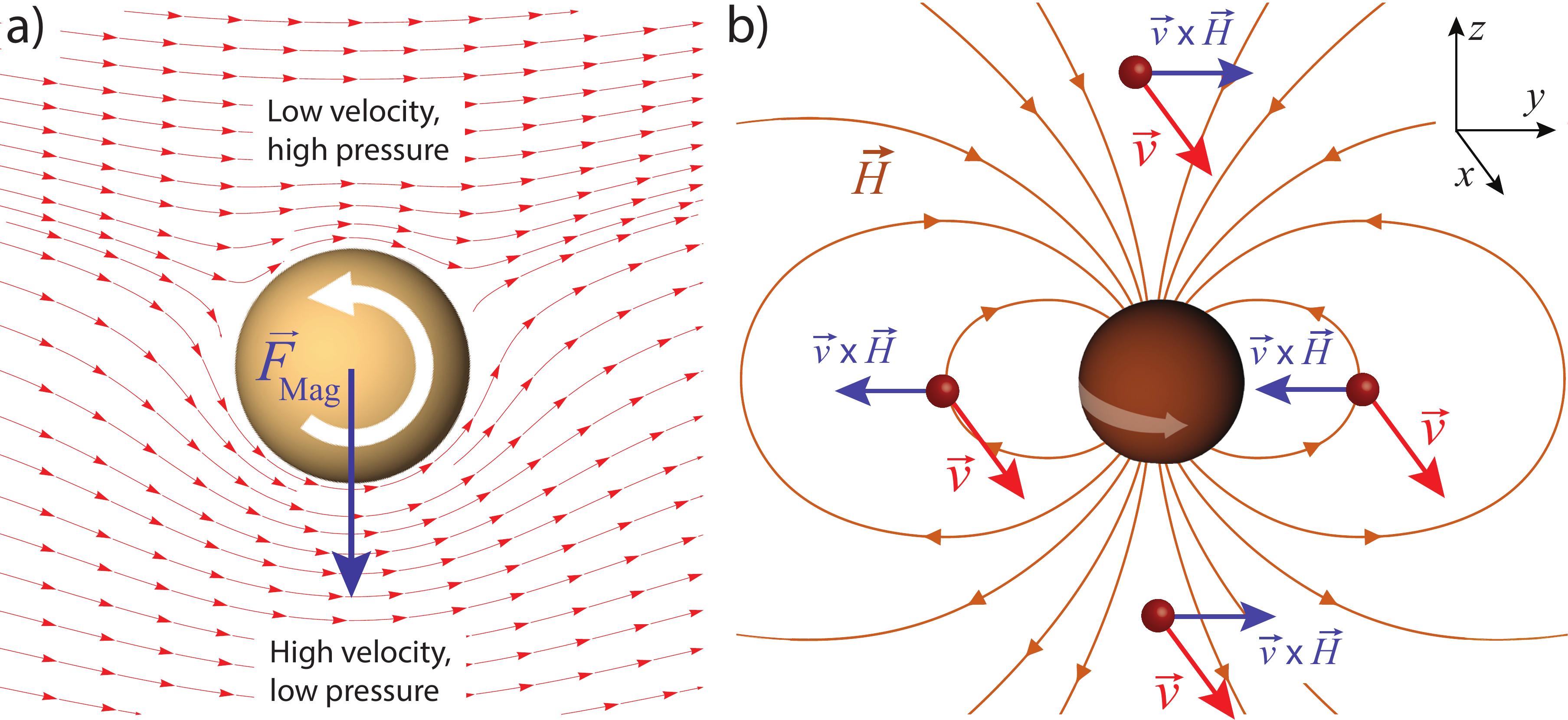}
\protect\protect\protect\protect\protect\caption{\label{fig:Magnus_GMDeflection}a) Magnus effect in fluid dynamics,
as viewed from a spinning body's frame: the body's rotation slows
down the flow which opposes the body's surface velocity, while speeding
it up otherwise, generating a pressure gradient and a net force $\vec{F}_{{\rm Mag}}$
on the body. b) A gravitational analogue of the Magnus effect? ---
due to its gravitomagnetic field $\vec{H}$, a spinning body deflects
particles of a cloud flowing around it, via the gravitomagnetic ``force''
$\vec{F}_{{\rm GM}}=M\vec{v}\times\vec{H}$. By naive application
of an action-reaction principle, a force on the body orthogonal to
its spin and velocity (like the Magnus force) might be expected. }
\end{figure}

The Magnus effect is well known in classical fluid dynamics: when
a spinning body moves in a fluid, a force orthogonal to the body's
velocity and spin acts on it. If the body spins with angular velocity
$\vec{\omega}$, moves with velocity $\vec{v}$, and the fluid density
is $\rho$, such force has the form (see e.g. \cite{rubinow_keller_1961,TsujiMorikawaMizuno1985})
\begin{equation}
\vec{F}_{{\rm Mag}}=\alpha\rho\vec{\omega}\times\vec{v}\,.
\end{equation}
(Here $\alpha$ is a a factor that differs according to the flow regime.\footnote{Its value is not generically established. According to theoretical
and experimental results, it is nearly a constant at low Reynolds
numbers \cite{rubinow_keller_1961,TsujiMorikawaMizuno1985,WattsFerrer1987},
but seemingly velocity dependent at higher Reynolds numbers \cite{WattsFerrer1987}.}) This effect is illustrated in Fig.~\ref{fig:Magnus_GMDeflection}.
It can, in simple terms, be understood from the fact that the fluid
circulation induced by the body's rotation decreases the flow velocity
on one side of the body while increasing it on the opposite side.
The Bernoulli equation then implies that a pressure differential occurs,
leading to a net force on the body~\cite{MunsunYoungOkiishi}. 

By its very own nature, the fluid-dynamical Magnus force hinges on
contact interactions between the spinning body and the fluid. Thus,
\textit{ordinary} Magnus forces \textit{cannot exist} in the interaction
between (i) a fluid and a spinning black hole (BH), or (ii) an ordinary
star and dark matter (DM) which only interacts with it via gravity.
However, in general relativity any form of energy gravitates and contributes
to the gravitational field of bodies. In particular, a spinning body
produces a ``gravitomagnetic field''~\cite{CiufoliniWheeler};
if the spinning body is immersed in a fluid, such field deflects the
fluid-particles in a direction orthogonal to their velocity, as illustrated
in Fig.~\ref{fig:Magnus_GMDeflection}b, seemingly leading to a nonzero
``momentum transfer'' to the fluid. The question then arises if
some backreaction on the body, in the form of a Magnus- (or anti-Magnus)-like
force --- in the sense of being orthogonal to the flow and to the
body's spin --- might arise. Indeed, the existence of such a force,
in the same direction of the Magnus effect of fluid dynamics, is strongly
suggested by numerical studies of nonaxisymmetric relativistic Bondi-Hoyle
accretion onto a Kerr BH~\cite{Font:1998sc}. These studies focused
on a fixed background geometry and studied the momentum imparted to
the fluid as it accretes or scatters from the BH. A theoretical argument
for the existence of such an effect has also been put forth in Ref.~\cite{Okawa:2014sxa},
based on the asymmetric accretion of matter around a spinning BH (i.e,
the absorption cross-section being larger for counter- than for corotating
particles) --- which is but another consequence of the gravitomagnetic
``forces'': these are attractive for counterrotating particles,
and repulsive for corotating ones, as illustrated in Fig.~\ref{fig:Magnus_GMDeflection}b
(for particles in the equatorial plane). Such argument leads however
to the prediction of an effect in the direction \emph{opposite} to
the Magnus effect (``anti-Magnus''), thus seemingly at odds with
the results in Ref.~\cite{Font:1998sc}. Very recently, and while
our work was being completed, there was also an attempt to demonstrate
the existence of what, in practice, would amount to such an effect,
based both on particle's absorption and on orbital precessions around
a spinning BH~\cite{Cashen:2016neh} (which, again, are gravitomagnetic
effects); a force in the direction opposite to the Magnus effect was
again suggested. These (conflicting) treatments are however based
on loose estimates, not on a concrete computation of the overall gravitomagnetic
force exerted by the spinning body on the surrounding matter. Moreover,
these are all indirect methods, in which one infers the motion of
the body by observing its effect on the cloud, trying then to figure
out the backreaction on the body (which, as we shall see, is problematic,
since the gravitomagnetic interactions, analogously to the magnetic
interactions, do not obey in general an action-reaction law).

One of the purposes of this work is to perform the first concrete
and rigorous calculation of this effect. We first take a direct approach
--- that is, we investigate this effect from the actual equations
of motion for spinning bodies in general relativity. These are well
established, and known as the Mathisson-Papapetrou (or Mathisson-Papapetrou-Dixon)
equations~\cite{Mathisson:1937zz,Papapetrou:1951pa,Dixon1964,Dixon:1970zza,Gralla:2010xg,Tulczyjew}.
We will show that a Magnus-type force is a fundamental part of the
spin-curvature force, which arises whenever a spinning body moves
in a medium with a relative velocity not parallel to its spin axis;
it has the \emph{same} direction as the Magnus force in fluid dynamics,
and depends only on the mass-energy current relative to the body,
and on the body's spin angular momentum. Then we also consider the
reciprocal problem, rigorously computing the force that the body exerts
on the surrounding matter (in the regime where such ``force'' is
defined), correcting and clarifying the earlier results in the literature.
These effects have a close parallel in electromagnetism, where an
analogous (anti) Magnus effect also arises. For this reason we will
start by electromagnetism --- and by the classical problem of a magnetic
dipole inside a current slab --- which will give us insight into the
gravitational case.

\subsection{Notation and conventions\label{sub:Notation-and-conventions}}

We use the signature $(-+++)$; $\epsilon_{\alpha\beta\sigma\gamma}\equiv\sqrt{-g}[\alpha\beta\gamma\delta]$
is the Levi-Civita tensor, with the orientation $[1230]=1$ (i.e.,
in flat spacetime, $\epsilon_{1230}=1$); $\epsilon_{ijk}\equiv\epsilon_{ijk0}$.
\textcolor{black}{Greek letters $\alpha$, $\beta$, $\gamma$, ...
denote 4D spacetime indices, running 0-3; Roman letters $i,j,k,...$
denote spatial indices, running 1-3}. The convention for the Riemann
tensor is $R_{\ \beta\mu\nu}^{\alpha}=\Gamma_{\beta\nu,\mu}^{\alpha}-\Gamma_{\beta\mu,\nu}^{\alpha}+...$
. $\star$ denotes the Hodge dual: $\star F_{\alpha\beta}\equiv\epsilon_{\alpha\beta}^{\ \ \ \mu\nu}F_{\mu\nu}/2$
for an antisymmetric tensor $F_{\alpha\beta}=F_{[\alpha\beta]}$.
Ordinary time derivatives are sometimes denoted by dot: $\dot{X}\equiv\partial X/\partial t$.

\subsection{Executive summary}

For the busy reader, we briefly outline here the main results of our
paper. A spinning body in a gravitational field is acted, in general,
by a covariant force $DP^{\alpha}/d\tau$ (the spin-curvature force),
deviating it from geodesic motion. Such force can be can be split
into the two components 
\begin{align}
 & \frac{DP^{\alpha}}{d\tau}=F_{{\rm Weyl}}^{\alpha}+F_{{\rm Mag}}^{\alpha}\,,\label{eq:FSpinCurvature}\\
 & F_{{\rm Weyl}}^{\alpha}\equiv-\mathcal{H}^{\alpha\beta}S_{\beta};\qquad F_{{\rm Mag}}^{\alpha}\equiv4\pi\epsilon_{\ \beta\sigma\gamma}^{\alpha}J^{\beta}S^{\sigma}U^{\gamma}\ ,\label{eq:FWeylMag}
\end{align}
where $U^{\alpha}$ is the body's 4-velocity, $S^{\alpha}$ its spin
angular momentum 4-vector, and $J^{\alpha}=-T^{\alpha\beta}U_{\beta}$
the mass-energy 4-current relative to the body. The force $F_{{\rm Weyl}}^{\alpha}$
is due to the magnetic part of the Weyl tensor, $\mathcal{H}_{\alpha\beta}=\star C_{\alpha\mu\beta\nu}U^{\mu}U^{\nu}$,
determined by the details of the system (boundary conditions, etc).
The force $F_{{\rm Mag}}^{\alpha}$, which, in the body's rest frame
reads $\vec{F}_{{\rm Mag}}=4\pi\vec{J}\times\vec{S}$, is what we
call a gravitational analogue to the Magnus force of fluid dynamics;
it arises whenever, relative to the body, there is a \emph{spatial}
mass-energy current $\vec{J}$ not parallel to $\vec{S}$. We argue
that (\ref{eq:FSpinCurvature}) is the force that has been attempted
to be indirectly computed in the literature \cite{Font:1998sc,Okawa:2014sxa,Cashen:2016neh},
from the effect of a moving BH (or spinning body) on the surrounding
matter. We base our claim on a rigorous computation of the reciprocal
force exerted by the body on the medium, in the cases where the problem
is well posed, and where an action-reaction law can be applied. $F_{{\rm Mag}}^{\alpha}$
and $F_{{\rm Weyl}}^{\alpha}$ are also seen to have direct analogues
in the force that an electromagnetic field exerts on a magnetic dipole.

The two components of the force are studied for spinning bodies in
(``slab'') toy models, and in some astrophysical setups. For quasi-circular
orbits around stationary axisymmetric spacetimes studied --- spherical
DM halos, BH accretion disks --- when $\vec{S}$ lies in the orbital
plane, the spin-curvature force takes the form $\vec{F}=A(r)\vec{S}\times\vec{v}$,
where the function $A(r)$ is specific to the system. Its Magnus component
is similar for all systems, whereas the Weyl component greatly differs.
The force $\vec{F}$ causes the orbits to oscillate, and to undergo
a secular precession, given by 
\[
\left\langle \frac{d\vec{L}}{dt}\right\rangle =\vec{\Omega}\times\vec{L};\qquad\vec{\Omega}=\frac{A(r)}{2\Pmass}\vec{S}\ .
\]
The effect might be detectable in some astrophysical settings, likely
candidates are: i) signature in the Milky Way galactic disk: stars
or BHs with spin axes nearly parallel to the galactic plane, should
be in average more distant from the plane than other bodies; ii) BH
binaries where one of the BHs moves in the others' accretion disk,
the secular precession might be detected in gravitational wave measurements
in the future, through its impact on the waveforms and emission directions.

In an universe filled with an homogeneous isotropic fluid, described
by the FLRW spacetime, representing the large scale structure of the
universe, which is conformally flat, we have that $\mathcal{H}^{\alpha\beta}=0\Rightarrow F_{{\rm Weyl}}^{\alpha}=0$,
and so the Magnus force $F_{{\rm Mag}}^{\alpha}$ is the only force
that acts on a spinning body. It reads, \emph{exactly}, 
\[
\vec{F}=-4\pi(\rho+p)(U^{0})^{2}\vec{v}\times\vec{S}
\]
It acts on any celestial body that moves with respect to the background
fluid with a velocity $\vec{v}\nparallel\vec{S}$, and might possibly
be observed in the motion of galaxies with large peculiar velocities
$\vec{v}$. Due to the occurrence of the factor $(\rho+p)$, this
force acts as a probe for the matter/energy content of the universe
(namely for the ratio $\rho/p$, and for the different dark energy
candidates). Any mater/energy content gives rise to such gravitational
Magnus force, except for dark energy \emph{if} modeled with a cosmological
constant ($\rho=-p$).

\section{Electromagnetic (anti) Magnus effect}

We start with a toy problem borrowed from the electromagnetic interaction.
Consider a magnetic dipole within a cloud of charged particles. Is
there a Magnus-type force?

The relativistic expression for the force exerted on a magnetic dipole,
of magnetic moment 4-vector $\mu^{\alpha}$, placed in a electromagnetic
field described by a Faraday tensor $F^{\alpha\beta}$, is~\cite{Dixon1964,Dixon:1970zza,Gralla:2010xg,Costa:2012cy}
\begin{equation}
{\displaystyle \frac{DP^{\alpha}}{d\tau}=B^{\beta\alpha}\mu_{\beta}\equiv F_{{\rm EM}}^{\alpha}};\qquad B_{\alpha\beta}\equiv\star F_{\alpha\mu;\beta}U^{\mu}\,,\label{eq:FEM_TT}
\end{equation}
where $P^{\alpha}$ is the particle's 4-momentum, $U^{\alpha}$ its
4-velocity, and $B_{\alpha\beta}$ is the ``magnetic tidal tensor''
\cite{FilipeCosta:2006fz,Costa:2016iwu} as measured in the particle's
rest frame. In the inertial frame momentarily comoving with the particle,
the space components of $F_{{\rm EM}}^{\alpha}$ yield the textbook
expression 
\begin{equation}
\vec{F}_{{\rm EM}}=\nabla(\vec{B}\cdot\vec{\mu})\,.\label{eq:FEMBook}
\end{equation}
Taking the projection orthogonal to $U^{\alpha}$ of the Maxwell field
equations $F_{\ \ ;\beta}^{\alpha\beta}=4\pi j^{\alpha}$, leads to
$B_{[\alpha\beta]}=\star F_{\alpha\beta;\gamma}U^{\gamma}/2-2\pi\epsilon_{\alpha\beta\sigma\gamma}j^{\sigma}U^{\gamma}$
(cf. Eq. (I.3a) in Table I of Ref.~\cite{Costa:2012cy}), where $j^{\alpha}$
is the current density 4-vector. Therefore 
\begin{equation}
B_{\alpha\beta}=B_{(\alpha\beta)}+\frac{1}{2}\star F_{\alpha\beta;\gamma}U^{\gamma}-2\pi\epsilon_{\alpha\beta\sigma\gamma}j^{\sigma}U^{\gamma}\,.
\end{equation}
Thus, the magnetic tidal tensor decomposes into three parts: its symmetric
part $B_{(\alpha\beta)}$, plus two antisymmetric contributions: the
current term $-2\pi\epsilon_{\alpha\beta\sigma\gamma}j^{\sigma}U^{\gamma}$,
and the term $\star F_{\alpha\beta;\gamma}U^{\gamma}/2$, which arises
when the fields are not covariantly constant along the particle's
worldline (it is related to the laws of electromagnetic induction,
as discussed in detail in \cite{Costa:2012cy}). The force (\ref{eq:FEM_TT})
can then be decomposed as 
\begin{align}
{\displaystyle } & F_{{\rm EM}}^{\alpha}=F_{{\rm Sym}}^{\alpha}+F_{{\rm Mag}}^{\alpha}+F_{{\rm ind}}^{\alpha}\,,\label{eq:FEM_Decomp}\\
 & F_{{\rm Sym}}^{\alpha}\equiv B^{(\alpha\beta)}\mu_{\beta}\,,\qquad F_{{\rm ind}}^{\alpha}\equiv-\frac{1}{2}\star F_{\ \ ;\gamma}^{\alpha\beta}U^{\gamma}\mu_{\beta}\,,\label{eq:FSym_Find}\\
 & F_{{\rm Mag}}^{\alpha}\equiv2\pi\epsilon_{\ \beta\sigma\gamma}^{\alpha}U^{\gamma}j^{\sigma}\mu^{\beta}\,.\label{eq:FMagEM}
\end{align}
Let $h_{\ \beta}^{\alpha}$ denote the space projector with respect
to $U^{\alpha}$ (projector orthogonal to $U^{\alpha}$), 
\begin{equation}
h_{\ \beta}^{\alpha}\equiv U^{\alpha}U_{\beta}+\delta_{\ \beta}^{\alpha}\,.\label{eq:Spaceprojector}
\end{equation}
Since the tensor $\epsilon_{\alpha\beta\sigma\gamma}U^{\gamma}$ automatically
projects spatially, in any of its indices, in fact only the projection
of $j^{\sigma}$ orthogonal to $U^{\gamma}$, $h_{\ \mu}^{\sigma}j^{\mu}$,
contributes to $F_{{\rm Mag}}^{\alpha}$. Physically, $h_{\ \mu}^{\sigma}j^{\mu}$
is the spatial charge current density \emph{as measured} \emph{in
the particle's rest frame}. In such frame, the time component of $F_{{\rm Mag}}^{\alpha}$
vanishes, and the space components read 
\begin{equation}
\vec{F}_{{\rm Mag}}=2\pi\vec{\mu}\times\vec{j}\ .\label{eq:FMagvecEM}
\end{equation}
This is a force orthogonal to $\vec{\mu}$ and to the spatial current
density $\vec{j}$, which we dub electromagnetic ``Magnus'' force.
If the magnetic dipole consists of a spinning, positively (and uniformly)
charged body, so that $\vec{\mu}\parallel\vec{S}$, the force $\vec{F}_{{\rm Mag}}$
has a direction opposite to the Magnus force of fluid dynamics (so
it is actually ``\emph{anti-Magnus''}). If the body is negatively
charged, so that $\vec{\mu}\parallel-\vec{S}$, the force points in
the same direction of a Magnus force.

\subsection{Example: The force exerted by a current slab on a dipole\label{sub:A-current-slab}}

\begin{figure}
\includegraphics[width=1\columnwidth]{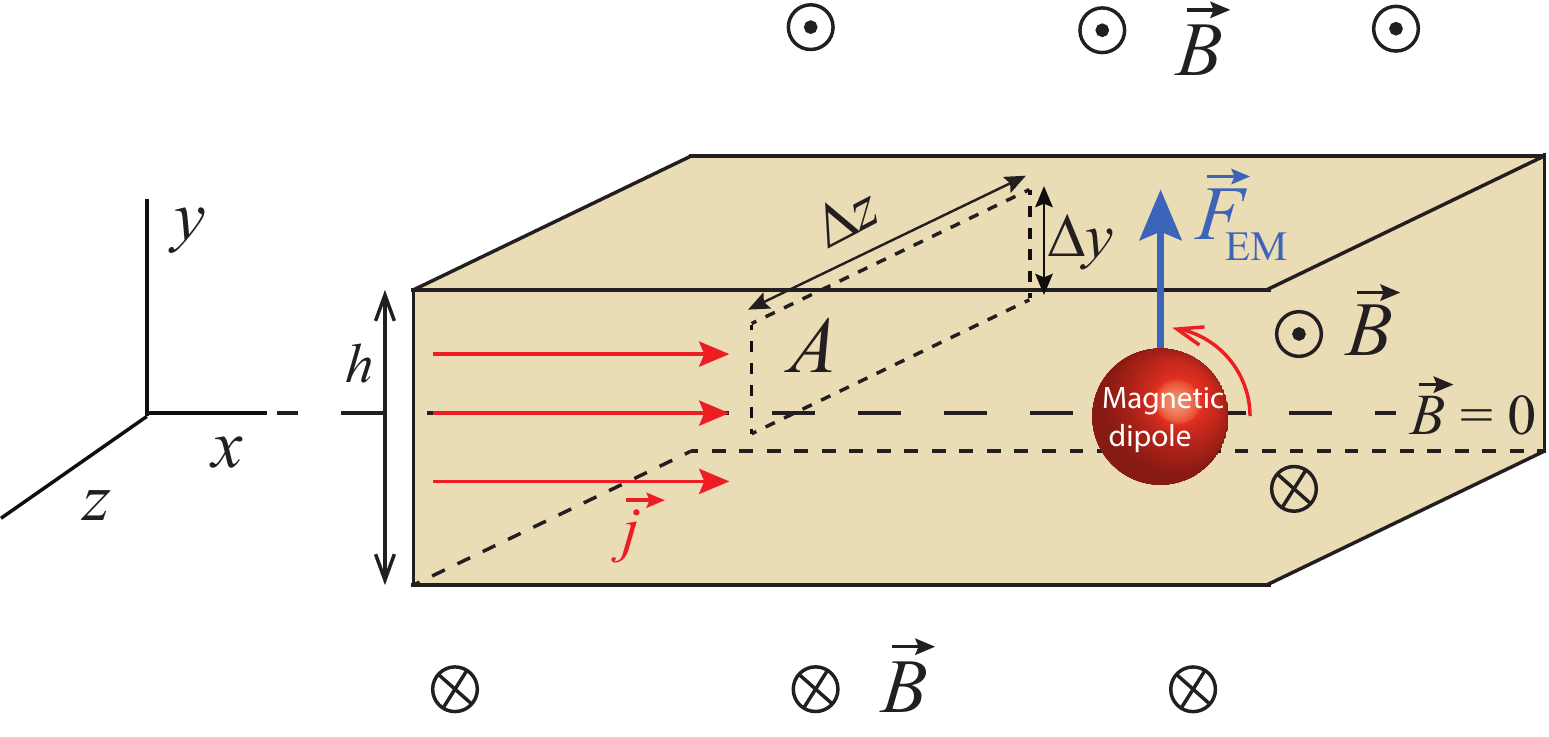}\protect\protect\caption{\label{fig:AntimagnusEM}A magnetic dipole $\vec{\mu}=\mu\vec{e}_{z}$
inside a semi-infinite cloud of charged particles flowing in the $\vec{e}_{x}$
direction. The cloud is infinite along the $x$ and $z$ directions,
but of finite thickness in the $y$ direction, contained within $-h/2\le y\le+h/2$.
The magnetic field $\vec{B}$ generated by the cloud points in the
positive $z$ direction for $y>0$, and in the negative $z$ direction
for $y<0$. $\vec{B}$ has a gradient inside the cloud, whose only
nonvanishing component is $B^{z,y}=4\pi j$. Due to that, a force
$\vec{F}_{{\rm EM}}=B^{z,y}\mu\vec{e}_{y}=4\pi j\mu\vec{e}_{y}$,
pointing \emph{upwards}, is exerted on the dipole. In this case $\vec{F}_{{\rm {\rm Sym}}}=\vec{F}_{{\rm Mag}}$,
so the force is twice the Magnus force: $\vec{F}_{{\rm EM}}=\vec{F}_{{\rm Sym}}+\vec{F}_{{\rm Mag}}=2\vec{F}_{{\rm Mag}}$.
Considering instead a cloud finite along $z$, infinite along $x$
and $y$, $\vec{F}_{{\rm Mag}}=2\pi j\mu\vec{e}_{y}$ remains the
same, but $\vec{F}_{{\rm Sym}}$ inverts its direction: $\vec{F}_{{\rm {\rm Sym}}}=-\vec{F}_{{\rm Mag}}$,
causing the total force to vanish: $\vec{F}_{{\rm EM}}=0$.}
\end{figure}

The induction component $F_{{\rm ind}}^{\alpha}$ has no gravitational
counterpart, as we shall see. Therefore, from now onwards we will
not consider it any further. To shed light on the components $F_{{\rm Mag}}^{\alpha}$
and $F_{{\rm Sym}}^{\alpha}$, we consider a simple stationary setup
(Exercise 5.14 of Ref.~\cite{GriffithsBook}): a semi-infinite cloud
of charged gas which is infinitely long ($x$ direction) and wide
($z$ direction), but of finite thickness $h$ in the $y$ direction,
contained between the planes $y=h/2$ and $y=-h/2$, see Fig.~\ref{fig:AntimagnusEM}.

Outside the slab, the field is uniform and has opposite directions
in either side~\cite{GriffithsBook}. The field at any point inside
the cloud is readily obtained by application of the Stokes theorem
to the stationary Maxwell-Ampère equation 
\begin{equation}
\nabla\times\vec{B}=4\pi\vec{j}\ .\label{eq:MaxwellAmpere}
\end{equation}
That is, let $A$ be a rectangle in the $z-y$ plane, as illustrated
in Fig. \ref{fig:AntimagnusEM}, with boundary $\partial A$ and normal
unit vector $\vec{n}$. By the Stokes theorem 
\begin{equation}
\varoint_{\partial A}\vec{B}\cdot d\vec{l}=\varoint_{A}\nabla\times\vec{B}\cdot\vec{n}dA=4\pi\varoint_{A}\vec{j}\cdot\vec{n}dA=4\pi\Delta z\Delta yj\,,\label{eq:Stokes}
\end{equation}
where we took, for the surface $A$, the orientation $\vec{n}\parallel\vec{j}$.
By the right-hand-rule and symmetry arguments, $\vec{B}$ is parallel
to the slab and orthogonal to $\vec{j}$, pointing in the positive
$z$ direction for $y>0$, in the negative $z$ direction for $y<0$,
and vanishing at $y=0$. Therefore $\varoint_{\partial A}\vec{B}\cdot d\vec{l}=B|_{y=\Delta y}\Delta z$,
and so 
\begin{equation}
B^{z}(y)=4\pi\Delta yj=4\pi yj\ .\label{eq:B}
\end{equation}
Consider now a magnetic dipole at rest inside the cloud (for instance,
the magnetic dipole moment of a spinning charged body), as depicted
in Fig. \ref{fig:AntimagnusEM}. The magnetic field (\ref{eq:B})
has a gradient inside the cloud, leading to a magnetic tidal tensor
$B^{\alpha\beta}$ (as measured by the dipole) whose only nonvanishing
component is $B^{zy}=B^{z,y}=4\pi j$. Therefore, the force exerted
on the dipole is, cf. Eq. (\ref{eq:FEM_TT}), 
\begin{equation}
\vec{F}_{{\rm EM}}=B^{ji}\mu_{j}\vec{e}_{i}=B^{zy}\mu_{z}\vec{e}_{y}=4\pi j\mu_{z}\vec{e}_{y}\,.\label{eq:ForceEM}
\end{equation}
It consists of the sum of the Magnus force plus the force $\vec{F}_{{\rm Sym}}$
($\vec{F}_{{\rm ind}}=0$ since the configuration is stationary):
$\vec{F}_{{\rm EM}}=\vec{F}_{{\rm Mag}}+\vec{F}_{{\rm Sym}}$, 
\begin{align}
 & \vec{F}_{{\rm Mag}}=2\pi\vec{\mu}\times\vec{j}=2\pi j(\mu_{z}\vec{e}_{y}-\mu_{y}\vec{e}_{z})\label{eq:FmagEMslab}\\
 & \vec{F}_{{\rm Sym}}=B^{(ji)}\mu_{j}\vec{e}_{i}=2\pi j(\mu_{z}\vec{e}_{y}+\mu_{y}\vec{e}_{z})\label{eq:Fsymslab}
\end{align}
Equations (\ref{eq:ForceEM})-(\ref{eq:Fsymslab}) yield the forces
for a fixed orientation of the slab (orthogonal to the $y$-axis),
and an arbitrary $\vec{\mu}$. This is of course physically equivalent
to considering instead a magnetic dipole $\vec{\mu}$ with \emph{fixed
direction}, and varying the orientation of the slab; in this framework,
taking $\vec{\mu}=\mu\vec{e}_{z}$, two notable cases stand out: 
\begin{enumerate}
\item \label{enu:EMcaseuz}Slab finite along $y$ axis, infinite along $x$
and $z$ (see Fig. \ref{fig:AntimagnusEM}). The Magnus and the ``symmetric''
forces \emph{are equal}: $\vec{F}_{{\rm Mag}}=\vec{F}_{{\rm Sym}}=2\pi j\mu\vec{e}_{y}$,
so there is a total force along the $y$ direction equaling twice
the Magnus force: $\vec{F}_{{\rm EM}}=2\vec{F}_{{\rm Mag}}=4\pi j\mu\vec{e}_{y}$. 
\item \label{enu:EMcaseuy}Slab finite along $z$ axis, infinite along $x$
and $y$ (slab orthogonal to $\vec{\mu}$). The Magnus force remains
the same as in case \ref{enu:EMcaseuz}; but $\vec{F}_{{\rm Sym}}$
changes to the exact \emph{opposite}, $\vec{F}_{{\rm Sym}}=-2\pi j\mu\vec{e}_{y}=-\vec{F}_{{\rm Mag}}$.
The total force on the dipole now vanishes: $\vec{F}_{{\rm EM}}=0$. 
\end{enumerate}
The results in case \ref{enu:EMcaseuy} follow\footnote{Equivalently, they follow from rotating the frame in Fig. \ref{fig:AntimagnusEM}
by $-\pi/2$ about $\vec{e}_{x}$ {[}which amounts to swapping $y\leftrightarrow z$
and changing the signs of the right-hand members of Eqs. (\ref{eq:ForceEM})-(\ref{eq:Fsymslab}){]}
while \emph{still} \emph{demanding} $\vec{\mu}=\mu\vec{e}_{z}$.} from noting that, for a slab orthogonal to the $z$ axis, $\vec{B}$
is along $y$ and given by $\vec{B}=-4\pi jz\vec{e}_{y}$, leading
to a magnetic tidal tensor with only nonvanishing component $B^{yz}=B^{y,z}=-4\pi j$,
thus causing $B^{(ij)}$ to globally change sign comparing to case
\ref{enu:EMcaseuz}. For other orientations of the slab/dipole, the
forces $\vec{F}_{{\rm Sym}}$ and $\vec{F}_{{\rm Mag}}$ are not collinear.
When $\vec{\mu}$ coincides with an eigenvector of the matrix $B^{(ij)}$,
they are actually orthogonal; in the slab in Fig. \ref{fig:AntimagnusEM}
(orthogonal to the $y$-axis), that is the case for $\vec{\mu}=\mu(\vec{e}_{y}+\vec{e}_{z})/\sqrt{2}$
and $\vec{\mu}=\mu(\vec{e}_{z}-\vec{e}_{y})/\sqrt{2}$ (the third
eigenvector of $B^{(ij)}$, $\vec{\mu}=\mu\vec{e}_{x}$, has zero
eigenvalue and leads to $\vec{F}_{{\rm Mag}}=\vec{F}_{{\rm {\rm Sym}}}=0$).

Notice that neither the field $\vec{B}$ at any point \emph{inside}
the cloud, nor its gradient, or the force (\ref{eq:ForceEM}), depend
on the precise width $h$ of the cloud; in particular, they remain
the same in the limit $h\rightarrow\infty$. The role of considering
(at least in a first moment) a finite $h$ is to fix the direction
of $\vec{B}$. Equation (\ref{eq:MaxwellAmpere}), together with the
problem's symmetries, then fully fix $\vec{B}$ via Eq. (\ref{eq:Stokes})
and, therefore, $\vec{F}_{{\rm Sym}}$. Taking the limit $h\rightarrow\infty$
in cases \ref{enu:EMcaseuz}-\ref{enu:EMcaseuy} above yields two
different ways of constructing an infinite current cloud, each of
them \emph{leading to a different} $\vec{B}$ and force on the dipole
(the situation is analogous to the ``paradoxes'' of the electric
field of a uniform, infinite charge distribution, or of the Newtonian
gravitational field of a uniform, infinite mass distribution, see
Sec. \ref{sub:Infinite-clouds}). Had one started with a cloud about
which all one is told is that it is infinite in \emph{all directions},
it would not be possible to set up the boundary conditions needed
to solve the Maxwell-Ampère equation (\ref{eq:MaxwellAmpere}), so
the question of which is the magnetic field (thus the force the dipole)
would have no answer.\footnote{This indeterminacy is readily seen noting that, given a solution of
Eq. (\ref{eq:MaxwellAmpere}), adding to it any solution of the homogeneous
equation $\nabla\times\vec{B}=0$ yields another solution of Eq. (\ref{eq:MaxwellAmpere}).}

\subsection{Reciprocal problem: The force exerted by the dipole on the slab\label{sub:EM_Reciprocal-probem}}

There have been attempts at understanding and quantifying the gravitational
analogue of the Magnus effect~\cite{Okawa:2014sxa,Cashen:2016neh}.
However, in these works, the force on the spinning body was inferred
from its effect on the cloud, by guessing its back reaction on the
body. Here we will start by computing it rigorously in the electromagnetic
analogue, i.e., the reciprocal of the problem considered above: the
force exerted by the magnetic dipole on the cloud. It is given by
the integral 
\begin{equation}
\vec{F}_{{\rm dip,cloud}}=\int_{{\rm cloud}}\vec{j}\times\vec{B}_{{\rm dip}}d^{3}x=\vec{j}\times\int_{{\rm cloud}}\vec{B}_{{\rm dip}}d^{3}x\ .\label{eq:F_dip_cloud}
\end{equation}
Consider a sphere completely enclosing the magnetic dipole, and let
$R$ be its radius; we may then split 
\begin{equation}
\vec{F}_{{\rm dip,cloud}}=\vec{j}\times\int_{r\le R}\vec{B}_{{\rm dip}}d^{3}x+\vec{j}\times\int_{r>R}\vec{B}_{{\rm dip}}d^{3}x\label{eq:Fdip_cloud_Split}
\end{equation}
The interior integral yields 
\begin{equation}
\int_{r\le R}\vec{B}_{{\rm dip}}d^{3}x=\frac{8\pi}{3}\vec{\mu}\ ,\label{eq:InteriorIntegral}
\end{equation}
as explained in detail in pp. 187-188 of \cite{Jackson:1998nia}.
The magnetic field in any region \emph{exterior} to the dipole is
(e.g. \cite{GriffithsBook,Jackson:1998nia}) 
\begin{equation}
\vec{B}_{{\rm dip}}|_{r>R}=-\frac{\vec{\mu}}{r^{3}}+\frac{3(\vec{\mu}\cdot\vec{r})\vec{r}}{r^{5}}\ .\label{eq:Bdip}
\end{equation}
For the setup in Fig. \ref{fig:AntimagnusEM} (slab orthogonal to
the $y$-axis, $\vec{\mu}=\mu\vec{e}_{z}$), and considering a spherical
coordinate system where $z^{2}/r^{2}=\cos^{2}\theta$ and the plane
$y=h/2$ is given by the equation $r=h/(2\sin\theta\sin\phi)$, the
exterior integral becomes 
\begin{eqnarray}
 &  & \int_{r>R}\vec{B}_{{\rm dip}}d^{3}x=2\mu\vec{e}_{z}\nonumber \\
 & \times & \int_{0}^{\pi}d\theta\int_{0}^{\pi}d\phi\int_{R}^{\beta}dr\frac{3\cos^{2}\theta-1}{r}\sin\theta=\frac{4}{3}\pi\mu\vec{e}_{z}\,,\label{eq:IntegralBy}
\end{eqnarray}
with $\beta=h/(2\sin\theta\sin\phi)$. Substituting Eqs. (\ref{eq:InteriorIntegral})
and (\ref{eq:IntegralBy}) into (\ref{eq:Fdip_cloud_Split}), and
comparing to Eq. (\ref{eq:ForceEM}), we see that 
\begin{equation}
\vec{F}_{{\rm dip,cloud}}=-4\pi j\mu\vec{e}_{y}=-\vec{F}_{{\rm EM}}\equiv-\vec{F}_{{\rm cloud,dip}}\ ,\label{eq:ForcedipoleCloudy}
\end{equation}
i.e., the force exerted by the dipole on the cloud indeed equals \emph{minus}
the force exerted by the cloud on the dipole. It is however important
to note that this occurs because one is dealing here with \emph{magnetostatics};
for general electromagnetic interactions \emph{do not obey} the action-reaction
law (in the sense of a reaction force equaling minus the action).
This is exemplified in Appendix \ref{sub:3rdLawMagnetism}. In particular
it is so for the interaction of the dipole with individual particles
of the cloud.

If one considers instead a slab orthogonal to the $z$ axis (contained
within $-h/2\le z\le+h/2$), and noting that the plane $z=h/2$ is
given by $r=h/(2\cos\theta)$, one obtains $\int_{r>R}\vec{B}_{{\rm dip}}d^{3}x=-(8\pi/3)\mu\vec{e}_{z}$,
which exactly cancels out the interior integral (\ref{eq:InteriorIntegral}),
leading to a zero force on the cloud: $\vec{F}_{{\rm dip,cloud}}=0$
(matching, again, its reciprocal).

Just like in the reciprocal problem, the results do not depend on
the width $h$ of the slabs, so taking the limit $h\rightarrow\infty$
of the slabs orthogonal to $y$ and to $z$ are two different ways
of obtaining equally infinite clouds, but on which very different
forces are exerted. Here the issue does not boil down to a problem
of boundary conditions for PDE's (as was the case for $\vec{B}$ in
Sec. \ref{sub:A-current-slab}); it comes about instead in another
fundamental mathematical principle (Fubini's theorem \cite{Fubini,PeterWalker}):
the multiple integral of a function which is not \emph{absolutely
convergent}, depends in general on the way the integration is performed.
This is discussed in detail in Appendix \ref{sec:Fubini_Theorem}.
It tells us that, like its reciprocal, $\vec{F}_{{\rm dip,cloud}}$
\emph{is not a well defined quantity for an infinite cloud}.

\section{Gravitational Magnus effect\label{sec:Gravitational-Magnus-effect}}

Contrary to idealized point (``monopole'') particles, ``real,''
extended bodies, endowed with a multipole structure, do not move along
geodesics in a gravitational field. This is because the curvature
tensor couples to the multipole moments of the body's energy momentum
tensor $T^{\alpha\beta}$ (much like in the way the electromagnetic
field couples to the multipole moments of the current 4-vector $j^{\alpha}$).
In a multiple scheme, the first correction to geodesic motion arises
when one considers pole-dipole spinning particles, i.e., particles
whose only multipole moments of $T^{\alpha\beta}$ relevant to the
equations of motion are the momentum $P^{\alpha}$ and the spin tensor
$S^{\alpha\beta}$ (see e.g. \cite{Dixon1964,Dixon:1970zza,Costa:2014nta}
for their definitions in a curved spacetime). In this case the equations
of motion that follow from the conservation laws $T_{\ \ \ ;\beta}^{\alpha\beta}=0$
are the so-called Mathisson-Papapetrou (or Mathisson-Papapetrou-Dixon)
equations~\cite{Mathisson:1937zz,Papapetrou:1951pa,Dixon1964,Dixon:1970zza,Gralla:2010xg,Tulczyjew}.
According to these equations, a spinning body experiences a force,
the so called \emph{spin-curvature force}, when placed in a gravitational
field. It is described by 
\begin{eqnarray}
\frac{DP^{\alpha}}{d\tau} & = & -\frac{1}{2}R_{\ \beta\mu\nu}^{\alpha}S^{\mu\nu}U^{\beta}\ \equiv\ F^{\alpha}\,,\label{eq:ForceDS}
\end{eqnarray}
where $U^{\alpha}=dx^{\alpha}/d\tau$ is the body's 4-velocity (that
is, the tangent vector to its center of mass worldline). This is a
physical, covariant force (as manifest in the covariant derivative
operator $D/d\tau\equiv U^{\alpha}\nabla_{\alpha}$), which causes
the body to deviate from geodesic motion. Under the so-called Mathisson-Pirani
\cite{Mathisson:1937zz,Pirani:1956tn} spin condition $S^{\alpha\beta}U_{\beta}=0$,
one may write $S^{\mu\nu}=\epsilon^{\mu\nu\tau\lambda}S_{\tau}U_{\lambda}$,
where $S^{\alpha}\equiv\epsilon_{\ \beta\mu\nu}^{\alpha}U^{\beta}S^{\mu\nu}/2$
is the spin 4-vector {[}whose components in an \emph{orthonormal}
frame comoving with the body are $S^{\alpha}=(0,\vec{S})${]}. Substituting
in (\ref{eq:ForceDS}), leads to~\cite{Costa:2012cy,FilipeCosta:2006fz},
\begin{equation}
F^{\alpha}=\frac{DP^{\alpha}}{d\tau}=-\mathbb{H}^{\beta\alpha}S_{\beta}\,,\label{eq:FG_MP}
\end{equation}
where 
\begin{equation}
\mathbb{H}_{\alpha\beta}\equiv\star R_{\alpha\mu\beta\nu}U^{\mu}U^{\nu}=\frac{1}{2}\epsilon_{\alpha\mu}^{\ \ \ \ \lambda\tau}R_{\lambda\tau\beta\nu}U^{\mu}U^{\nu}\,,\label{eq:Hab}
\end{equation}
is the ``gravitomagnetic tidal tensor'' (or ``magnetic part''
of the Riemann tensor, e.g. \cite{Dadhich:1999eh}) as measured by
an observer comoving with the particle. Using the decomposition of
the Riemann tensor in terms of the Weyl ($C_{\alpha\beta\gamma\delta}$)
and Ricci tensors (e.g. Eq. (2.79) of Ref.~\cite{EllisMaartensMacCallum}),
\begin{align}
R_{\ \ \gamma\delta}^{\alpha\beta} & =C_{\ \ \gamma\delta}^{\alpha\beta}+2\delta_{[\gamma}^{[\alpha}R^{\beta]}{}_{\delta]}-\frac{1}{3}R\delta_{[\gamma}^{\alpha}\delta_{\delta]}^{\beta}\ ,\label{Riemann decomp-R}
\end{align}
we may decompose $\mathbb{H}_{\alpha\beta}$ as 
\begin{equation}
\mathbb{H}_{\alpha\beta}=\mathbb{H}_{(\alpha\beta)}+\mathbb{H}_{[\alpha\beta]}=\mathcal{H}_{\alpha\beta}+\frac{1}{2}\epsilon_{\alpha\beta\sigma\gamma}U^{\gamma}R^{\sigma\lambda}U_{\lambda}\ ,\label{eq:H_ab_Decomp-1}
\end{equation}
where the symmetric tensor $\mathcal{H}_{\alpha\beta}=\mathbb{H}_{(\alpha\beta)}$
is the \emph{magnetic part of the Weyl tensor}, $\mathcal{H}_{\alpha\beta}\equiv\star C_{\alpha\mu\beta\nu}U^{\mu}U^{\nu}$.
Using the Einstein field equations 
\begin{equation}
R_{\mu\nu}=8\pi(T_{\mu\nu}-\frac{1}{2}g_{\mu\nu}T_{\,\,\,\alpha}^{\alpha})+\Lambda g_{\alpha\beta}\ ,\label{eq:EinsteinField}
\end{equation}
this becomes (cf. e.g. Eq. (I.3b) of Table I of \cite{Costa:2012cy})
\begin{equation}
\mathbb{H}_{\alpha\beta}=\mathbb{H}_{(\alpha\beta)}+\mathbb{H}_{[\alpha\beta]}=\mathcal{H}_{\alpha\beta}-4\pi\epsilon_{\alpha\beta\sigma\gamma}U^{\gamma}J^{\sigma}\,,\label{eq:H_ab_Decomp}
\end{equation}
where $J^{\alpha}\equiv-T^{\alpha\beta}U_{\beta}$ is the mass/energy
current 4-vector as measured by an observer of 4-velocity $U^{\alpha}$
(comoving with the particle, in this case). We thus can write 
\begin{equation}
F^{\alpha}=-\mathcal{H}^{\alpha\beta}S_{\beta}+4\pi\epsilon_{\ \beta\sigma\gamma}^{\alpha}J^{\beta}S^{\sigma}U^{\gamma}=F_{{\rm Weyl}}^{\alpha}+F_{{\rm Mag}}^{\alpha}\ ,\label{eq:Fdecomp}
\end{equation}
where 
\begin{align}
F_{{\rm Mag}}^{\alpha} & \equiv4\pi\epsilon_{\ \beta\sigma\gamma}^{\alpha}U^{\gamma}J^{\beta}S^{\sigma}\ ,\label{eq:FMagGrav}\\
F_{{\rm Weyl}}^{\alpha} & \equiv-\mathcal{H}^{\alpha\beta}S_{\beta}\ .\label{eq:FWeyl}
\end{align}
Since the tensor $\epsilon_{\alpha\beta\sigma\gamma}U^{\gamma}$ automatically
projects spatially (in any of its free indices), only the projection
of $J^{\beta}$ orthogonal to $U^{\gamma}$, $h_{\ \mu}^{\beta}J^{\mu}$
{[}see Eq. (\ref{eq:Spaceprojector}){]}, contributes to $F_{{\rm Mag}}^{\alpha}$.

Equations (\ref{eq:Fdecomp})-(\ref{eq:FWeyl}) thus tell us that
the spin-curvature force splits into two parts: $F_{{\rm Weyl}}^{\alpha}$,
which is due to the magnetic part of the Weyl tensor, and is analogous
(to some extent) to the ``symmetric force'' $F_{{\rm Sym}}^{\alpha}$
of electromagnetism, Eq. (\ref{eq:FSym_Find}). The second part is
$F_{{\rm Mag}}^{\alpha}$ which is nonvanishing whenever, relative
to the body, there is a \emph{spatial} mass-energy current $h_{\ \mu}^{\beta}J^{\mu}$
not parallel to $S^{\alpha}$. In the body's rest frame, we have 
\begin{equation}
\vec{F}_{{\rm Mag}}=4\pi\vec{J}\times\vec{S}\,,\label{eq:FMagvec}
\end{equation}
thus $F_{{\rm Mag}}^{\alpha}$ is what one would call a gravitational
analogue of the Magnus effect in fluid dynamics, since

\noindent \textbf{i)} it arises whenever the body rotates and moves
in a medium with a relative velocity not parallel to its spin axis
(that is, when there is a spatial mass-energy current density $\vec{J}$
\emph{relative to the body}, such that $\vec{S}\nparallel\vec{J}$);

\noindent \textbf{ii)} the force is orthogonal to both the axis of
rotation of the body (i.e. to $\vec{S}$) and to the current density
$\vec{J}$, like in an ordinary Magnus effect; moreover, it points
precisely in \emph{the same direction} of the latter.\footnote{This is always so if the parallelism drawn is between $\vec{J}$ and
the flux vector of fluid dynamics. If the analogy is based instead
on the velocity of the fluid relative to the body, the gravitational
and ordinary Magnus effects have the same direction for a perfect
fluid obeying the weak energy condition (cf. Sec. \ref{sub:FLWR}
and Eq. (\ref{eq:FLWRCov}) below), but otherwise it is not necessarily
so (e.g., an imperfect fluid conducting heat gives rise to mass currents
nonparallel to the fluid's velocity).}

Notice that Eq. (\ref{eq:Fdecomp}) is a fully general equation that
can be applied to any system, and that the ``Magnus force'' $F_{{\rm Mag}}^{\alpha}$
depends only on $U^{\alpha}$, $S^{\alpha}$, and the local $J^{\alpha}$,
and not on any further detail of the system. The force $F_{{\rm Weyl}}^{\alpha}$,
by contrast, strongly depends on the details of the system (as exemplified
in Sec. \ref{sub:Toy-model-revisited} below). This can be traced
back to the fact that $F_{{\rm Mag}}^{\alpha}$ comes from the Ricci
part of the curvature, totally fixed by the energy-momentum tensor
$T^{\alpha\beta}$ of the \emph{local} sources via the Einstein Eqs.
(\ref{eq:EinsteinField}), whereas the Weyl tensor describes the ``free
gravitational field,'' which does not couple to the sources via algebraic
equations, only through \emph{differential} ones (the differential
Bianchi identities \cite{MaartensBasset1997,EllisMaartensMacCallum}),
being thus determined not by the value of $T^{\alpha\beta}$ at a
point, but by conditions elsewhere \cite{EllisMaartensMacCallum}.

Note also that, in general, $\vec{F}_{{\rm Mag}}$ is not the total
force in the direction orthogonal to $\vec{S}$ and $\vec{J}$; $F_{{\rm Weyl}}^{\alpha}$
may also have a component along it\footnote{Its behavior however (unlike the part $F_{{\rm Mag}}^{\alpha}$) is
not what one would expect from a gravitational analogue of the Magnus
effect, namely (a) it is not determined, nor does it depend on $\vec{J}$
and $\vec{S}$ in the way one would expect from a Magnus effect and
(b) it is not necessarily nonzero when $\vec{J}\times\vec{S}\ne0$.
For this reason we argue that only $F_{{\rm Mag}}^{\alpha}$ should
be cast as a gravitational ``Magnus effect.''}.

\subsection{Post-Newtonian approximation\label{sub:Post-Newtonian-approximation}}

Up to here, we used no approximations in the description of the gravitational
forces. For most astrophysical systems, however, no exact solutions
of the Einstein field equations are known; in these cases we use the
post-Newtonian (PN) approximation to general relativity. This expansion
can be framed in different --- equivalent --- ways; namely, by counting
powers of $c$ \cite{Damour:1990pi,WillPoissonBook} or in terms of
a dimensionless parameter~\cite{Misner:1974qy,Costa:2016iwu,Will:1993ns,Kaplan:2009,Binietal1994}.
Here we will follow the latter, which consists of making an expansion
in terms of a small \emph{dimensionless} parameter $\epsilon$, such
that $U\sim\epsilon^{2}$ and $v\lesssim\epsilon$, where $U$ is
(minus) the Newtonian potential, and $v$ is the velocity of the bodies
(notice that, for bodies in bounded orbits, $v\sim\sqrt{U}$). In
terms of ``forces,'' the Newtonian force $m\nabla U$ is taken to
be of zeroth PN order (0PN), and each factor $\epsilon^{2}$ amounts
to a unity increase of the PN order. Time derivatives increase the
degree of smallness of a quantity by a factor $\epsilon$; for example,
$\partial U/\partial t\sim Uv\sim\epsilon U$. The 1PN expansion consists
of keeping terms up to $\keeping\Os{4}$ in the equations of motion~\cite{Will:1993ns}.
This amounts to considering a metric of the form \cite{Damour:1990pi,Kaplan:2009}
\begin{align}
g_{00} & =-1+2w-2w^{2}+\Os{6}\nonumber \\
g_{i0} & =\mathcal{A}_{i}+\Os{5};\qquad g_{ij}=\delta_{ij}\left(1+2U\right)+\Os{4}\,,\label{eq:PNmetric}
\end{align}
where $\vec{\mathcal{A}}$ is the ``gravitomagnetic vector potential''
and the scalar $w$ consists of the sum of $U$ plus \emph{nonlinear}
terms of order $\epsilon^{4}$, $w=U+\Os{4}$. For the computation
of the space part of the force (\ref{eq:FG_MP}), the components $\mathbb{H}_{ij}$,
$\mathbb{H}_{0i}$, of the gravitomagnetic tidal tensor (\ref{eq:Hab})
are needed. Using $U^{\alpha}=U^{0}(1,\vec{v})$ and the 1PN Christoffel
symbols in e.g. Eqs. (8.15) of Ref.~\cite{WillPoissonBook},\footnote{Identifying, in the notation therein, $w\rightarrow U+\Psi$, $\mathcal{A}_{i}\rightarrow-4U_{i}$.}
they read 
\begin{align}
\mathbb{H}_{ij}= & -\frac{1}{2}\epsilon_{i}^{\ lk}\mathcal{A}_{k,lj}-\epsilon_{ij}^{\ \ k}\dot{U}_{,k}+2\epsilon_{i}^{\ km}v_{k}U_{,jm}\nonumber \\
 & -\epsilon_{ij}^{\ \ m}U_{,km}v^{k}+\O{5}{2}\,,\label{eq:HijPN}\\
\mathbb{H}_{0i}= & \epsilon_{ij}^{\ \ l}\dot{U}_{,l}v^{j}+\frac{1}{2}\epsilon_{j}^{\,\,\,lk}\mathcal{A}_{k,li}v^{j}+\epsilon_{\ ji}^{l}U_{,lk}v^{j}v^{k}\ (=\O{4}{2})\,,\label{eq:Hi0PN}
\end{align}
where dot denotes ordinary time derivative, $\partial/\partial t$.
Equation (\ref{eq:HijPN}) is a generalization of Eq. (3.41) of Ref.~\cite{Damour:1990pi}
for nonvacuum, and for the general case that the observer measuring
the tensor $\mathbb{H}_{\alpha\beta}$ moves (i.e., $\vec{v}\ne0$).
It is useful to write $\mathbb{H}_{ij}$ in terms of the gravitoelectric
($\vec{G}$) and gravitomagnetic ($\vec{H}$) fields, defined by~\cite{Costa:2016iwu,Kaplan:2009,Damour:1990pi}
\begin{equation}
\vec{G}=\nabla w-\dot{\vec{\mathcal{A}}}+\O{6}{1}\,,\quad\vec{H}=\nabla\times\vec{\mathcal{A}}+\O{5}{1}\ .\label{eq:GEMfieldsPN}
\end{equation}
The reason for these denominations is that these fields play in gravity
a role analogous to the electric and magnetic fields.\footnote{Namely comparing the geodesic equation $d^{2}x^{i}/dt^{2}=F_{{\rm I}}^{i}/m$
{[}$F_{{\rm I}}^{i}$ given by Eq. (\ref{eq:FI}){]} with the Lorentz
force, and comparing Einstein's field equations in e.g. Eqs. (3.22)
of Ref.~\cite{Damour:1990pi} with the Maxwell equations.} One has then 
\begin{align}
\mathbb{H}_{ij} & =-\frac{1}{2}H_{i,j}-\epsilon_{ijk}\dot{G^{k}}+2\epsilon_{i}^{\ km}v_{k}G_{j,m}\nonumber \\
 & -\epsilon_{ij}^{\ \ m}G_{k,m}v^{k}+\O{5}{2}\ .\label{eq:HijGEMPN}
\end{align}
Noting that the orthogonality relation $S_{\alpha}U^{\alpha}=0$ implies
$S_{0}=-S_{i}v^{i}=\Os{1}$, it follows, from Eqs. (\ref{eq:FG_MP})
and (\ref{eq:Hi0PN}), that $F^{j}=-\mathbb{H}^{ij}S_{i}-\mathbb{H}^{0j}S_{0}=-\mathbb{H}^{ij}S_{i}+\Os{5}$,
and so the 1PN spin-curvature force reads
\begin{align}
F^{j} & =\frac{1}{2}H^{i,j}S_{i}-(\vec{S}\times\dot{\vec{G}})^{j}-2\epsilon^{ikm}v_{k}G_{\ ,m}^{j}S_{i}\nonumber \\
 & -\epsilon^{jim}G_{k,m}v^{k}S_{i}+\Os{5}\ .\label{eq:FGPN}
\end{align}
Its Magnus and Weyl components, Eqs. (\ref{eq:FMagGrav})-(\ref{eq:FWeyl}),
are
\begin{align}
F_{{\rm Mag}}^{i}= & \ 4\pi\epsilon_{\ jk}^{i}S^{k}(T^{0j}-\rho v^{j})+\Os{5}\ ,\label{eq:FMagPN}\\
F_{{\rm Weyl}}^{i}= & \ -\mathcal{H}^{ij}S_{j}+\Os{5}\label{eq:FWeylPN0}\\
= & \ \frac{1}{2}H^{(i,j)}S_{j}-2\epsilon_{\ km}^{(i}G^{j),m}v^{k}S_{j}+\Os{5}\ ;\label{eq:FWeylPN}
\end{align}
where for $\rho$ one can take the mass/energy density as measured
either in the body's rest frame, or in the PN background frame (the
distinction is immaterial in Eq. (\ref{eq:FMagPN}), to the accuracy
at hand). Notice that $T^{0j}-\rho v^{j}=h_{\ \beta}^{j}J^{\beta}$
{[}see Eq. (\ref{eq:Spaceprojector}){]} is indeed the spatial mass-energy
current with respect to the body's rest frame ($T^{0j}$, in turn,
yields the spatial mass-energy current as measured\emph{ }in the PN
frame).

To obtain the coordinate acceleration of a spinning test body, we
first note that, under the Mathisson-Pirani spin condition, the relation
between the particle's 4-momentum $P^{\alpha}$ and its 4-velocity
is (e.g. \cite{Costa:2014nta}) $P^{\alpha}=mU^{\alpha}+S^{\alpha\beta}a_{\beta}$,
where $m\equiv-P^{\alpha}U_{\alpha}$ is the proper mass, which is
a constant, $a^{\alpha}\equiv DU^{\alpha}/d\tau$ the \emph{covariant}
acceleration, and the term $S^{\alpha\beta}a_{\beta}$ is the so-called
``hidden momentum'' \cite{Gralla:2010xg,Costa:2012cy,Costa:2014nta}.
In the post-Newtonian regime one can neglect\footnote{That actually amounts to pick, among the infinite solutions allowed
by the (degenerate) Mathisson-Pirani spin condition, the ``non-helical''
one (avoiding the spurious helical solutions) \cite{Costa:2012cy,Costa:2017kdr}.
\emph{For such solution}, the acceleration comes, at leading order,
from the force $F^{\alpha}$; and so the term $D(S^{\alpha\beta}a_{\beta})/d\tau$
is always of higher PN order than $F^{\alpha}$ (for details, see
Sec. 3.1 of the Supplement in \cite{Costa:2012cy}). It is also quadratic
in spin, and, as such, arguably to be neglected at pole-dipole order
\cite{Tulczyjew,Costa:2012cy,Costa:2017kdr}.} the hidden momentum, leading to the acceleration equation $ma^{\alpha}\simeq DP^{\alpha}/d\tau\equiv F^{\alpha}$.
Using $DU^{\alpha}/d\tau=d^{2}x^{\alpha}/d\tau^{2}+\Gamma_{\beta\gamma}^{\alpha}U^{\beta}U^{\gamma}$
and $d/d\tau=(dt/d\tau)d/dt$, where $t$ is the coordinate time,
one gets, after some algebra, 
\begin{equation}
m\frac{d^{2}x^{i}}{dt^{2}}=F_{{\rm I}}^{i}+F^{i}+\O{5}{1}\ ,\label{eq:CoordAccelForce}
\end{equation}
where 
\begin{equation}
F_{{\rm I}}^{i}=m\left[\frac{dx^{i}}{dt}\Gamma_{\beta\gamma}^{0}-\Gamma_{\beta\gamma}^{i}\right]\frac{dx^{\beta}}{dt}\frac{dx^{\gamma}}{dt}
\end{equation}
is the inertial ``force'' already present in the geodesic equation
for a \emph{nonspinning} \emph{point} particle: $d^{2}x^{i}/dt^{2}=F_{{\rm I}}^{i}/m$
(cf. e.g. Eq. (8.14) of \cite{WillPoissonBook}). Using, again, the
1PN Christoffel symbols in Eqs. (8.15) of \cite{WillPoissonBook},
yields 
\begin{equation}
\frac{\vec{F}_{{\rm I}}}{m}=(1+v^{2}-2U)\vec{G}+\vec{v}\times\vec{H}-3\dot{U}\vec{v}-4(\vec{G}\cdot\vec{v})\vec{v}+\O{6}{1}\ .\label{eq:FI}
\end{equation}
Equation (\ref{eq:CoordAccelForce}) is a general expression for the
\emph{coordinate} acceleration of a spinning particle in a gravitational
field, accurate to 1PN order.

\subsection{A cloud ``slab''\label{sub:Toy-model-revisited}}

\subsubsection{The Magnus force on spinning objects}

\begin{figure}
\includegraphics[width=1\columnwidth]{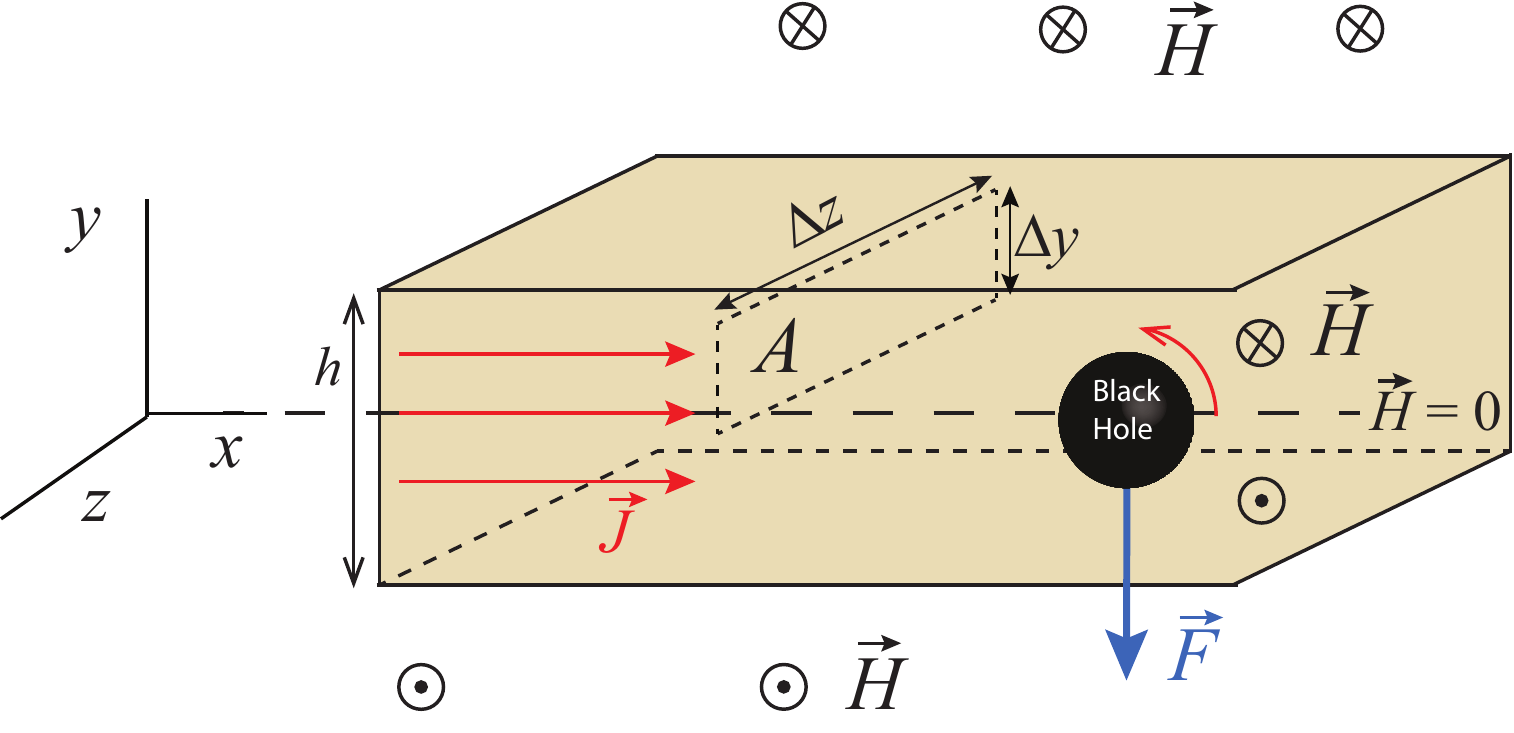}\protect\protect\protect\protect\protect\caption{\label{fig:MagnusGrav}A spinning body (e.g., a BH), with $\vec{S}=S\vec{e}_{z}$,
inside a massive cloud flowing in the $\vec{e}_{x}$ direction. The
cloud is infinite along $x$ and $z$, and finite along the $y$-axis,
contained within $-h/2\le y\le h/2$. The vector $\vec{H}$ is the
gravitomagnetic field generated by the cloud; it points in the negative
(positive) $z$ direction for $y>0$ ($<0$). It has a gradient inside
the cloud, whose only nonvanishing component is $H^{z,y}=-16\pi J$;
due to that, a spin-curvature force $\vec{F}=(1/2)H^{z,y}S\vec{e}_{y}=-8\pi JS\vec{e}_{y}$,
pointing \emph{downwards}, is exerted on the body. In this case $\vec{F}_{{\rm Mag}}=\vec{F}_{{\rm Weyl}}$,
so the total force is twice the Magnus force: $\vec{F}=\vec{F}_{{\rm Mag}}+\vec{F}_{{\rm Weyl}}=2\vec{F}_{{\rm Mag}}$.
Considering instead a cloud finite along $z$, and infinite along
$x$ and $y$, $\vec{F}_{{\rm Mag}}=-4\pi JS\vec{e}_{y}$ remains
the same, but $\vec{F}_{{\rm Weyl}}$ changes to the exact opposite:
$\vec{F}_{{\rm Weyl}}=-\vec{F}_{{\rm Mag}}$, causing the total spin-curvature
force to vanish: $\vec{F}=0$.}
\end{figure}

Before moving on to more realistic scenarios, we start by investigating
the gravitational Magnus force, in the PN approximation, for the gravitational
analogue of the electromagnetic system in Sec.~\ref{sub:A-current-slab}.
In particular, we consider a spinning body (for example, a BH) inside
a medium flowing in the $\vec{e}_{x}$ direction, that we assume to
be infinitely long and wide (in the $x$ and $z$ directions), but
of finite thickness $h$ ($y$ direction), contained within the planes
$-h/2\le y\le h/2$. The system is depicted in Fig. \ref{fig:MagnusGrav}.

The Einstein field equations yield a gravitational analogue to the
Maxwell-Ampère law (\ref{eq:MaxwellAmpere}), as we shall now see.
For the metric (\ref{eq:PNmetric}), the Ricci tensor component $R_{0i}=(\nabla\times H)_{i}/2-2\dot{\vec{G}}+\O{5}{2}$,
where $\vec{H}$ is the gravitomagnetic field as defined by Eq. (\ref{eq:GEMfieldsPN}).
On the other hand, from the Einstein equations (\ref{eq:EinsteinField}),
we have that $R_{0i}=8\pi T_{0i}+\O{5}{2}$. Equating the two expressions,
and taking the special case of stationary setups, we have (cf. e.g.
Eq. (2.6d) of Ref.~\cite{Kaplan:2009}) 
\begin{equation}
\nabla\times\vec{H}=-16\pi\vec{J}\ ,\label{eq:CurlHstationary}
\end{equation}
where we noted that $T^{0i}=J^{i}+\O{5}{2}$, and $J^{\alpha}=-T^{\alpha\beta}u_{\beta}$
is the mass/energy current as measured by the reference observers
$u^{\alpha}=u^{0}\delta_{0}^{\alpha}$ {[}at rest in the coordinate
system of (\ref{eq:PNmetric}){]}. This equation resembles very closely
Eq. (\ref{eq:MaxwellAmpere}). For a system analogous to that in Fig.~\ref{fig:AntimagnusEM}
--- a cloud of matter passing through a spinning body --- an entirely
analogous reasoning to that leading to Eq. (\ref{eq:B}) applies here
to obtain the gravitomagnetic field 
\begin{equation}
\vec{H}=H^{z}(y)\vec{e}_{z}=-16\pi yJ\vec{e}_{z}\,.\label{eq:HSlab}
\end{equation}
This solution is formally similar to the magnetic field in Eq. (\ref{eq:B}),
apart from the different factor and sign. For a spinning body at rest
in a stationary gravitational field, the spin-curvature force, Eq.
(\ref{eq:FGPN}), reduces to 
\begin{equation}
F^{i}=\frac{1}{2}H^{j,i}S_{j}\qquad\Leftrightarrow\qquad\vec{F}=\dfrac{1}{2}\nabla(\vec{H}\cdot\vec{S})\,,\label{eq:FGStationary}
\end{equation}
(cf. e.g. Eq. (4) of \cite{Wald:1972sz}), similar to the dipole force
(\ref{eq:ForceEM}). Hence, due to the gradient of $\vec{H}$, whose
only nonvanishing component $H^{i,j}$ is $H^{z,y}=-16\pi J$, a force
$\vec{F}$ is exerted on a spinning body at rest inside the cloud,
given by 
\begin{equation}
\vec{F}=-8\pi JS_{z}\vec{e}_{y}\ .\label{eq:FGcloud}
\end{equation}
It is thus along the $y$ direction, pointing downwards, in the \emph{same
direction} of an ordinary Magnus effect (and opposite to the electromagnetic
analogue). This force consists of the sum of the Magnus force plus
the Weyl force: $\vec{F}=\vec{F}_{{\rm Mag}}+\vec{F}_{{\rm Weyl}}$,
\begin{align}
 & \vec{F}_{{\rm Mag}}=4\pi\vec{J}\times\vec{S}=4\pi J(S_{y}\vec{e}_{z}-S_{z}\vec{e}_{y})\,,\label{eq:FMagGravSlab}\\
 & \vec{F}_{{\rm Weyl}}=-\mathcal{H}^{ij}\mu_{j}\vec{e}_{i}=-4\pi J(S_{z}\vec{e}_{y}+S_{y}\vec{e}_{z})\,.\label{eq:FWeylGravSlab}
\end{align}
Here, $\mathcal{H}^{ij}=\mathbb{H}^{(ij)}=-H^{(i,j)}/2$, and its
nonvanishing components are $\mathcal{H}^{zy}=\mathcal{H}^{yz}=4\pi J$.
Again, Eqs. (\ref{eq:FGcloud})-(\ref{eq:FWeylGravSlab}) yield the
forces for a fixed orientation of the slab (orthogonal to the $y$-axis),
and an arbitrary $\vec{S}$. Of course, this is physically equivalent
to considering instead a body with \emph{fixed} \emph{spin direction},
and varying the orientation of the slab; choosing $\vec{S}=S\vec{e}_{z}$,
one can make formally similar statements to those in Sec. \ref{sub:A-current-slab},
by replacing $\vec{F}_{{\rm Sym}}$ by $\vec{F}_{{\rm Weyl}}$. Namely,
the two notable cases arise: 
\begin{enumerate}
\item \label{enu:GravcaseSz}Cloud finite along the $y$-axis, infinite
along $x$ and $z$ (Fig. \ref{fig:MagnusGrav}). The Magnus force
$\vec{F}_{{\rm Mag}}$ equals the Weyl force: $\vec{F}_{{\rm Mag}}=\vec{F}_{{\rm Weyl}}=-4\pi JS\vec{e}_{y}$,
so there is a total force downwards which is twice the Magnus force:
$\vec{F}=2\vec{F}_{{\rm Mag}}=-8\pi JS\vec{e}_{y}$. 
\item \label{enu:GravcaseSy}Cloud finite along $z$, infinite along $x$
and $y$ (i.e., slab orthogonal to $\vec{S}$). The Magnus force $\vec{F}_{{\rm Mag}}$
remains the same as in case \ref{enu:GravcaseSz}; the Weyl force
is now exactly \emph{opposite} to the Magnus force: $\vec{F}_{{\rm Weyl}}=4\pi JS\vec{e}_{y}=-\vec{F}_{{\rm Mag}}$,
so the total spin-curvature force vanishes: $\vec{F}=\vec{F}_{{\rm Mag}}+\vec{F}_{{\rm Weyl}}=0$. 
\end{enumerate}
In case \ref{enu:GravcaseSy} we noted that, for a slab orthogonal
to the $z$ axis, $\vec{H}=16\pi Jz\vec{e}_{y}$, and so the magnetic
part of the Weyl tensor $\mathcal{H}^{ij}$ \emph{changes sign} comparing
to the setup in Fig. \ref{fig:MagnusGrav}: $\mathcal{H}^{zy}=\mathcal{H}^{yz}=-4\pi J$.
For other orientations of the slab/$\vec{S}$, the Weyl and Magnus
forces are not parallel. When $\vec{S}$ coincides with an eigenvector
of the magnetic part of the Weyl tensor $\mathcal{H}^{ij}$, $\vec{F}_{{\rm Weyl}}\propto\vec{S}$,
being therefore orthogonal to $\vec{F}_{{\rm Mag}}$. For the cloud
in Fig. \ref{fig:MagnusGrav} (orthogonal to the $y$-axis), this
is the case for $\vec{S}=S(\vec{e}_{y}+\vec{e}_{z})/\sqrt{2}$ and
$\vec{S}=S(\vec{e}_{z}-\vec{e}_{y})/\sqrt{2}$ (the third eigenvector
of $\mathcal{H}^{ij}$, $\vec{S}=S\vec{e}_{x}$, has zero eigenvalue
and leads to $\vec{F}_{{\rm Mag}}=\vec{F}_{{\rm {\rm Weyl}}}=0$).
Cases \ref{enu:GravcaseSz}-\ref{enu:GravcaseSy} sharply illustrate
the contrast between the two parts of the spin curvature force: on
the one hand the Magnus force $\vec{F}_{{\rm Mag}}$, which depends
only on $\vec{S}$ and on the local mass-density current $\vec{J}$,
and is therefore the same regardless of the boundary; and, on the
other hand, the Weyl force, which \emph{is determined} by the details
of the system, namely the direction along which this cloud model has
a finite width $h$. Similarly to the electromagnetic case, neither
$\vec{H}$ at any point inside the cloud (or its gradient $H^{i,j}$),
nor $\vec{F}_{{\rm Weyl}}$, depend on the precise value of $h$;
the role of its finiteness boils down to fixing the direction of $\vec{H}$.
Equation (\ref{eq:CurlHstationary}), together with the problem's
symmetries, then fully fix $\vec{H}$ (analogously to the situation
for $\vec{B}$ in Sec. \ref{sub:A-current-slab}). One can then say
that, in this example, the magnetic part of the Weyl tensor, $\mathcal{H}_{ij}=\mathbb{H}_{(ij)}$
(and therefore $\vec{F}_{{\rm Weyl}}$), is fixed by the boundary,
whereas antisymmetric part of the gravitomagnetic tidal tensor, $\mathbb{H}_{[ij]}$,
depends only on the local mass current density $\vec{J}$, cf. Eq.
(\ref{eq:H_ab_Decomp}).

In general one is interested in the total force $\vec{F}=\vec{F}_{{\rm Mag}}+\vec{F}_{{\rm Weyl}}$
(for it is what determines the body's motion); the dependence of $\vec{F}_{{\rm Weyl}}$
on the details/boundary conditions of the system shown by the results
above hints at the importance of appropriately modeling the astrophysical
systems of interest.

\subsubsection{The force exerted by the body on the cloud\label{sub:PN_slab_Reciprocal_Fporce}}

Previous approaches in the literature attempted to compute the gravitational
Magnus force by inferring it from its reciprocal -- the force exerted
by the body on the cloud~\cite{Okawa:2014sxa,Cashen:2016neh}. Unfortunately,
these attempts were not based on concrete computations of such force,
but on estimates which are either not complete (and thereby misleading)
or rigorous, and turn out in fact to yield incorrect conclusions (see
Sec. \ref{sub:Infinite-clouds} and Appendix~\ref{sub:Gravity interaction with individual particles of the cloud}
below for details). In this section we shall rigorously compute, in
the framework of the PN approximation, the ``force'' exerted by
the spinning body on the cloud for the setups considered above.

In the first PN approximation, the geodesic equation for a point particle
of coordinate velocity $\vec{v}=d\vec{x}/dt$ can be written as $d\vec{v}/dt=\vec{F}_{{\rm I}}/m$,
with $\vec{F}_{{\rm I}}$ given by Eq. (\ref{eq:FI}). This equation
exhibits formal similarities with the Lorentz force law; namely the
gravitomagnetic ``force'' $m\vec{v}\times\vec{H}$, analogous to
the magnetic force $q\vec{v}\times\vec{B}$. The total \emph{gravitomagnetic}
force exerted by the spinning body on the cloud is the sum of the
force exerted in each of its individual particles, given by the integral
\begin{align}
\vec{F}_{{\rm body,cloud}} & =\int_{{\rm cloud}}\vec{J}\times\vec{H}_{{\rm body}}d^{3}x\label{eq:FbodyCloud}\\
 & =\vec{J}\times\int_{r\le R}\vec{H}_{{\rm body}}d^{3}x+\vec{J}\times\int_{r>R}\vec{H}_{{\rm body}}d^{3}x\nonumber 
\end{align}
where $\vec{H}_{{\rm body}}$ is the gravitomagnetic field generated
by the spinning body. Equation (\ref{eq:CurlHstationary}), formally
similar to (\ref{eq:MaxwellAmpere}) up to a factor $-4$, implies
that\footnote{This can be shown by steps analogous to those in pp. 187-188 of Ref.~\cite{Jackson:1998nia},
replacing therein the magnetic vector potential $\vec{A}$ by the
gravitomagnetic vector potential $\mathcal{\vec{A}}_{{\rm body}}(x)=-4\int\vec{J}_{{\rm body}}(x')/|\vec{x}-\vec{x}'|d^{3}\vec{x}'.$} 
\begin{equation}
\int_{r\le R}\vec{H}_{{\rm body}}d^{3}x=-\frac{16\pi}{3}\vec{S}\label{eq:InteriorIntGarv}
\end{equation}
and that the exterior gravitomagnetic field is (cf. e.g. \cite{CiufoliniWheeler,Costa:2015hlh})
\begin{equation}
\vec{H}_{{\rm body}}|_{r>R}=2\frac{\vec{S}}{r^{3}}-6\frac{(\vec{S}\cdot\vec{r})\vec{r}}{r^{5}}\ ,\label{eq:GMfieldSpinning}
\end{equation}
analogous, up to a factor -2, to (\ref{eq:InteriorIntegral}) and
(\ref{eq:Bdip}), respectively. Therefore, for $\vec{S}=S\vec{e}_{z}$,
and a slab finite in the $y$ direction (contained within $-h/2<y<h/2$),
as depicted in Fig. \ref{fig:MagnusGrav}, an integration analogous
to (\ref{eq:IntegralBy}) leads to 
\[
\vec{F}_{{\rm body,cloud}}=8\pi JS\vec{e}_{y}=-\vec{F}\equiv-\vec{F}_{{\rm cloud,body}}\ ,
\]
i.e., minus the force exerted by the slab on the body, Eq. (\ref{eq:FGcloud}),
satisfying an action-reaction law. For a slab finite in the $z$ direction
(contained within $-h/2<z<h/2$), like in the electromagnetic analogue
the force vanishes: $\vec{F}_{{\rm body,cloud}}=0$, matching its
reciprocal.

Several remarks must however be made on this result. First we note
that, unlike the spin-curvature force $\vec{F}\equiv\vec{F}_{{\rm cloud,body}}$
exerted by the cloud on the body (which is a physical, covariant force),
the gravitomagnetic ``force'' $m\vec{v}\times\vec{H}$, that (when
summed over all particles of the cloud) leads to $\vec{F}_{{\rm body,cloud}}$,
is an \emph{inertial force}, i.e., a fictitious force (in fact $\vec{H}$
is but twice the vorticity of the reference observers, see e.g. \cite{Costa:2016iwu,Costa:2015hlh}).
Moreover, an integration in the likes of Eq. (\ref{eq:FbodyCloud})
is not possible in a strong field region, for the sum of vectors at
different points is not well defined. Such integrations make sense
only in the context of a PN approximation, which requires a Newtonian
potential such that $U\ll1$ everywhere within the region of integration.
This requires a body with a radius such that $R\gg m$ (and spinning
slowly), so that the field is weak even in its interior regions, which
precludes in particular the case of BHs or compact bodies. (It does
not even make sense to talk about an overall force on the cloud in
these cases). In addition to that, the interior integral $\int_{r\le R}\vec{J}\times\vec{H}_{{\rm body}}d^{3}x$
obviously only makes sense if the cloud is made of dark matter or
some other exotic matter that is able to permeate the body; otherwise
$\vec{J}=0$ for $r\le R$, and so such integral would be zero.\footnote{Still that will not lead to a mismatch between action and reaction
{[}comparing to $\vec{F}\equiv\vec{F}_{{\rm cloud,body}}$ as given
by Eq. (\ref{eq:FGcloud}){]}, because in that case the mass current
around the body would not be uniform and along $x$ {[}that would
violate the PN continuity equation $\partial\rho/\partial t=-\nabla\cdot\vec{J}+\Os{5}${]},
but instead one would have a continuous flow around the body, as described
by fluid dynamics, accordingly changing $\int_{r>R}\vec{J}\times\vec{H}_{{\rm body}}d^{3}\vec{x}$.} Finally, it should be noted that, although for these \emph{stationary}
setups the action $\vec{F}_{{\rm cloud,body}}$ equals minus the reaction
$\vec{F}_{{\rm body,cloud}}$, in general dynamics the gravitomagnetic
interactions, just like magnetism, \emph{do not} \emph{obey} the action-reaction
law (contrary to the belief in some literature). This is due to the
momentum exchange between the matter and the gravitational field.
In particular it is so, \emph{at leading order}, for the spin-orbit
interaction of the spinning body with individual particles of the
cloud, as discussed in detail in Appendix \ref{sub:Gravity interaction with individual particles of the cloud}.

\subsubsection{Infinite clouds\label{sub:Infinite-clouds}}

\emph{In the framework of} \emph{the post-Newtonian approximation},
the situation with infinite clouds is analogous to that in electromagnetism
discussed in Sec. \ref{sub:A-current-slab}. Taking the limit $h\rightarrow\infty$
in the cases of a slab contained within $-h/2\le y\le h/2$ (case
\ref{enu:GravcaseSz} above), or $-h/2\le z\le h/2$ (case \ref{enu:GravcaseSy}),
are two different ways of constructing an infinite cloud, each of
them leading to a different gravitomagnetic field $\vec{H}$ inside,
a different Weyl force $\vec{F}_{{\rm Weyl}}$ (only the Magnus force
$\vec{F}_{{\rm Mag}}$ is the same in both cases), and thus to a different
total spin-curvature force $\vec{F}$ exerted on a spinning body.
(Again, notice that none of these quantities depends on the precise
value of the slab's width $h$, but only in the direction along which
the slab was initially taken to be finite, cf. Eqs. (\ref{eq:HSlab}),
(\ref{eq:FGcloud})-(\ref{eq:FWeylGravSlab})). The same applies to
the reciprocal force, $\vec{F}_{{\rm body,cloud}}$, exerted by the
body on the cloud. This manifests that, just like in the electromagnetic
case, these are \emph{not well defined quantities for an infinite
(in all directions) cloud}. If one had started with a cloud about
which all one is told is that it is infinite in \emph{all directions},
the questions of which is $\vec{F}\equiv\vec{F}_{{\rm cloud,body}}$
and $\vec{F}_{{\rm body,cloud}}$ would simply have no answer. This
is down to the same fundamental mathematical principles at stake in
the electromagnetic problem: in the case of $\vec{F}_{{\rm cloud,body}}$,
to the impossibility of setting up the boundary conditions required
to solve Eq. (\ref{eq:CurlHstationary}); and, in the case of $\vec{F}_{{\rm body,cloud}}$,
to the implications of Fubini's theorem, discussed in Appendix \ref{sec:Fubini_Theorem}.
The situation is moreover analogous to the ``paradox'' concerning
the Newtonian gravitational field of an infinite homogeneous matter
distribution, which likewise is not well defined, and is a well known
difficulty in Newtonian cosmology (see e.g. \cite{VICKERS2009,McCrea1955,Raifeartaigh2017,EinsteinPrinciple,EinsteinRelativityl,EllisMaartensMacCallum}
and references therein).

This means that the problem of the force exerted on a spinning body
by an infinite homogeneous cloud (or its reciprocal) cannot be solved
in the context of a PN approximation, and in particular in the framework
of an analogy with electromagnetism. Recently, an attempt to find
$\vec{F}\equiv\vec{F}_{{\rm cloud,}{\rm body}}$ (cast therein as
``gravitomagnetic dynamical friction'') for such a cloud has been
presented \cite{Cashen:2016neh}; a result was inferred from an estimate
of the reciprocal force $\vec{F}_{{\rm body,cloud}}$. However, not
only the correct answer is actually that the force is not well defined
for the problem and framework therein, but also the estimate obtained
has a direction \emph{opposite} to the Magnus effect, which is at
odds with the result from the exact relativistic theory (where the
problem is well posed, see Sec. \ref{sub:FLWR} below), and even with
the result obtained from a PN computation for the setting at stake:
therein a stellar cloud with spherical boundary is considered, with
arbitrarily large radius $R$. The limit $R\rightarrow\infty$ yields
yet another way of constructing an infinite cloud. The force exerted
by the body on such cloud, $\vec{F}_{{\rm body,cloud}}=\vec{J}\times\int_{r<R}\vec{H}_{{\rm body}}d^{3}x$,
is obtained from (\ref{eq:InteriorIntGarv}), and reads, regardless
of the value of $R$, $\vec{F}_{{\rm body,cloud}}=-16\pi\vec{J}\times\vec{S}/3$.
Hence, a naive\footnote{In rigor an action-reaction law cannot be employed here, for such
setup is not stationary (see in this respect Appendix \ref{sec:Action-reaction-law}).
The actual force exerted on a spinning body with velocity $\vec{v}$
at \emph{any point} inside the sphere is given by Eq. (\ref{eq:FuniformHalo}).
It thus differs by a factor $4/3$ from $-\vec{F}_{{\rm body,cloud}}$.} application of an action-reaction principle leads to a force on the
body parallel to $\vec{J}\times\vec{S}$, in the \emph{same direction}
of the Magnus effect (But, again, such result is irrelevant, for the
problem is not well posed in this framework).

On the other hand, general relativity (in its exact form), unlike
electromagnetism, or Newtonian and PN theory, has no problem with
an infinite universe filled everywhere with a fluid of constant density;
in fact this is precisely the case of the FLRW solution, which is
the standard cosmological model, and where the spin-curvature force
exerted on a spinning body is well defined, as we shall see in Sec.
\ref{sub:FLWR} below.

\section{Magnus effect in dark matter halos\label{sub:Dark-matter-halo}}

Consider a dark matter halo with a spherically symmetric density profile
$\rho(r)$, with arbitrary radial dependence. Here (by contrast with
the example in Sec. \ref{sub:Toy-model-revisited}) we will not base
our analysis in the test particle's center of mass frame, but instead
consider a particle moving in the static background with velocity
$\vec{v}$, see Fig. \ref{fig:DMHalo}. To compute the spin-curvature
force acting on it, we start by computing the gravitoelectric field
$\vec{G}$ and its derivatives \emph{inside the halo}. To lowest order
(which is the accuracy needed for the 1PN spin-curvature force), $\vec{G}$
amounts to the Newtonian field 
\begin{equation}
\vec{G}=-\frac{M(r)}{r^{3}}\vec{r}\equiv M(r)\vec{\mathcal{G}}\,,\label{eq:G_Halo}
\end{equation}
where $\vec{\mathcal{G}}\equiv-\vec{r}/r^{3}$ is the Newtonian field
of a point mass per unit mass and
\begin{equation}
M(r)=4\pi\int_{0}^{r}r^{2}\rho(r)dr\label{eq:M(r)}
\end{equation}
is the mass enclosed inside a sphere of radius $r$. It follows that
\begin{equation}
G_{i,j}=M(r)\mathcal{G}_{i,j}-4\pi\rho\frac{r_{i}r_{j}}{r^{2}}\,.\label{eq:GijHalo}
\end{equation}
Since the source is static, $\vec{H}=0=\dot{\vec{G}}$; therefore,
by Eq. (\ref{eq:HijGEMPN}), the gravitomagnetic tidal tensor $\mathbb{H}_{ij}$
as measured by a body/observer of velocity $\vec{v}$ reduces here
to $\mathbb{H}_{ij}=2\epsilon_{i}^{\ km}v_{k}G_{j,m}-\epsilon_{ij}^{\ \ m}G_{k,m}v^{k}$.
Splitting into symmetric and antisymmetric parts, one gets, after
some algebra, 
\begin{align}
 & \mathcal{H}_{ij}=\mathbb{H}_{(ij)}=2\frac{A(r)-4\pi\rho}{r^{2}}(\vec{v}\times\vec{r})_{(i}r_{j)}\,,\label{eq:HWeylDMHalo}\\
 & \mathbb{H}_{[ij]}=\frac{1}{2}\epsilon_{ijl}\epsilon^{lkm}\mathbb{H}_{km}=4\pi\rho\epsilon_{ijl}v^{l}\,,\label{eq:HijantiHalo}
\end{align}
where
\begin{equation}
A(r)\equiv\frac{3M(r)}{r^{3}}\ .\label{eq:ArHalo}
\end{equation}
The spin-curvature force on the body, $F^{i}=-\mathbb{H}^{ji}S_{j}$,
reads then, cf. Eqs. (\ref{eq:FMagPN})-(\ref{eq:FWeylPN0}) (notice
that, for a static source, $T^{0i}=0$) 
\begin{align}
 & \vec{F}=\vec{F}_{{\rm Weyl}}+\vec{F}_{{\rm Mag}}\,,\label{eq:FGHalo}\\
 & F_{{\rm Weyl}}^{i}=-\mathcal{H}^{ij}S_{j}\,,\qquad\vec{F}_{{\rm Mag}}=4\pi\rho\vec{S}\times\vec{v}\,,\label{eq:FDecompHalo}
\end{align}
with $\mathcal{H}_{ij}$ given by (\ref{eq:HWeylDMHalo}).

\subsection{Spherical, uniform dark matter halo\label{sec:Spherical,-static-uniform}}

Let us start by considering a spherical DM halo of constant density
$\rho=\rho_{0}$, which, although unrealistic, is useful as a toy
model. It follows from Eqs. (\ref{eq:M(r)}) and (\ref{eq:ArHalo})
that $M(r)=4\pi\rho_{0}r^{3}/3$ and $A(r)=4\pi\rho_{0}$, therefore,
by Eqs. (\ref{eq:HWeylDMHalo}) and (\ref{eq:FDecompHalo}), the magnetic
part of the Weyl tensor, and the Weyl force, vanish \emph{for all
$\vec{v}$}: $\mathcal{H}_{ij}=\mathbb{H}_{(ij)}=0\Rightarrow\vec{F}_{{\rm Weyl}}=0$.
The gravitomagnetic tidal tensor reduces to its antisymmetric part,
$\mathbb{H}_{ij}=\mathbb{H}_{[ij]}$, and the total force reduces
to the Magnus force, cf. Eq. (\ref{eq:FGHalo}),
\begin{equation}
\vec{F}=\vec{F}_{{\rm Mag}}=4\pi\rho_{0}\vec{S}\times\vec{v}\ .\label{eq:FuniformHalo}
\end{equation}
This equation tells us that \emph{any} spinning body moving inside
such halo suffers a Magnus force. It is (to dipole order) \emph{the
only physical force} acting on the body, deviating it from geodesic
motion. It can also be seen from Eqs. (\ref{eq:CoordAccelForce})-(\ref{eq:FI})
that $\vec{F}_{{\rm Mag}}/m$ is, to leading PN order, the total coordinate
acceleration in the direction orthogonal to $\vec{v}$ .

\subsection{Realistic halos\label{sub:Isothermal-dark-matter}}

The simplistic model above can be improved to include more realistic
density profiles.

\emph{Power law profiles (}$\rho\propto r^{-\gamma}$\emph{)}---In
some literature (e.g. \cite{BinneyTremaine,Bertone:2005hw}) models
of the form $\rho(r)=Kr^{-\gamma}$ are proposed, where $K$ is a
$r$ independent factor. The condition that the mass (\ref{eq:M(r)})
inside a sphere of radius $r$ be finite requires $\gamma<3$; in
this case we have
\begin{equation}
A(r)=12\pi\frac{Kr^{-\gamma}}{3-\gamma}=12\pi\frac{\rho(r)}{3-\gamma}\ .\label{eq:ArPower}
\end{equation}
For $\gamma=2$, this yields the \emph{isothermal }profile $\rho(r)=K/r^{2}$,
leading to $A(r)=12\pi\rho(r)$, and to a constant orbital velocity
$v=\sqrt{Gr}=2\sqrt{K\pi}$. This is consistent with the observed
flat rotation curves of some galaxies, and is known to accurately
describe at least an intermediate region of the Milky Way DM halo
\cite{BinneyTremaine}. Values $1\le\gamma\le1.5$ have also been
suggested \cite{Bertone:2005hw,Burkert:1995yz}, based on numerical
simulations, for the inner regions of spiral galaxies like the Milky
Way.

\emph{Pseudo-isothermal density profile.}---Consider a density profile
$\rho(r)$ given by~\cite{Burkert:1995yz} 
\begin{figure}
\includegraphics[width=0.4\textwidth]{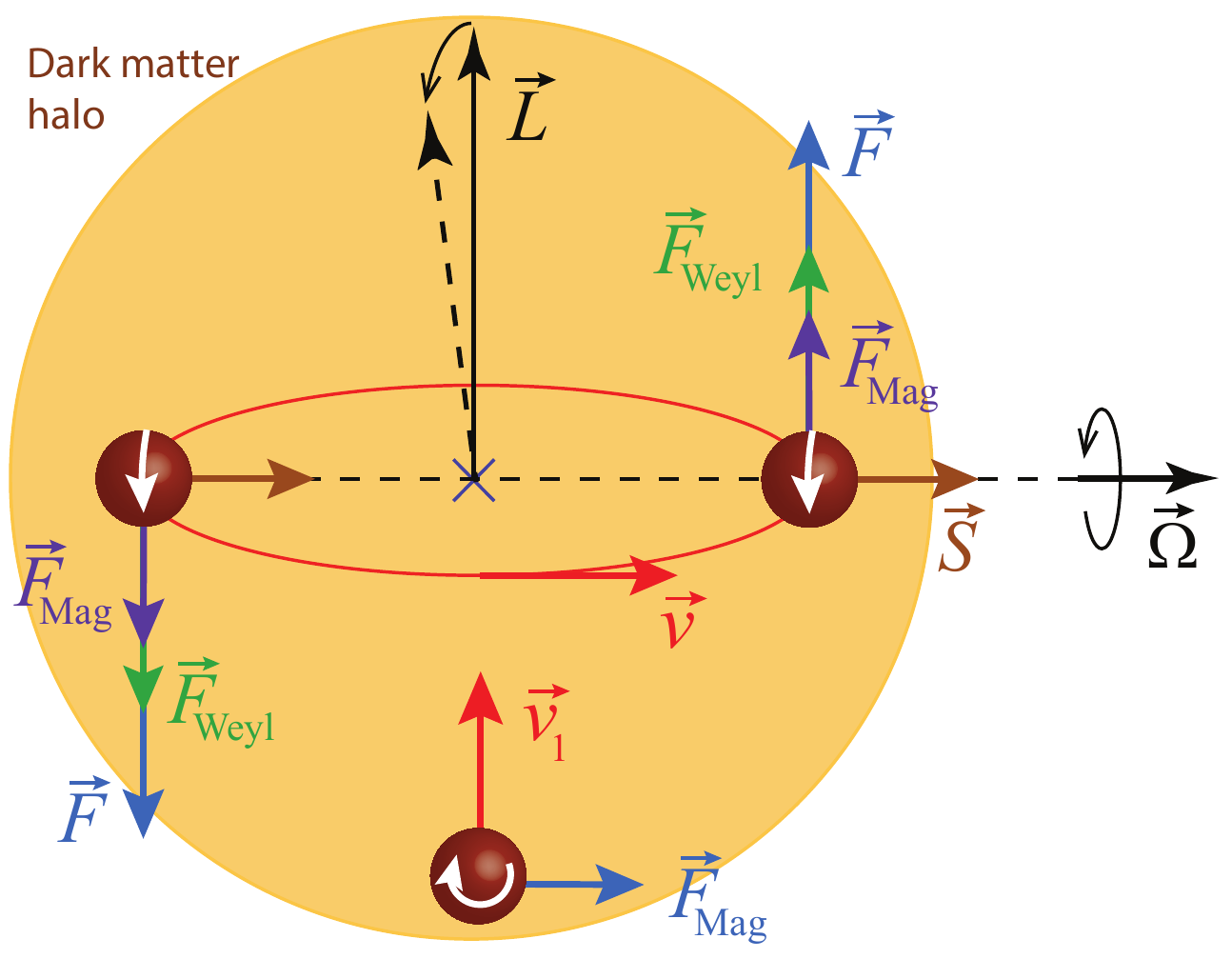} \protect\protect\protect\protect\caption{\label{fig:DMHalo}Spinning bodies moving in a ``pseudo-isothermal''
DM halo. For a body in quasi-circular orbits, with spin lying in the
orbital plane, the Magnus ($\vec{F}_{{\rm Mag}}$) and Weyl ($\vec{F}_{{\rm Weyl}}$)
forces are parallel. The total force is of the form $\vec{F}=A(r)\vec{S}\times\vec{v}$,
pointing outwards the orbital plane on one half of the orbit, and
inwards the other half; this ``torques'' the orbit, leading to a
secular orbital precession $\vec{\Omega}$. For a body moving radially
towards the center of the halo, $\vec{F}_{{\rm Weyl}}=0$, and so
the total force exerted on it reduces to the Magnus force: $\vec{F}=\vec{F}_{{\rm Mag}}=4\pi\rho\vec{S}\times\vec{v}_{1}$.
Generically $\vec{F}_{{\rm Weyl}}$ and $\vec{F}_{{\rm Mag}}$ have
different directions. If the halo's density was uniform, $\vec{F}_{{\rm Weyl}}=0\Rightarrow\vec{F}=\vec{F}_{{\rm Mag}}$
for all particles.}
\end{figure}

\begin{equation}
\rho(r)=\frac{\rho_{0}}{1+\frac{r^{2}}{r_{{\rm c}}^{2}}}\,,\label{eq:Isothermal}
\end{equation}
where $r_{{\rm c}}$ is the \emph{core radius}. For $r\gg r_{{\rm c}}$,
the velocity of the circular orbits becomes nearly constant, whilst
at the same time not diverging at $r=0$ (as is the case for the isothermal
profile, $\rho(r)\propto r^{-2}$). From Eqs. (\ref{eq:Isothermal}),
(\ref{eq:M(r)}), and (\ref{eq:ArHalo}), we have 
\begin{equation}
A(r)=12\pi\rho_{0}\frac{r_{{\rm c}}^{2}}{r^{2}}\left[1-\frac{r_{{\rm c}}}{r}\arctan\left(\frac{r}{r_{{\rm c}}}\right)\right]\,.\label{eq:ArIsothermal}
\end{equation}
Notice that $A(r)>0$ for all $r$.

Substitution of the expressions for $\rho(r)$ and $A(r)$ in (\ref{eq:HWeylDMHalo})-(\ref{eq:HijantiHalo}),
(\ref{eq:FDecompHalo}), yield, for each model, the gravitomagnetic
tidal tensor $\mathbb{H}_{ij}$ as measured by the body moving with
velocity $\vec{v}$, and the spin-curvature force exerted on it. Comparing
with the situation for the uniform halo highlights the contrast between
the two components of the spin-curvature force (and the dependence
of the Weyl force on the details of the system): $\vec{F}_{{\rm Mag}}$
remains formally the same (for it depends only on the local density
$\rho$ and on $\vec{v}$), whereas $\vec{F}_{{\rm Weyl}}$ is now
generically nonzero. It is different for each model, and has generically
a different direction from $\vec{F}_{{\rm Mag}}$. The Weyl force
vanishes remarkably when (at some instant) $\vec{v}\parallel\vec{r}$.
Hence, if one takes a particle with initial radial velocity, initially
one has, exactly, $\vec{F}=\vec{F}_{{\rm Mag}}$; and afterwards the
spin-curvature force will consist on $\vec{F}_{{\rm Mag}}$ plus a
smaller correction $\vec{F}_{{\rm Weyl}}$ due to the nonradial component
of the velocity that the particle gains due to the force's own action.

\subsection{Objects on quasi-circular orbits}

We shall now consider the effect of the spin-curvature force ($\vec{F}_{{\rm Mag}}+\vec{F}_{{\rm Weyl}}$)
exerted on test bodies on (quasi-) circular orbits within the DM halo.
The evolution equation for the spin vector of a spinning body reads,
in an \emph{orthonormal} system of axes tied to the PN background
frame (i.e., to the basis vectors of the coordinate system in (\ref{eq:PNmetric});
this is a frame anchored to the ``distant stars'') \cite{Misner:1974qy,CiufoliniWheeler,Costa:2015hlh}
\begin{equation}
\frac{d\vec{S}}{dt}=\vec{\Omega}_{{\rm s}}\times\vec{S}\ ;\qquad\ \vec{\Omega}_{{\rm s}}=-\frac{1}{2}\vec{v}\times\vec{a}+\frac{3}{2}\vec{v}\times\vec{G}\,,\label{eq:SpinPrecession}
\end{equation}
where the first term is the Thomas precession and the second the geodetic
(or de Sitter) precession. Since the only force present is the spin-curvature
force, then $\vec{a}=\vec{F}/m$, and the Thomas precession is negligible
to first PN order. So, in what follows, $\vec{\Omega}_{{\rm s}}\approx3\vec{v}\times\vec{G}/2$.
Without loss of generality, let us assume the orbit to lie in the
$xy$-plane. Two notable cases to consider are the following.

\subsubsection{\noindent Spin orthogonal to the orbital plane ($\vec{S}=S^{z}\vec{e}_{z}$)\label{sub:Spin-orthogonal-to}}

\noindent In this case $\vec{\Omega}_{{\rm s}}\parallel\vec{S}$,
and so $d\vec{S}/dt=0$, i.e., the components of the spin vector are
constant along the orbit (so it remains along $\vec{e}_{z}$). The
Magnus and Weyl forces are 
\begin{equation}
\vec{F}_{{\rm Mag}}=-4\pi\rho\frac{(\vec{S}\cdot\vec{L})}{mr}\vec{e}_{r}\,,\quad\vec{F}_{{\rm Weyl}}=\vec{F}_{{\rm Mag}}+A(r)\frac{(\vec{S}\cdot\vec{L})}{mr}\vec{e}_{r}\,,
\end{equation}
where $\vec{L}=m\vec{r}\times\vec{v}$ is (to lowest order) the orbital
angular momentum (see e.g. \cite{CiufoliniWheeler}). All the forces
are radial. For $A(r)<4\pi\rho$, $\vec{F}_{{\rm Weyl}}$ points in
the same direction of $\vec{F}_{{\rm Mag}}$, which resembles case
\ref{enu:GravcaseSz} of the slab in Sec. \ref{sub:Toy-model-revisited}.
For $A(r)>4\pi\rho$, which is the case in all the models considered
in Sec. \ref{sub:Isothermal-dark-matter}, $\vec{F}_{{\rm Weyl}}$
points in the direction \emph{opposite} to $\vec{F}_{{\rm Mag}}$,
resembling case \ref{enu:GravcaseSy} of the slab. As for the total
force $\vec{F}$, it points in the same direction of $\vec{F}_{{\rm Mag}}$
for $A(r)<8\pi\rho$, which, for the power law profiles $\rho\propto r^{-\gamma}$
in Sec. \ref{sub:Isothermal-dark-matter}, is the case for $\gamma<1.5$;
it vanishes when $A(r)=8\pi\rho$; and it points in opposite direction
to $\vec{F}_{{\rm Mag}}$ when $A(r)>8\pi\rho$, which is the case
for $\gamma>1.5$. The pseudo-isothermal profile (\ref{eq:Isothermal})
realizes all the three cases, having an interior region where $\vec{F}\parallel\vec{F}_{{\rm Mag}}$,
whereas $\vec{F}\parallel-\vec{F}_{{\rm Mag}}$ for large $r$. The
orbital effect of $\vec{F}$ amounts to a change in the effective
gravitational attraction.

\subsubsection{\noindent Spin parallel to the orbital plane ($S^{z}=0$) }

\noindent In this case Eq. (\ref{eq:SpinPrecession}) tells us that
$\vec{S}$ precesses, but remains always in the plane; since $\vec{\Omega}_{{\rm s}}$
is constant, this equation yields (taking, \emph{initially}, $\vec{S}=S\vec{e}_{x}$)
\begin{equation}
\vec{S}=S\cos(\Omega_{{\rm s}}t)\vec{e}_{x}+S\sin(\Omega_{{\rm s}}t)\vec{e}_{y}\ .\label{eq:Spinevol}
\end{equation}
To the accuracy needed for Eqs. (\ref{eq:FDecompHalo}), $\vec{v}\approx v(-\sin\phi\vec{e}_{x}+\cos\phi\vec{e}_{y})$,
with $\phi=\omega t$, where $\omega$ is the orbital angular velocity.
Therefore 
\begin{equation}
\vec{S}\times\vec{v}=vS\cos(\phi-\Omega_{{\rm s}}t)\vec{e}_{z}=vS\cos[(\omega-\Omega_{{\rm s}})t]\vec{e}_{z}\ .\label{eq:Sxv}
\end{equation}
The Magnus, Weyl, and total forces then read 
\begin{align}
 & \vec{F}_{{\rm Mag}}=4\pi\rho Sv\cos[(\omega-\Omega_{{\rm s}})t]\vec{e}_{z}\,,\nonumber \\
 & \vec{F}_{{\rm Weyl}}=[A(r)-4\pi\rho]Sv\cos[(\omega-\Omega_{{\rm s}})t]\vec{e}_{z}\,,\nonumber \\
 & \vec{F}=A(r)\vec{S}\times\vec{v}=A(r)Sv\cos[(\omega-\Omega_{{\rm s}})t]\vec{e}_{z}\,.\label{eq:FtotHalo}
\end{align}
All these forces are thus along the direction orthogonal to the orbital
plane. The situation is inverted comparing to the case in Sec. \ref{sub:Spin-orthogonal-to}
above: $\vec{F}_{{\rm Weyl}}$ points in the same direction of $\vec{F}_{{\rm Mag}}$
for $A(r)>4\pi\rho$, and in opposite direction for $A(r)<4\pi\rho$.
For all the models considered in Sec. \ref{sub:Isothermal-dark-matter}
(the pseudo-isothermal, and those of the form $\rho\propto r^{-\gamma}$,
with $0\le\gamma<3$), we have $A(r)>4\pi\rho$, cf. Eqs. (\ref{eq:ArIsothermal}),
(\ref{eq:ArPower}); so both $\vec{F}_{{\rm Weyl}}$ and the total
force $\vec{F}$ point in the same direction as $\vec{F}_{{\rm Mag}}$
(the latter condition requiring only $A(r)>0$), see Fig. \ref{fig:DMHalo}. 

\noindent The force $\vec{F}$ causes the spinning body to oscillate
(in the $\vec{e}_{z}$ direction) along the orbit, perturbing the
circular motion. The coordinate acceleration orthogonal to the orbital
plane is, from Eqs. (\ref{eq:CoordAccelForce})-(\ref{eq:FI}), $\ddot{z}=F^{z}/\Pmass+G^{z}$.
$G^{z}$ is the component of the gravitational field along $z$, that
is acquired when the body oscillates out of the plane. Making a first
order Taylor expansion about $z=0$, we have $G^{z}\simeq G_{z,z}|_{z=0}z\equiv G_{z,z}z$.
The general solution, for $G_{z,z}<0$ and $G_{z,z}\ne-\Delta\omega^{2}$,
is 
\begin{equation}
z(t)=c_{1}\cos(\sqrt{-G_{z,z}}t)+c_{2}\sin(\sqrt{-G_{z,z}}t)+Z\cos(\Delta\omega t)\ ,\label{eq:z(t)}
\end{equation}
where 
\begin{equation}
Z\equiv-\frac{S}{\Pmass}\frac{A(r)v}{G_{z,z}+\Delta\omega^{2}}=\frac{S}{\Pmass}\frac{A(r)v}{\Omega_{{\rm s}}(2\omega-\Omega_{{\rm s}})}\ ,\label{eq:Z}
\end{equation}
$c_{1}$ and $c_{2}$ are arbitrary integration constants, $\Delta\omega\equiv\omega-\Omega_{{\rm s}}$
and $r$ is the radius of the fiducial circular geodesic. In the second
equality in (\ref{eq:Z}) we noted, from Eq. (\ref{eq:GijHalo}),
that $G_{z,z}=-G/r=-\omega^{2}$. Noticing, moreover, from Eqs. (\ref{eq:G_Halo}),
(\ref{eq:ArHalo}), that 
\begin{equation}
A(r)=3\omega^{2}=3\frac{v^{2}}{r^{2}}\label{eq:Awv}
\end{equation}
 and $\Omega_{{\rm s}}=3\omega^{3}r^{2}/2$, we can re-write (\ref{eq:Z})
as a function of the orbital velocity ($v=\omega r$) only,
\begin{equation}
Z=\frac{S}{\Pmass}\frac{1}{v\left[1-\frac{3}{4}v^{2}\right]}\ .\label{eq:Zv}
\end{equation}
The first two terms of Eq. (\ref{eq:z(t)}) are independent of the
spin-curvature force (if $\vec{F}=0$, they simply describe the $z$
oscillations of a circular orbit lying off the $xy$-plane), so $c_{1}$
and $c_{2}$ essentially set up the initial inclination of the orbit.
Two natural choices of these constants stand out (analogous to those
first found in Ref.~\cite{Karpov:2003bn}, for orbits around BHs).

\emph{Constant amplitude regime:} $c_{1}=c_{2}=0$. In this case $z(t)=Z\cos(\Delta\omega t)$,
yielding a ``bobbing'' motion of frequency $\Delta\omega$ and constant
amplitude $Z$. They may be seen as an orbit which is inclined relative
to the fiducial geodesic, and whose plane precesses with the frequency
of the geodetic precession ($\Omega_{{\rm s}}$). 
\begin{figure*}
\includegraphics[width=2\columnwidth]{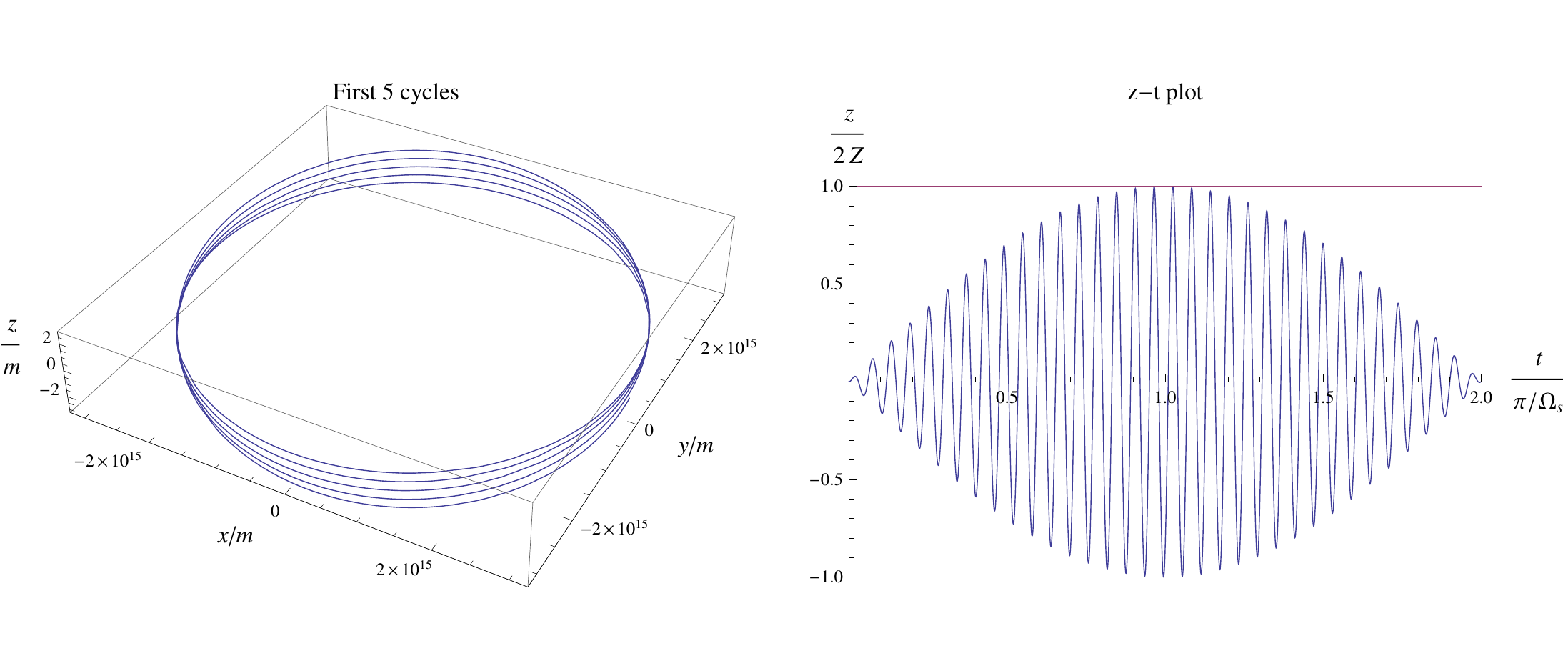}\protect\protect\protect\protect\caption{\label{fig:Haloplot}Numerical 1PN results for quasi-circular orbits
in a pseudo-isothermal DM halo, with $\rho_{0}=10^{8}M_{\odot}{\rm pc}^{-3}$,
$r_{{\rm c}}=0.02\,{\rm kpc}$ (typical of satellite galaxies \cite{Barausse:2014tra}),
and $r=8r_{{\rm c}}$, case in which $v=0.15c$. The test body has
the Sun's mass $m=M_{\odot}$, and an initial spin vector $\vec{S}|_{{\rm in}}=S\vec{e}_{x}$,
with $S=0.5m^{2}$. Left panel: three-dimensional plot of the orbit,
showing the orbital precession $\vec{\Omega}\propto\vec{S}$ in Eq.
(\ref{eq:SecularPrecession}) (i.e., about $\vec{e}_{x}$, initially).
Right panel: plot of $z(t)/2Z$, for $t\in[0,2\pi/\Omega_{{\rm s}}]$;
the numerical result agrees well with the (simplified) analytical
result (\ref{eq:Beating}). It shows clearly the modulation by the
spin precession $\vec{\Omega}_{{\rm s}}$: the orbital precession
$\vec{\Omega}$ causes $z$ oscillations of initially increasing amplitude,
reaching its peak $z=2Z$ at $t=\pi/\Omega_{{\rm s}}$, corresponding
to the maximum inclination of the orbital plane. At that point the
direction of $\vec{S}$ (thus of $\vec{\Omega}$) becomes inverted
relative to the initial one, so the orbital inclination (and the oscillation
amplitude) starts decreasing.}
\end{figure*}

\emph{``Beating'' regime: }one starts with the same initial data
of a circular orbit in the $xy$-plane: $z(0)=\dot{z}(0)=0$, implying
$c_{2}=0$, $c_{1}=-Z$. Using the trigonometric identity $\cos(b)-\cos(a)=2\sin\left[\frac{a+b}{2}\right]\sin\left[\frac{a-b}{2}\right]$,
Eq. (\ref{eq:z(t)}) becomes 
\begin{equation}
z(t)=2Z\sin\left[\frac{2\omega-\Omega_{{\rm s}}}{2}t\right]\sin\left[\frac{\Omega_{{\rm s}}}{2}t\right]\ .\label{eq:Beating}
\end{equation}
This corresponds to a rapid oscillatory motion of frequency $(2\omega-\Omega_{{\rm s}})/2$
(close to the orbital frequency $\omega$), modulated by a sinusoid
of frequency $\Omega_{{\rm s}}/2$ (half the frequency of the spin
precession), and of peak amplitude $2Z$. In spite of the simplifying
approximations made in its derivation, Eq. (\ref{eq:Beating}) shows
very good agreement with the numerical results plotted in the right
panel of Fig. \ref{fig:Haloplot}. As shown by Eqs. (\ref{eq:Z})-(\ref{eq:Zv}),
$Z$ is proportional to the ratio $S/\Pmass$, known as the test body's
``Møller radius''~\cite{Moller1949}; it is the minimum size an
extended body can have in order to have finite spin without violating
the dominant energy condition \cite{Costa:2014nta,Moller1949}. Since
$v<1$, we see from Eq. (\ref{eq:Zv}) that $Z$ is always larger
than such radius.

The force (\ref{eq:FtotHalo}) originates also a precession of the
orbital plane. Recalling that (to lowest order) $\vec{L}=m\vec{r}\times\vec{v}$,
\begin{equation}
\frac{d\vec{L}}{dt}=m\vec{r}\times\frac{d\vec{v}}{dt}=\vec{r}\times\vec{F}=-A(r)\vec{r}\times(\vec{v}\times\vec{S})\,,
\end{equation}
where we substituted $d\vec{v}/dt\equiv d^{2}\vec{x}/dt^{2}$ from
Eqs. (\ref{eq:CoordAccelForce})-(\ref{eq:FI}) (noting that $\vec{G}\times\vec{r}=0$),
and $A(r)$ is given by Eq. (\ref{eq:ArHalo}). Using the vector identity
$\vec{r}\times(\vec{v}\times\vec{S})=(\vec{r}\times\vec{v})\times\vec{S}+(\vec{S}\times\vec{r})\times\vec{v}$,
we have 
\begin{equation}
\frac{d\vec{L}}{dt}=A(r)\left[\frac{1}{\Pmass}\vec{S}\times\vec{L}-(\vec{S}\times\vec{r})\times\vec{v}\right]\,.\label{eq:Orbprec1}
\end{equation}
The first term is fixed along the orbit, and is already in a precession
form. The second term must be averaged along the orbit, in order to
extract the secular effect. First we note, from (\ref{eq:SpinPrecession}),
that $\Omega_{{\rm s}}=3vG/2\sim\omega\epsilon^{2}$, so typically
$\Omega_{{\rm s}}\ll\omega$, and, therefore, along \emph{one} orbit,
the spin vector is nearly constant. So, for averaging along an orbit,
we may approximate $\vec{S}\simeq S\vec{e}_{x}$. It follows that
$\left\langle (\vec{S}\times\vec{r})\times\vec{v}\right\rangle =-Srv\left\langle \sin^{2}\phi\right\rangle \vec{e}_{y}=\vec{S}\times\vec{L}/(2\Pmass)$,
leading to the secular orbital precession 
\begin{equation}
\left\langle \frac{d\vec{L}}{dt}\right\rangle =\vec{\Omega}\times\vec{L};\qquad\vec{\Omega}=\frac{A(r)}{2\Pmass}\vec{S}\ .\label{eq:SecularPrecession}
\end{equation}
So we are led to the interesting result that the orbit precesses about
the direction of the spin vector $\vec{S}$. This can be simply understood
from Fig. \ref{fig:DMHalo}: since $\vec{S}$ is nearly constant along
\emph{one} orbit, the force (\ref{eq:FtotHalo}) points in the positive
$\vec{e}_{z}$ direction for nearly half of the orbit, and in the
opposite direction in the other half; this ``torques'' the orbit,
causing it to precess. The effect is clear in the numerical results
in the left panel of Fig. \ref{fig:Haloplot}. This precession is,
of course, not independent from the oscillations studied above; in
fact, it is the origin of the beating regime of Eq. (\ref{eq:Beating}),
which may be seen as follows. Multiplying the angular velocity $\Omega$
of rotation of the orbital plane by $r$, yields the ``rotational
velocity'' of the orbit; this precisely matches {[}under the same
assumption $\omega\gg\Omega_{{\rm s}}$ that leads to Eq. (\ref{eq:SecularPrecession}){]}
the initial slope of the function $2Z\sin(\Omega_{{\rm s}}t/2)$ that
modulates (\ref{eq:Beating}): 
\begin{equation}
\Omega_{{\rm s}}Z\simeq\Omega r\ .\label{eq:Zslope}
\end{equation}
So, the increase in the amplitude of the oscillations in Fig. \ref{fig:Haloplot}
(these rapid oscillations are the variation of $z$ along each orbit,
notice) is the reflex of the orbital precession $\vec{\Omega}$. Now,
such orbital rotation does not go on forever \emph{in the same sense},
because $\vec{S}$ itself undergoes the precession in Eq. (\ref{eq:SpinPrecession}),
which means that after a time $t=\pi/\Omega_{{\rm s}}$ the direction
of the spin vector is reversed. Likewise $\vec{\Omega}$ and $\langle d\vec{L}/dt\rangle$
are reversed (before one full revolution about $\vec{S}$ is completed
if $\Omega<\Omega_{{\rm s}}$, as is usually the case), and this is
why the amplitude in Eq. (\ref{eq:Beating}) is modulated by the geodetic
spin precession $\Omega_{{\rm s}}$.

In Fig. \ref{fig:Haloplot} numerical results are plotted for a test
body with the Sun's mass $\Pmass=M_{\odot}$ and $S=0.5m^{2}$, in
a pseudo-isothermal DM halo typical of a satellite galaxy (corresponding
to much larger DM densities than those typical of the Milky Way, which
makes them more suitable to illustrate the effects described above).
Such results are obtained by numerically solving the system of equations
formed by $md^{2}\vec{x}/dt^{2}=m\vec{G}+\vec{F}+\vec{F}_{{\rm D}}$
together with Eq. (\ref{eq:SpinPrecession}), with $\vec{F}$ as given
by Eqs. (\ref{eq:FGHalo})-(\ref{eq:FDecompHalo}), (\ref{eq:HWeylDMHalo}),
(\ref{eq:ArIsothermal}), and $\vec{G}$ given by Eqs. (\ref{eq:G_Halo}),
(\ref{eq:M(r)}), (\ref{eq:ArIsothermal}). The term $\vec{F}_{{\rm D}}$
is the \emph{dynamical friction} force
\begin{equation}
\vec{F}_{{\rm D}}=-40\pi\rho\Pmass^{2}\left({\rm erf}(X)-\frac{2X}{\sqrt{\pi}}e^{-X^{2}}\right)\frac{\vec{v}}{v^{3}}\,,\quad X=\frac{v}{v_{{\rm circ}}}\,,\label{eq:Fd-1}
\end{equation}
which is here included. It follows from Eq. (8.6) of \cite{BinneyTremaine},
or Eq. (3) of \cite{Pani:2015qhr}, by taking $\lambda=10$ and $\sqrt{2}\sigma=v_{{\rm circ}}\equiv$
velocity of the circular orbit at $r$, cf. \cite{BinneyTremaine}.
(The impact of $\vec{F}_{{\rm D}}$, in the case of the motion in
Fig. \ref{fig:Haloplot}, turns out to be unnoticeable.)

\subsubsection{Particular examples in the Milky Way DM Halo\label{sub:Particular-examples-in-MW}}

\emph{Pseudo-isothermal profile.}---The DM density at the solar system
position is about $0.01M_{\odot}/{\rm pc}^{3}=10^{-21}{\rm kg\,m}^{-3}$
\cite{Read:2014qva,Gilmore:1996pr}; the Sun's distance from the center
of the Milky Way is $r_{\odot}=8\,{\rm kpc}$. The core radius $r_{{\rm c}}$
of the Milky Way DM halo is about $1\,{\rm kpc}$. Assuming the pseudo-isothermal
density profile in Eq. (\ref{eq:Isothermal}), this means that $\rho_{0}=0.7M_{\odot}{\rm pc}^{-3}$.
The velocity $v$ of the quasi-circular orbits is obtained from Eqs.
(\ref{eq:ArIsothermal}), (\ref{eq:Awv}). For a body orbiting at
$r=r_{\odot}$, the peak amplitude in Eqs. (\ref{eq:Zv})-(\ref{eq:Beating})
then reads 
\begin{equation}
2Z=3.5\times10^{3}\frac{S}{\Pmass}\ .\label{eq:ZMilkyWay}
\end{equation}
The time to reach it (``beating'' half period) is however very long:
$t_{{\rm peak}}\approx\pi/\Omega_{{\rm s}}=10^{14}\,{\rm yr}$ ($10^{4}\times$
age of the universe); this corresponds to $\omega/(2\Omega_{{\rm s}})=10^{6}$
laps around the center. Noting that the initial of slope of the function
$2Z\sin(\Omega_{{\rm s}}t/2)$ that modulates Eq. (\ref{eq:Beating})
is $Z\Omega_{{\rm s}}$, the maximum amplitude actually reached within
the age of the universe ($\equiv t_{0}$) is
\begin{align}
 & 2Z_{{\rm today}}\equiv2Z\sin\left[\frac{\Omega_{{\rm s}}t_{0}}{2}\right]\approx Z\Omega_{{\rm s}}t_{0}\label{eq:Ztoday}\\
 & \ \approx\Omega rt_{0}=\frac{3v^{2}t_{0}}{2r}\frac{S}{\Pmass}\label{eq:Ztodayv}
\end{align}
where in the second approximate equality we used (\ref{eq:Zslope}),
and in the last equality we used (\ref{eq:SecularPrecession}), (\ref{eq:Awv}).
Hence, for the setting above, 
\begin{figure*}
\includegraphics[width=2\columnwidth]{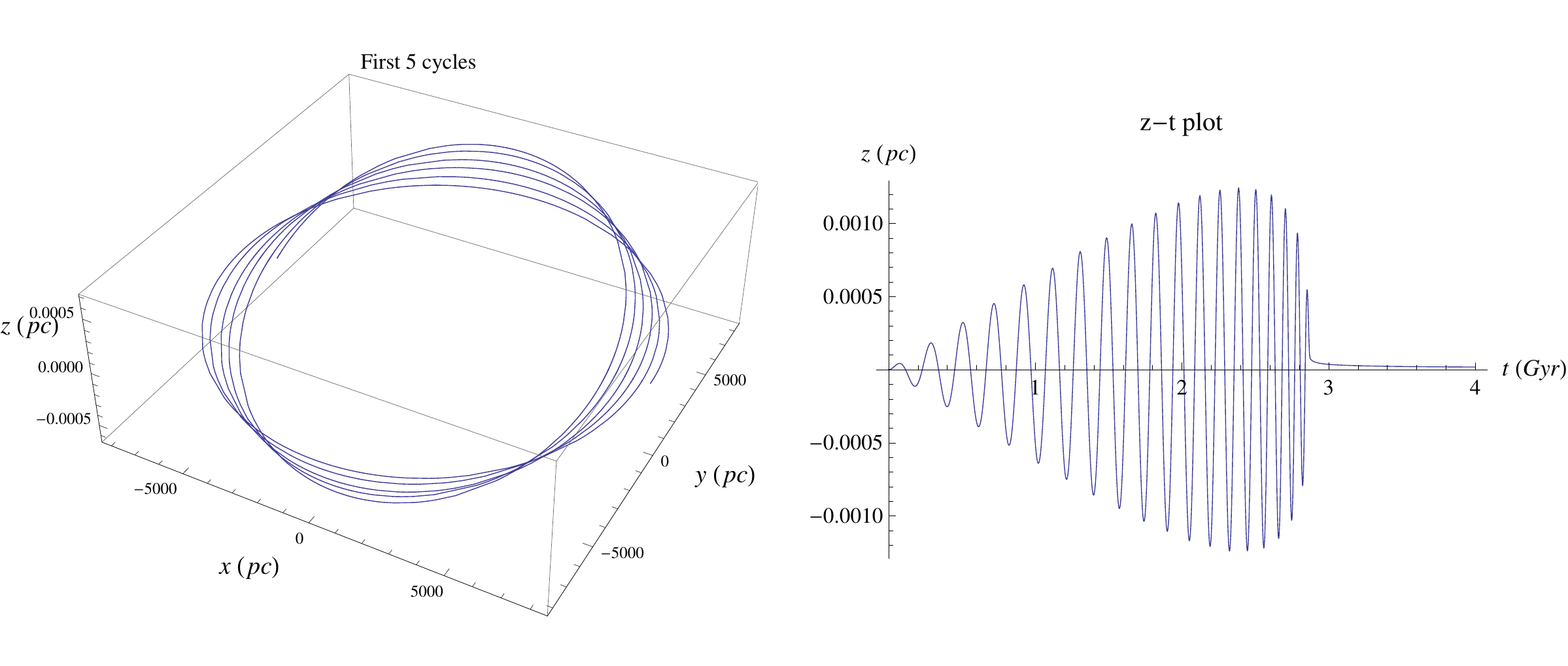} \protect\protect\caption{\label{fig:SatelliteOrbit}Orbits of a satellite galaxy in the Milky
way DM halo (numerical 1PN results). Due to their disk-like shape,
galaxies have a large Møller radius $S/m$, making them especially
suitable test bodies for the effects under study. The satellite is
assumed to be a scale reduced version of the Milky Way, of diameter
2.5${\rm kpc}$, and orbiting at a distance 8${\rm kpc}$ (= Sun's
distance) from the Halo's center. The peak amplitude predicted by
Eqs. (\ref{eq:Z})-(\ref{eq:Beating}) is $2Z\approx60\,{\rm pc}$;
but it is never reached, due to the very strong dynamical friction.
Still a peak of about $10^{-3}\,{\rm pc}$ is reached after about
$2.2\,{\rm Gyr}$ ($\simeq1/5$ the age of the universe).}
\end{figure*}
\begin{equation}
2Z_{{\rm today}}=1.4\times10^{-4}Z=0.25\frac{S}{\Pmass}\ .\label{eq:ZtodayIsothermalSun}
\end{equation}
Both $Z$ and the angular velocity $\Omega$ of orbital precession,
Eq. (\ref{eq:SecularPrecession}), are proportional to the body's
Møller radius $S/\Pmass$. For a Sun-like star ($\Pmass=M_{\odot}$,
$S\approx0.2\Pmass^{2}$~\cite{KommetAl}, $S/m=3\times10^{2}\,{\rm m}$),
the effect is very small: the secular orbital precession (\ref{eq:SecularPrecession})
inclines the orbit by about $1.6\,{\rm m}$ per lap, with peak amplitude
$2Z=10^{6}\,{\rm m}$ (about 1400 times smaller that the Sun's diameter),
and $2Z_{{\rm today}}\sim10^{2}\,{\rm m}$. Larger or more massive
bodies will typically have a larger Møller radius, thereby yielding
more interesting numbers; but, on the other hand, for large $\Pmass$,
the dynamical friction force $\vec{F}_{{\rm D}}$, Eq. (\ref{eq:Fd-1})
(which is proportional to $\Pmass^{2}$), becomes also important.
From the known objects moving in the Milky Way's DM halo, those with
largest Møller radius are (due to their size and flattened shape)
Milky Way's satellite galaxies. Consider first, for a comparison (still
at\footnote{It is nearly the case for the Canis Major Dwarf, and nearly twice
that value for the Sagittarius dwarf.} $r\simeq r_{\odot}$), a hypothetical satellite galaxy with diameter
$\simeq2.5\,{\rm kpc}$, and assume it to be a ``scale reduced''
version of the Milky Way (diameter $55\,{\rm kpc}$, mass $\Pmass_{{\rm MW}}=10^{12}M_{\odot}$,
angular momentum $S_{{\rm MW}}=2.6\times10^{31}\,{\rm m}^{2}$), rotating
with the same velocity. Since $S\propto mv_{{\rm rot}}R$, this yields
$S_{{\rm sat}}/S_{{\rm MW}}\sim(2.5/55)^{4}$, $\Pmass_{{\rm sat}}/\Pmass_{{\rm MW}}\sim(2.5/55)^{3}$,
leading to a Møller radius $S_{{\rm sat}}/\Pmass_{{\rm sat}}\sim0.02\,{\rm pc}$.
In this case the orbit inclines at an initial rate of $4\times10^{12}\,{\rm m}$
per orbit ($2\times10^{4}\,{\rm m}$ per year), and the peak amplitude,
as predicted by Eqs. (\ref{eq:Z})-(\ref{eq:Beating}), (\ref{eq:ZMilkyWay})
would now be $2Z\simeq60{\rm pc}$. Such large peak value however
is \emph{never reached}, due to the damping action of $\vec{F}_{{\rm D}}$;
the numerical results shown in Fig.~\ref{fig:SatelliteOrbit} show
that a peak of about $0.001\,{\rm pc}$ (i.e., about 10 times the
radius of the solar system), is reached within about $2.2{\rm Gyr}$
(a fifth of the age of the universe), after which the orbit and its
oscillations pronouncedly decay. As a concrete example in the Milky
Way DM Halo, we take the Large Magellanic Cloud (the largest satellite
galaxy), located at $r=48\,{\rm kpc}$ from the MW center. It has
mass $\Pmass_{{\rm LMC}}\approx10^{10}M_{\odot}$, diameter $\approx4.3\,{\rm kpc}$,
and rotational velocity $v_{{\rm rot}}\approx9\times10^{4}\,{\rm m\,s^{-1}}$
\cite{vanderMarel:2013jza}, from which we estimate a Møller radius
$S_{{\rm LMC}}/\Pmass_{{\rm LMC}}\sim0.3\,{\rm pc}$. We find a gradual
inclination of the orbit of about $\sim4\times10^{-9}\,{\rm mas}{\rm yr}^{-1}$
(or $3\times10^{4}\,{\rm m}\,{\rm yr}^{-1}$); this is far beyond
the current observational accuracy, since the uncertainty in the LMC's
proper motion is presently much larger ($\sim10^{-2}{\rm mas}{\rm yr}^{-1}$
\cite{vanderMarel:2013jza}). The peak amplitude predicted by Eqs.
(\ref{eq:Z})-(\ref{eq:Beating}), (\ref{eq:ZMilkyWay}) is $2Z\approx1\,{\rm kpc}$
(which, again, is not reached due to dynamical friction). Numerical
simulations (similar to those in Fig. \ref{fig:SatelliteOrbit}) show
that an effective peak of about $10^{-3}\,{\rm pc}$ is reached within
$2.5\,{\rm Gyr}$.

\emph{Power law profiles.---} For the models of the form $\rho(r)=Kr^{-\gamma}$
in Sec. \ref{sub:Isothermal-dark-matter}, substituting (\ref{eq:ArPower})
in (\ref{eq:Awv}), (\ref{eq:SecularPrecession}), it follows from
Eqs. (\ref{eq:Zv}) and (\ref{eq:Ztodayv}) that 
\begin{align}
 & 2Z\approx\frac{2S}{\Pmass v}=\frac{S}{\Pmass}\left[\frac{3-\gamma}{K\pi}\right]^{1/2}r^{\gamma/2-1}\ ,\label{eq:PowerDMZ}\\
 & 2Z_{{\rm today}}\approx\Omega rt_{0}=\frac{S}{\Pmass}\frac{6\pi t_{0}K}{3-\gamma}r^{1-\gamma}\ ,\label{eq:PowerDMMW}
\end{align}
where $K$ is determined from the value of $\rho(r_{\odot})$ which
we assume, for all models, $\rho(r_{\odot})\approx0.01M_{\odot}{\rm pc}^{-3}$
\cite{Bertone:2005hw,Read:2014qva,Gilmore:1996pr}. For $\gamma=1$,
$Z_{{\rm today}}\approx0.08S/m$ is approximately constant, and $Z$
decreases with $r$ as $Z\propto r^{-1/2}$; for bodies orbiting at
$r=r_{\odot}$, one has $Z\approx2.3\times10^{3}S/m$. That is, the
peak/present time amplitudes are, respectively (at $r=r_{\odot}$),
somewhat larger/smaller than those for the pseudo-isothermal profile,
Eqs. (\ref{eq:ZMilkyWay}), (\ref{eq:ZtodayIsothermalSun}). For $1<\gamma<2$,
it follows from Eqs. (\ref{eq:PowerDMMW}) that both $Z$ and $Z_{{\rm today}}$
decrease with $r$. The isothermal case, $\gamma=2$, yields a $Z=1.6\times10^{3}S/m$
approximately independent of $r$, and $Z_{{\rm today}}\propto r^{-1}$.
At $r=r_{\odot}$, $Z_{{\rm today}}=0.15S/m$; thus $Z$ is slightly
smaller and $Z_{{\rm today}}$ slightly larger that in the pseudo-isothermal
profile. However, contrary to the pseudo-isothermal case (where $Z_{{\rm today}}$
reaches a maximum $\approx0.26S/m$ at $r\approx1.5{\rm kpc}$), $Z_{{\rm today}}$
increases steeply as one approaches the halo center, approaching the
peak value $Z$ {[}cf. Eq. (\ref{eq:Ztodayv}){]}.

\emph{Inside the galactic disk.}---The above are results taking into
account DM only; so they apply to orbits \emph{outside} the galactic
disk. Within the disk, the density of baryonic matter, in the vicinity
of the Sun, is about $\rho_{{\rm b}}\approx0.1M_{\odot}{\rm pc}^{-3}$~\cite{BinneyTremaine,Read:2014qva},
i.e. one order of magnitude larger than that of DM. This leads to
an enlarged effect. The field produced by the disk is a complicated
problem (see Sec. \ref{sec:Disks} below). The analysis of a simple
model in Sec.~\ref{sub:Miyamoto-Nagai-disks} reveals however that,
just for an order of magnitude estimate, the force caused by the disk
can be taken as the corresponding Magnus force $\vec{F}_{{\rm Mag}}$,
and its contribution to the orbital precession as $\Omega_{{\rm b}}\sim F_{{\rm Mag}}/(vm)\sim4\pi\rho_{{\rm b}}S/m$.
Assuming, for DM, the pseudo-isothermal profile (\ref{eq:Isothermal}),
leads to $2Z_{{\rm today}}\approx\Omega rt_{0}\sim2S/m$, cf. Eq.
(\ref{eq:Ztodayv}) {[}here $\Omega\equiv\Omega_{{\rm b}}+\Omega_{{\rm DM}}$,
with $\Omega_{{\rm DM}}$ given by Eqs. (\ref{eq:SecularPrecession}),
(\ref{eq:ArIsothermal}){]}.

More importantly, the galactic disk might reveal a signature of the
orbital precession (\ref{eq:SecularPrecession}): BHs or stars with
spin axes nearly parallel to the galactic plane are, on average, more
distant from the plane than other bodies, by a distance of order $\sim4Z_{{\rm today}}/\pi$.\footnote{Note that the time scale for formation and flattening of the galactic
disk is much shorter than that of the orbital precession ($2\pi/\Omega$).} This effect might be observable. The most precise map of the sky
is expected to be given by the Gaia mission \cite{Gaia}, able to
measure angles of about $2\times10^{-11}$ rads. Therefore, on test
bodies whose distance $d$ from Gaia (i.e., from the Earth) is such
that $d\lesssim4Z_{{\rm today}}/(\pi2\times10^{-11})$, the effect
would be within the angular resolution. To be concrete, consider a
giant star like Antares; it has radius $R_{{\rm ant}}\sim10^{3}R_{\odot}$,
mass $\Pmass_{{\rm ant}}\sim12M_{{\rm \odot}}$, surface rotational
velocity $v_{{\rm rot}}=7\times10^{-5}$. For simplicity, assume it
to be uniform and rotate rigidly, leading to a Møller radius $S_{{\rm ant}}/\Pmass_{{\rm ant}}=2\times10^{7}\,{\rm m}$
(five orders of magnitude larger than that of the Sun). Assuming the
pseudo-isothermal profile, this yields a present time amplitude $2Z_{{\rm today}}\sim2S/m=4\times10^{7}\,{\rm m}$.
Giants of this type, with spin axes nearly parallel to the galactic
plane, should on average be farther from the plane than others, by
about $\sim3\times10^{7}\,{\rm m}$ ($\sim10^{-5}\times$ Antares's
diameter). Considering the density value at $r=r_{\odot}$, their
maximum allowed distance from the Earth (so that the effect can be
detected) is then $d_{{\rm max}}\approx0.04\,{\rm kpc}$, which is
not far from the order of magnitude of Antares' actual distance ($d=0.17\,{\rm kpc}$),
and of other large stars. Thus, albeit small, the effect on such stars
is close to the angular resolution. The matter density (baryonic and
DM) increases however as one approaches the galaxy center; for stars
along the line connecting the solar system to the center, the angle
that the effect subtends on the GAIA spacecraft is $\theta\approx Z_{{\rm today}}/d$,
with $d=r_{\odot}-r$. For DM models of the type $\rho_{{\rm DM}}\propto r^{-\gamma}$
with $\gamma>1$, the angle $\theta$ increases with decreasing $r$
(after initially decreasing, and bouncing), cf. Eq. (\ref{eq:PowerDMMW}).
Considering the isothermal profile ($\gamma=2$), and taking into
account DM only, $\theta$ enters GAIA's resolution for $r\lesssim4{\rm pc}$.
The Magnus signature on the galactic disk might thus serve as a test
for such models. Independently of such DM models, the baryonic matter
is known to reach high densities in the galaxy's inner regions; using
$\rho_{{\rm b}}(r)$ as given in Eq. (2) of \cite{Genzel2003}, we
have that, for $r\lesssim1{\rm pc}$, the baryonic matter alone is
sufficient for $\theta$ to enter GAIA's resolution.

\section{Magnus effect in accretion disks\label{sec:Disks}}

The gravitational Magnus effect due to DM is limited by its typically
very low density. Accretion disks around BHs or stars provide mediums
with relatively much higher densities, where the effect can be more
significant, possibly within the reach of near future observational
accuracy.

\subsection{Orders of magnitude for a realistic density profile\label{sub:Orders-of-magnitude}}

The standard model for relativistic thin disks is the Novikov-Thorne
model \cite{NovikovThorne1973}, which generalizes the Shakura-Sunyaev
\cite{Shakura:1972te} model to include relativistic corrections.
According to such model the disk is divided into different regions,
the outer and more extensive of them being well approximated by the
Newtonian counterpart. The density of the later reads (in the equatorial
plane) \cite{Shakura:1972te} 
\begin{equation}
\rho=\frac{f_{{\rm Edd}}^{11/20}}{\tilde{r}^{15/8}}\alpha^{-7/10}\left[\frac{M_{\odot}}{M_{{\rm BH}}}\right]^{\frac{7}{10}}\left[1-\sqrt{\frac{6}{\tilde{r}}}\right]^{\frac{11}{20}}1\times10^{6}{\rm kg}\,{\rm m}^{-3}\,,\label{eq:rhorealistic}
\end{equation}
where $\tilde{r}\equiv r/M_{{\rm BH}}$, $M_{{\rm BH}}$ is the mass
of the central BH, $\tilde{r}_{{\rm in}}\equiv r_{{\rm in}}/M_{{\rm BH}}$,
$\alpha$ the ``viscosity parameter'' and $f_{{\rm Edd}}$ Eddington's
ratio for mass accretion (e.g. \cite{Barausse:2014tra}). The density
(\ref{eq:rhorealistic}) leads to a Magnus force $\vec{F}_{{\rm Mag}}=4\pi\rho\vec{S}\times\vec{v}$
(cf. Eq. (\ref{eq:FMagPN})) of interesting magnitude, compared with
other relevant forces.

Comparing with the Newtonian gravitational force $\Pmass\vec{G}_{{\rm BH}}$
exerted on the body by the central BH, we have $F_{{\rm Mag}}/(\Pmass\|\vec{G}_{{\rm BH}}\|)\sim r^{2}\rho vS/(\Pmass M_{{\rm BH}})$.
Different estimates can be made. Let us consider the case of a binary
of BHs with similar masses $\Pmass\sim M_{{\rm BH}}$; in this case
$F_{{\rm Mag}}/(\Pmass\|\vec{G}_{{\rm BH}}\|)\sim\tilde{S}r^{2}\rho v$,
where $\tilde{S}\equiv S/m^{2}$ (for a fast spinning BH $\tilde{S}\lesssim1$;
for extended bodies it could be much larger). Now we need an estimate
for $v$ (the velocity of the ``test'' body with respect to the
disk of the ``source'' BH); it can be taken has having the magnitude\footnote{\label{fn:v_v2}Except for the case that the ``test'' body is much
smaller and as such can be in a circular orbit corotating with the
source's disk, and the latter is moreover mostly gravitationally driven
(i.e., not very affected by hydrodynamics), the test body will not
comove with the matter on the disk. In general the orbit will be eccentric
relative to the center of the disk; so it will have a velocity $v$
relative to the matter in the disk typically within the same order
of magnitude of its velocity relative to the central BH. It is also
so for counterrotating, or for unbound orbits.} $v\sim\sqrt{M_{{\rm BH}}/r}=\tilde{r}^{-1/2}$ . Then {[}converting
${\rm kg}\,{\rm m}^{-3}$ to geometrized units, and using $M_{{\rm BH}}=M_{\odot}(M_{{\rm BH}}/M_{\odot})${]},
\begin{equation}
\frac{F_{{\rm Mag}}}{\Pmass\|\vec{G}_{{\rm BH}}\|}\sim\tilde{S}\mathcal{F}(M_{{\rm BH}})\left[1-\sqrt{\frac{6}{\tilde{r}}}\right]^{\frac{11}{20}}\tilde{r}^{-3/8}\ \label{eq:FmagG}
\end{equation}
where 
\[
\mathcal{F}(M_{{\rm BH}})\equiv1.6\times10^{-15}f_{{\rm Edd}}^{11/20}\alpha^{-7/10}\left[\frac{M_{{\rm BH}}}{M_{\odot}}\right]^{\frac{13}{10}}\ .
\]

Let us now compare the magnitude of $\vec{F}_{{\rm Mag}}$ with the
spin-orbit force $\vec{F}_{{\rm SO}}$ exerted on the ``test'' body
due to its spin $\vec{S}$ (given by Eq. (\ref{eq:FBH}) below). Assume
it to move, relative to the central BH, with velocity $\sim v$ (i.e,
of the same order of magnitude of that relative to the matter in the
disk, see footnote~\ref{fn:v_v2}) so that $\vec{F}_{{\rm SO}}\sim M_{{\rm BH}}vS/r^{3}$.
It follows that 
\begin{equation}
\frac{F_{{\rm Mag}}}{F_{{\rm SO}}}\sim\frac{\rho r^{3}}{M_{{\rm BH}}}=\mathcal{F}(M_{{\rm BH}})\left[1-\sqrt{\frac{6}{\tilde{r}}}\right]^{\frac{11}{20}}\tilde{r}^{9/8}\ .\label{eq:FmagFSO}
\end{equation}

Comparing with the magnitude of the spin-spin force $F_{{\rm SS}}\sim S_{{\rm BH}}S/r^{4}$
\cite{Wald:1972sz}, 
\begin{figure*}
\includegraphics[width=1\columnwidth]{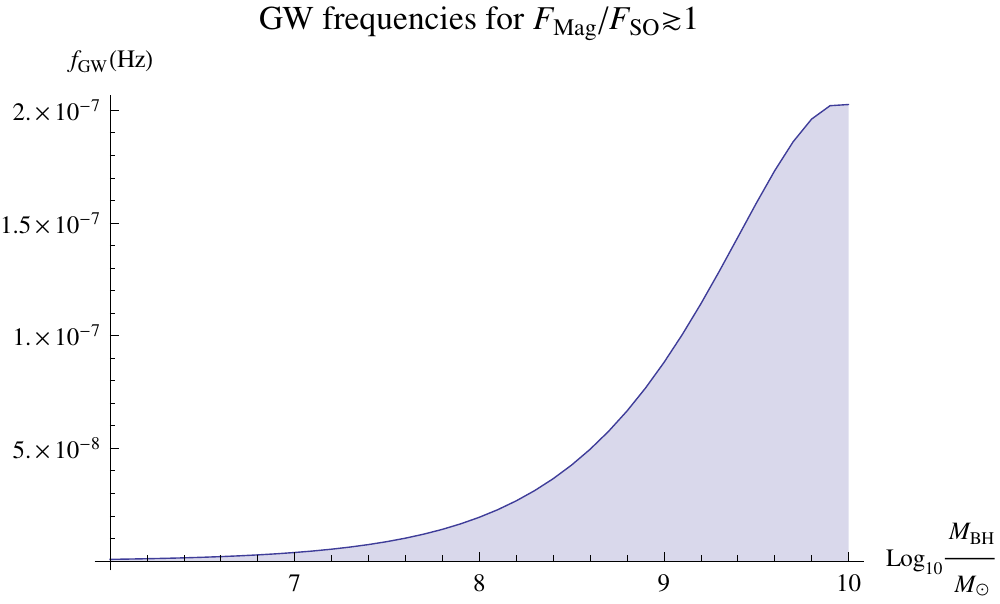}\includegraphics[width=1\columnwidth]{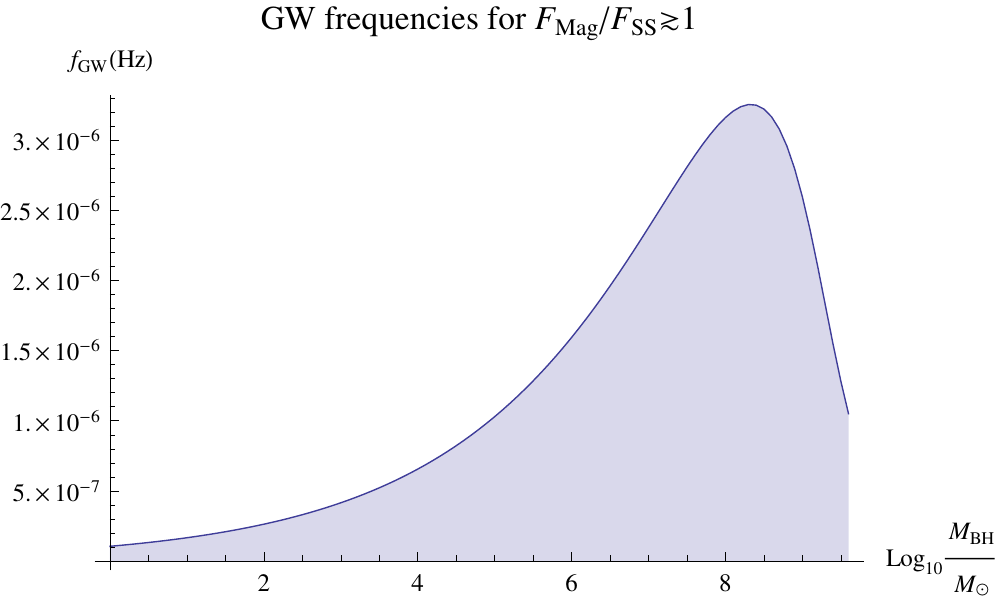}\protect\caption{\label{fig:fmax}Range of frequencies of the emitted gravitational
radiation for binary systems in which the Magnus force $F_{{\rm Mag}}$
is larger than $F_{{\rm SO}}$ and $F_{{\rm SS}}$, for $\alpha=0.01$,
$f_{{\rm Edd}}=0.2$ {[}see Eq. (\ref{eq:rhorealistic}){]} and $S_{{\rm BH}}=0.1M_{{\rm BH}}^{2}$.
The solid curves represent, as functions of the central BH's mass
($M_{{\rm BH}}$), the frequencies for which $F_{{\rm Mag}}\sim F_{{\rm SO}}$
and $F_{{\rm Mag}}\sim F_{{\rm SS}}$. In the shadowed regions it
holds, respectively, $F_{{\rm Mag}}\gtrsim F_{{\rm SO}}$ and $F_{{\rm Mag}}\gtrsim F_{{\rm SS}}$.
The Magnus force tends to be the leading spin dependent force at low
frequencies and for SMBHs, namely within the band of pulsar timing
arrays ($10^{-9}{\rm Hz}\lesssim f_{{\rm GW}}\lesssim10^{-7}{\rm Hz}$).
The frequency for which $F_{{\rm Mag}}\sim F_{{\rm SS}}$ lies moreover
just below (or within, depending on $\alpha$, $f_{{\rm Edd}}$, and
$S_{{\rm BH}}$) the LISA band.}
\end{figure*}
 
\begin{equation}
\frac{F_{{\rm Mag}}}{F_{{\rm SS}}}\sim\frac{\rho vr^{4}}{S_{{\rm BH}}}=\frac{\rho vr^{4}}{\tilde{S}_{{\rm BH}}M_{{\rm BH}}^{2}}=\frac{\mathcal{F}(M_{{\rm BH}})}{\tilde{S}_{{\rm BH}}}\left[1-\sqrt{\frac{6}{\tilde{r}}}\right]^{\frac{11}{20}}\tilde{r}^{13/8}\ ,\label{eq:FmagFSS}
\end{equation}
where $\tilde{S}_{{\rm BH}}\equiv S_{{\rm BH}}/M_{{\rm BH}}^{2}$.

Finally, let us compare the magnitude of $\vec{F}_{{\rm Mag}}$ with
that of the ``orbit-orbit'' gravitomagnetic forces $\vec{F}_{{\rm OO}}=\vec{v}_{1}\times\vec{H}_{1{\rm trans}}$
in the binary; that is, the force exerted on the ``test'' body (dub
it body 2) due to the gravitomagnetic field $\vec{H}_{1{\rm trans}}$
generated by the \emph{translational} motion of the ``source'' (body
1), with respect to \emph{the binary center of mass frame}. This is
of interest in this context for being an effect that has already been
detected to very high accuracy in binaries (relative uncertainty of
about $10^{-3}$, in observations of the Hulse-Taylor pulsar \cite{Nordtvedt:1988vt}).
It is moreover typically larger than its spin-spin and spin-orbit
counterparts. The translational gravitomagnetic field is given by
Eq. (\ref{eq:Htrans}) ($\vec{H}_{1}=\vec{H}_{1{\rm trans}}$ therein);
so $F_{{\rm OO}}\sim\Pmass M_{{\rm BH}}v_{1}v_{2}/r^{2}$, which we
may take as $F_{{\rm OO}}\sim\Pmass M_{{\rm BH}}v^{2}/r^{2}$ (see
footnote \ref{fn:v_v2}). Considering moreover $m\sim M_{{\rm BH}}$,
we have 
\begin{equation}
\frac{F_{{\rm Mag}}}{F_{{\rm OO}}}\sim\frac{\rho\tilde{S}r^{2}}{v}=\tilde{S}\mathcal{F}(M_{{\rm BH}})\left[1-\sqrt{\frac{6}{\tilde{r}}}\right]^{\frac{11}{20}}\tilde{r}^{5/8}\ ,\label{eq:FMagFOO}
\end{equation}
where, again, we used $v\sim\sqrt{M_{{\rm BH}}/r}$.

All the four ratios (\ref{eq:FmagG})-(\ref{eq:FMagFOO}) increase
with $M_{{\rm BH}}$, and depend also on $r$; the ratio to the Newtonian
force decreases with $r$, whereas all the others increase with $r$.
Choosing, from the range of values in \cite{Barausse:2014tra}, $\alpha\sim0.01$,
$f_{{\rm Edd}}\sim0.2$, and considering supermassive BHs with $M_{{\rm BH}}\sim10^{9}M_{\odot}$,
$F_{{\rm Mag}}$ starts being larger than $F_{{\rm SO}}$ for $r\gtrsim81M_{{\rm BH}}$.
Assuming the central black hole to be fast spinning (unfavorable case)
with e.g. $\tilde{S}_{{\rm BH}}\sim0.1$, we have that $F_{{\rm Mag}}\gtrsim F_{{\rm SS}}$
for $r\gtrsim17M_{{\rm BH}}$. Assuming moreover that the ``test''
black hole is also fast spinning (favorable scenario) with e.g. $\tilde{S}=0.5$,
$F_{{\rm Mag}}\gtrsim F_{{\rm OO}}$ for $r\gtrsim6\times10^{3}M_{{\rm BH}}$.
Even comparing with the Newtonian force, the magnitude of $F_{{\rm Mag}}$
is interesting: for $r\lesssim2\times10^{4}M_{{\rm BH}}$, we have
$F_{{\rm Mag}}/(m\|\vec{G}\|)\gtrsim10^{-4}$; this is the same order
of magnitude of the leading 1PN corrections (which are of fractional
order $\epsilon^{2}\sim M_{{\rm BH}}/r\sim10^{-4}$, see Sec. \ref{sub:Post-Newtonian-approximation}).

These comparisons are relevant for binary systems, due to the impact
of spin effects (both spin-orbit and spin-spin) in the emitted gravitational
radiation, which is expected to be observed in the near future \cite{Hannam:2013pra,LangHughes2006,Vecchio:2003tn,Lang:2007ge,Stavridis:2009ys,Schmidt:2014iyl,Berti:2004bd}.
There are different existing/proposed detectors, operating at different
frequencies (see e.g. \cite{Moore:2014lga}). The frequency $f_{{\rm GW}}$
of the emitted gravitational radiation is approximately related to
the Kepler orbital angular velocity $\omega$ by $f_{{\rm GW}}=\omega/\pi$.
Since $\omega\approx M_{{\rm BH}}^{1/2}r^{-3/2}=M_{{\rm BH}}^{-1}\tilde{r}^{-3/2}$,
one may eliminate either $M_{{\rm BH}}$ or $\tilde{r}$ from Eqs.
(\ref{eq:rhorealistic})-(\ref{eq:FMagFOO}) above. Eliminating $\tilde{r}$
{[}by substituting $\tilde{r}\rightarrow(\omega M_{{\rm BH}})^{-2/3}${]}
one obtains $\rho(M_{{\rm BH}},\omega)$, and all the ratios above,
as functions of $M_{{\rm BH}}$ and $\omega$. We are especially interested
in the ratios to the spin-spin and spin-orbit forces; they both decrease
with $\omega$; hence, solving for $\omega$ the equalities $F_{{\rm Mag}}/F_{{\rm SO}}=1$
and $F_{{\rm Mag}}/F_{{\rm SS}}=1$ yields, as a function of $M_{{\rm BH}}$,
the maximum orbital angular velocity $\omega_{{\rm max}}(M_{{\rm BH}})$
{[}and thus the maximum gravitational wave frequency $f_{{\rm GWmax}}(M_{{\rm BH}})${]}
allowed in order to have $F_{{\rm Mag}}$ larger than $F_{{\rm SS}}$
or $F_{{\rm SO}}$. These curves are plotted in Fig. \ref{fig:fmax},
for $\alpha\sim0.01$, $f_{{\rm Edd}}\sim0.2$, and $\tilde{S}_{{\rm BH}}=0.1$.
They tell us that the Magnus force is more important at low frequencies,
and for supermassive black holes. Within the band of groundbased detectors
such as LIGO ($10{\rm Hz}\lesssim f_{{\rm GW}}\lesssim10^{3}{\rm Hz}$)
it is much smaller than $F_{{\rm SO}}$ and $F_{{\rm SS}}$. Within
the band of the spacebased LISA \cite{Audley:2017drz} ($10^{-5}{\rm Hz}\lesssim f_{{\rm GW}}\lesssim1{\rm Hz}$),
we have that $F_{{\rm Mag}}\ll F_{{\rm SO}}$. Fixing the frequency
at the most favorable value $f_{{\rm GW}}=10^{-5}{\rm Hz}$ (i.e.,
fixing $\omega\sim3\times10^{-5}{\rm s^{-1}}$), and plotting the
corresponding ratio $F_{{\rm Mag}}/F_{{\rm SO}}=\rho(M_{{\rm BH}})/\omega^{2}$
(not shown in Fig. \ref{fig:fmax}), one sees that $F_{{\rm Mag}}/F_{{\rm SO}}\lesssim10^{-2}$.
On the other hand, Fig. \ref{fig:fmax} shows also that the frequency
for which $F_{{\rm Mag}}\sim F_{{\rm SS}}$ lies just below the LISA
band. In fact, the magnitude of $F_{{\rm Mag}}$ is already comparable
to $F_{{\rm SS}}$ within LISA's band (for $f_{{\rm GW}}\sim10^{5}{\rm Hz}$,
$F_{{\rm Mag}}$ reaches a maximum $F_{{\rm Mag}}\sim0.2F_{{\rm SS}}$
for $M\sim10^{7.5}M_{\odot}$). Moreover, mild deviations in the disk
parameters from the conservative choice above (e.g., $\alpha\sim0.005$,
$f_{{\rm Edd}}\sim0.8$ \cite{Shakura:1972te}), or simply considering
a central black hole with smaller spin (e.g. $\tilde{S}_{{\rm BH}}\sim0.03$)
are sufficient to make $F_{{\rm Mag}}$ of the same order of magnitude
as $F_{{\rm SS}}$ within such band. Since LISA is expected to be
sensitive to spin-spin effects \cite{LangHughes2006,Berti:2004bd},
this suggests that there might be prospects of detecting as well the
Magnus effect. The lowest frequency planned detectors are pulsar timing
arrays (PTAs, see e.g. \cite{Hobbs2009,Detweiler1979,Sazhin,Moore:2014eua,Moore:2014lga,Arzoumanian2016}),
of band $10^{-9}{\rm Hz}\lesssim f_{{\rm GW}}\lesssim10^{-7}{\rm Hz}$,
and which in the future are expected to detect waves from individual
SMBH binary sources \cite{Hobbs2009,Moore:2014eua}. Within this band,
Fig. \ref{fig:fmax}~shows that $F_{{\rm Mag}}$ can be the leading
spin effect.

\subsection{Miyamoto-Nagai disks\label{sub:Miyamoto-Nagai-disks}}

Although to compute the Magnus force (\ref{eq:FMagPN}) all one needs
to know is the disk's local density $\rho$ and the relative velocity
$\vec{v}$ of the test body, in order to determine the body's motion,
one needs the total spin-curvature force $\vec{F}=\vec{F}_{{\rm Mag}}+\vec{F}_{{\rm Weyl}}$;
that requires knowledge of the gravitational field produced by the
sources (disk plus central BH), since $\vec{F}_{{\rm Weyl}}$ depends
on it. This is however a complicated problem. There is an extensive
literature on the fields of disks, from exact solutions~\cite{Semerak:2012dw,Morgan:1969jr,Lemos:1988vf,Lemos:1993uy,Bicak:1993xat,Espitia:2001cj,Bicak:1993zz,NishidaEriguchiLanza}),
to perturbative \cite{Cizek:2017wzr,Kotlarik:2018nbd}, PN~\cite{Jaranowski:2014yva},
and Newtonian~\cite{Miyamoto:1975zz,Baes:2008td,Vogt:2009sy,Witzany:2015yqa}
solutions. Even though the formalism in Sec.~\ref{sec:Gravitational-Magnus-effect}
(by being exact) could in principle be used to treat the exact problem,
most exact solutions in the literature are not practical or suitable
for our problem, since they are either nonanalytical~\cite{NishidaEriguchiLanza,Kotlarik:2018nbd},
or describe the field only outside the disk~\cite{Semerak:2012dw},
or are not realistic models of astrophysical systems~\cite{Semerak:2012dw,Bicak:1993zz,Morgan:1969jr,Lemos:1988vf,Lemos:1993uy,Bicak:1993xat,Espitia:2001cj}.
In this context the Newtonian solutions provide the more treatable
examples for us. According to Eq. (\ref{eq:FWeylPN}), to compute
$\vec{F}_{{\rm Weyl}}$ (and thus $\vec{F}$) to leading PN order,
only the Newtonian and gravitomagnetic ($\vec{H}$) fields of the
source are needed. By considering a Newtonian field, one is ignoring
the gravitomagnetic fields (frame-dragging) produced by the rotation
of the disk (and of the central BH); this would be accurate if either
the disk was static, or composed of counterrotating streams of matter,
or the test body moves considerably faster relative to the disk than
the disk's average rotational velocity (so that one can have $vG\gg H$).
One might argue that none of these is a realistic assumption --- the
disk is (at least in part) gravitationally driven, so it must rotate,
with a velocity of the same order of magnitude of the velocity of
orbiting test bodies. But still it is no less realistic than most
exact solutions --- which are precisely static \cite{Semerak:2012dw,Morgan:1969jr,Lemos:1988vf,Lemos:1993uy,Bicak:1993xat,Espitia:2001cj}
and/or composed of streams of matter flowing in opposite directions
\cite{Morgan:1969jr,Lemos:1988vf,Lemos:1993uy,Bicak:1993xat,Espitia:2001cj,Bicak:1993zz}.
More realistic solutions, where $\vec{H}$ is taken into account,
are found in PN theory; the field is however very complicated already
at 1PN, and not obtained analytically (e.g. \cite{Jaranowski:2014yva}).
So, here, just to illustrate the basic features of the spin-curvature
force produced by the disk, we consider one of the simplest Newtonian
\emph{3D} models,\footnote{There are also 2D models of thin-disks such as those by Kusmin-Toomre
\cite{Miyamoto:1975zz,Baes:2008td,Vogt:2009sy}; they are however
unsuitable for studying the spin-curvature force, for having singular
tidal tensors along the disk.} the Miyamoto-Nagai disks \cite{Miyamoto:1975zz,Baes:2008td,Vogt:2009sy},
also called the ``inflated'' Kusmin model \cite{Vogt:2009sy}. The
Newtonian potential is 
\begin{equation}
U(\varrho,z)=\frac{M_{{\rm disk}}}{\sqrt{\varrho^{2}+(\sqrt{z^{2}+b^{2}}+a)^{2}}}+\frac{M_{{\rm BH}}}{r}\equiv U_{{\rm disk}}+U_{{\rm BH}}\label{eq:PotentialDisk}
\end{equation}
where $\varrho^{2}=x^{2}+y^{2}$, $M_{{\rm disk}}$ is the disk's
total mass, and $b$ and $a$ are constants with dimensions of length.
The ratio $b/a$ is a measure of the flatness of the disk \cite{Miyamoto:1975zz}.
The Laplace equation $\nabla^{2}U=\nabla^{2}U_{{\rm disk}}=-4\pi\rho$
yields the disk's density, Eq. (5) of \cite{Miyamoto:1975zz}. The
Weyl force is obtained from Eq. (\ref{eq:FWeylPN}), which reads here
$F_{{\rm Weyl}}^{i}=2\epsilon_{\ km}^{(i}G^{j),m}v^{k}S_{j}$, where
$\vec{G}$ is, to the accuracy needed for this expression, the sum
of the Newtonian fields produced by the disk and the central BH, $\vec{G}\simeq\nabla U_{{\rm disk}}+\nabla U_{{\rm BH}}$.
It can be split into the Weyl forces due to the disk and due to the
central BH, which read explicitly, in the equatorial plane 
\begin{equation}
\vec{F}_{{\rm Weyl}}=\vec{F}_{{\rm Weyldisk}}+\vec{F}_{{\rm BH}}\label{eq:FWeyldiskBH}
\end{equation}
\begin{align}
F_{{\rm Weyldisk}}^{j}=\  & \frac{M_{{\rm disk}}}{\left[r^{2}+(a+b)^{2}\right]^{\frac{3}{2}}}\left[(\vec{S}\times\vec{v})^{j}|_{j\ne z}-\epsilon_{\ ik}^{j}S^{i}v^{k}|_{i\ne z}\right.\nonumber \\
 & \left.-\frac{3r^{j}(\vec{S}\times\vec{v})\cdot\vec{r}+3(\vec{r}\cdot\vec{S})(\vec{v}\times\vec{r})^{j}}{r^{2}+\left(a+b\right)^{2}}\right]\nonumber \\
 & +\frac{(a+b)M_{{\rm disk}}}{b\left[r^{2}+(a+b)^{2}\right]^{\frac{3}{2}}}\left[\delta_{z}^{j}(\vec{S}\times\vec{v})^{z}+S^{z}\epsilon^{jkz}v_{k}\right]\label{eq:FWeyldisk}
\end{align}
\begin{equation}
\vec{F}_{{\rm BH}}=-\frac{3M_{{\rm BH}}}{r^{3}}\left[\vec{v}\times\vec{S}+\frac{2\vec{r}[(\vec{v}\times\vec{r})\cdot\vec{S}]}{r^{2}}+\frac{(\vec{v}\cdot\vec{r})\vec{S}\times\vec{r}}{r^{2}}\right]\label{eq:FBH}
\end{equation}
(Notice that $\vec{F}_{{\rm BH}}$ is the well-known expression for
the spin-orbit part of the spin-curvature force exerted by a BH on
a spinning body, e.g. Eq. (44) of \cite{Wald:1972sz}).

\subsubsection*{Quasi-circular orbits}

We shall now consider the effect of the spin-curvature force ($\vec{F}_{{\rm Mag}}+\vec{F}_{{\rm Weyl}}$)
produced by the disk on test bodies on (quasi-) equatorial circular
orbits around the central object. This demands the central object
to be much more massive than the test body, $M_{{\rm BH}}\gg\Pmass$.
We also consider the test body to be a BH, in order to preclude surface
effects (such as an ordinary Magnus effect), and ensure that the motion
is gravitationally driven. In the equatorial plane $\varrho=r$, thus
the disk's density $\rho=-\nabla^{2}U/4\pi$, that follows from (\ref{eq:PotentialDisk}),
is 
\begin{equation}
\rho=\frac{M_{{\rm disk}}}{4\pi\left[r^{2}+\left(a+b\right)^{2}\right]^{3/2}}\left[3-\frac{3r^{2}}{r^{2}+\left(a+b\right)^{2}}+\frac{a}{b}\right]\label{eq:rhoDisk}
\end{equation}
(cf. Eq. (5) of \cite{Miyamoto:1975zz}). As in Sec. \ref{sub:Isothermal-dark-matter},
there are two notable cases to consider.

\emph{Spin parallel to the symmetry axis} ($\vec{S}=S^{z}\vec{e}_{z}$).
Equation (\ref{eq:SpinPrecession}) tells us that, in this case, the
components of $\vec{S}$ are constant. The Magnus, Weyl and total
force due to the disk, $\vec{F}_{{\rm disk}}\equiv\vec{F}_{{\rm Mag}}+\vec{F}_{{\rm Weyldisk}}$,
are 
\begin{align*}
\vec{F}_{{\rm Mag}} & =-4\pi\rho\frac{(\vec{S}\cdot\vec{L})}{mr}\vec{e}_{r}\,,\\
\vec{F}_{{\rm Weyldisk}} & =\frac{(\vec{S}\cdot\vec{L})M_{{\rm disk}}}{mr\left[r^{2}+\left(a+b\right)^{2}\right]^{3/2}}\left[\frac{3r^{2}}{r^{2}+\left(a+b\right)^{2}}+\frac{a}{b}\right]\vec{e}_{r},\\
\vec{F}_{{\rm disk}} & =\frac{3(\vec{S}\cdot\vec{L})M_{{\rm disk}}}{mr\left[r^{2}+\left(a+b\right)^{2}\right]^{3/2}}\left[\frac{2r^{2}}{r^{2}+\left(a+b\right)^{2}}-1\right]\vec{e}_{r},
\end{align*}
with $\rho$ as given by Eq. (\ref{eq:rhoDisk}). So the Magnus and
Weyl forces are both radial, but have \emph{opposite} directions.
This resembles case \ref{enu:GravcaseSy} of the slab model of Sec.
\ref{sub:Toy-model-revisited}, but now the resulting force $\vec{F}_{{\rm disk}}$
is not zero. It has, for $r^{2}<(a+b)^{2}$, the same direction of
the Magnus force, and opposite direction for $r>(a+b)^{2}$; in any
case it is of qualitative different nature from $\vec{F}_{{\rm Mag}}$
or $\vec{F}_{{\rm Weyl}}$ in that it lacks the important $a/b$ term
(that can be very large for highly flattened disks). Since the forces
are radial, the orbital effect amounts to a change in the gravitational
attraction --- for $r>(a+b)^{2}$, $\vec{F}_{{\rm disk}}$ is repulsive
(attractive) when $\vec{S}$ is parallel (antiparallel) to the orbital
angular momentum $\vec{L}$; and the other way around for $r<(a+b)^{2}$.
Its relative magnitude compared to the Newtonian gravitational force
produced by the disk is $F_{{\rm disk}}/(\Pmass G)\sim vS/(r\Pmass)$. 

The effect is important in connection to the measurements of the gravitation
radiation emitted by binary systems, namely in mass estimates. These
are affected \cite{Berti:2004bd} by the spin-orbit ($\vec{F}_{{\rm SO}}\equiv\vec{F}_{{\rm BH}}$)
and spin-spin $(\vec{F}_{{\rm SS}}$) forces. $\vec{F}_{{\rm SO}}\equiv\vec{F}_{{\rm BH}}$
is given by Eq. (\ref{eq:FBH}), and like $\vec{F}_{{\rm disk}}$
it is parallel to the symmetry axis; $\vec{F}_{{\rm SS}}$ {[}not
taken into account in Eq. (\ref{eq:FBH}){]} is given by e.g. Eq.
(24) of \cite{Wald:1972sz} (it is parallel to the symmetry axis if
the spin of the central BH is along $\vec{e}_{z}$). As we have seen
in Sec. \ref{sub:Orders-of-magnitude} using a realistic density profile,
the Magnus force $F_{{\rm Mag}}$ is generically larger than both
$F_{{\rm SO}}$ and $F_{{\rm SS}}$ in systems emitting GW's within
the band of pulsar timing arrays, and is comparable to $F_{{\rm SS}}$
in the lowest part of LISA's band. In the latter, in particular, the
impact of $F_{{\rm SS}}$ in the mass measurement accuracy is significant
\cite{Berti:2004bd}; hence that of $F_{{\rm Mag}}$ (and $F_{{\rm disk}}$)
might likewise be.\\

\emph{Spin parallel to the orbital plane} ($S^{z}=0$). In this case
Eqs. (\ref{eq:SpinPrecession})-(\ref{eq:Spinevol}) tell us that
the spin vector $\vec{S}$ precesses, but remains in the plane. The
Magnus, Weyl ($\vec{F}_{{\rm Weyldisk}}+\vec{F}_{{\rm BH}}=\vec{F}_{{\rm Weyl}}$),
and total spin-curvature force, $\vec{F}=\vec{F}_{{\rm Mag}}+\vec{F}_{{\rm Weyl}}$,
are, from Eqs. (\ref{eq:FMagPN}), (\ref{eq:FWeyldiskBH})-(\ref{eq:rhoDisk}),
\begin{align}
 & \vec{F}_{{\rm Weyldisk}}=\frac{M_{{\rm disk}}}{\left[r^{2}+(a+b)^{2}\right]^{3/2}}\left[\frac{3r^{2}}{r^{2}+\left(a+b\right)^{2}}+\frac{a}{b}\right]\vec{S}\times\vec{v},\label{eq:FWeyldiskBob}\\
 & \vec{F}_{{\rm Mag}}=4\pi\rho\vec{S}\times\vec{v},\qquad\qquad\vec{F}_{{\rm BH}}=3\frac{M_{{\rm BH}}}{r^{3}}\vec{S}\times\vec{v},\label{eq:FBHBob}\\
 & \vec{F}=A(r)\vec{S}\times\vec{v};\quad A(r)\equiv\frac{(3+2a/b)M_{{\rm disk}}}{\left[r^{2}+(a+b)^{2}\right]^{3/2}}+3\frac{M_{{\rm BH}}}{r^{3}},\label{eq:FGSCbobbings}
\end{align}
with $\rho$ given by (\ref{eq:rhoDisk}). It is remarkable that all
the components of the force are parallel. In particular, for large
$a/b$ (thin disks), $\vec{F}_{{\rm Mag}}$ and $\vec{F}_{{\rm Weyldisk}}$
are \emph{qualitatively} similar. This resembles case \ref{enu:GravcaseSz}
of the slab model in Sec. \ref{sub:Toy-model-revisited}. Since $\vec{S}\times\vec{v}=vS\cos[(\omega-\Omega_{{\rm s}})t]\vec{e}_{z}$,
cf. Eq. (\ref{eq:Sxv}), the force (\ref{eq:FGSCbobbings}), similarly
to its counterpart in the DM halo of Sec. \ref{sub:Dark-matter-halo},
causes the spinning body to bob up and down (in the $\vec{e}_{z}$
direction) along the orbit. It leads also to a secular orbital precession,
which is again of the form (\ref{eq:SecularPrecession}), 
\begin{equation}
\vec{\Omega}=\frac{A(r)}{2\Pmass}\vec{S}\ ,\label{eq:SecPrecess2}
\end{equation}
where $A(r)$ is now given by Eq. (\ref{eq:FGSCbobbings}), leading
to a much larger precession. Unfortunately here we are unable to derive
an analytical expression for the oscillations along $z$ caused by
$\vec{\Omega}$ in the likes of Eq. (\ref{eq:Beating}) of Sec. \ref{sub:Isothermal-dark-matter}.
This because the first order Taylor expansion $G^{z}\simeq G_{\ ,z}^{z}|_{z=0}z$
made therein is here a bad approximation to the true value of $G^{z}$
when the body is outside the equatorial plane, due to the rapidly
varying derivative $G_{z,z}$ at the equatorial plane. This causes
Eq. (\ref{eq:Beating}) to fail, which is made clear by numerical
simulations. Still one can devise rough, but robust estimates of the
peak orbital inclination and oscillation amplitude. As explained in
Sec. \ref{sub:Isothermal-dark-matter} and caption of Fig. \ref{fig:Haloplot},
since the orbital precession $\vec{\Omega}$ is proportional to $\vec{S}$,
it is constrained by the spin precession $\vec{\Omega}_{{\rm s}}$
(Eq. (\ref{eq:SpinPrecession})), because after a time interval $t=\pi/\Omega_{{\rm s}}\equiv t_{{\rm peak}}$
the direction of $\vec{S}$, and thus of $\vec{\Omega}$ and $\langle d\vec{L}/dt\rangle$,
become inverted relative to the initial ones, so the inclination stops
increasing and starts decreasing. Approximating the inclination angle
$\alpha$ by $\Omega t$, one may estimate the peak inclination angle
and oscillation amplitude by
\begin{equation}
\alpha_{{\rm peak}}\sim\Omega t_{{\rm peak}}=\frac{\pi\Omega}{\Omega_{{\rm s}}};\qquad z_{{\rm peak}}\sim r\alpha_{{\rm peak}}=\frac{\pi r\Omega}{\Omega_{{\rm s}}}.\label{eq:aproxamplitude}
\end{equation}
Testing first the validity of these estimates in the problem of Sec.
\ref{sub:Isothermal-dark-matter}, we notice that therein $z_{{\rm peak}}$
differs from the precise result $2Z\simeq2\Omega r/\Omega_{{\rm s}}$
by a factor $\pi/2$ (corresponding to the error in approximating
the peak of a sinusoidal function by a first order Taylor expansion
at $t=0$). For the present problem, these estimates are validated
by numerical results \emph{assuming the force expressions }(\ref{eq:FWeyldiskBH})-(\ref{eq:FBH}).

It should be stressed that Eq. (\ref{eq:SecularPrecession}), with
$A(r)$ as given by (\ref{eq:FGSCbobbings}), assumes the orbit to
lie near the equator, since the force expressions (\ref{eq:FWeyldiskBH})-(\ref{eq:FBH}),
(\ref{eq:FWeyldiskBob})-(\ref{eq:FGSCbobbings}), are for the equatorial
plane. The precession $\vec{\Omega}$ will however gradually incline
the orbit; as the inclination increases, the body will be in contact
with the disk's higher density regions for shorter periods of time,
so Eq. (\ref{eq:SecularPrecession}) will gradually become a worse
approximation (it is a peak value). From relations (\ref{eq:aproxamplitude})
we see that, when $\Omega\ll\Omega_{{\rm s}}$, the peak inclination
angle is small, so the orbit remains, on the whole, close to the equatorial
plane. Otherwise, the approximation remains acceptable after several
orbits if $\Omega\ll\omega$. Noting that $\Omega_{{\rm s}}=3Gv/2\approx3M_{{\rm BH}}^{3/2}/(2r^{5/2})$
and $\omega=\sqrt{G/r}=M_{{\rm BH}}^{1/2}r^{-3/2}$, and since $S<\Pmass^{2}$
(for BHs), $M_{{\rm disk}}<M_{{\rm BH}}$, $r>2M_{{\rm BH}}$, and
we are assuming $\Pmass\ll M_{{\rm BH}}$, we have that, for not too
large $a/b$, both $\Omega\ll\Omega_{{\rm s}}$ and $\Omega\ll\omega$
are satisfied. The computation of the precise precession for an arbitrary
inclination can be done using the general expression for the force
as given in Eq. (\ref{eq:FWeylPN}), with $\vec{G}=\nabla U$ given
by Eq. (\ref{eq:PotentialDisk}).

An important conclusion that can directly be extrapolated to more
realistic models, is that the orbital precession caused by the disk
has the order of magnitude $\Omega\sim F_{{\rm Mag}}/(v\Pmass)\sim\rho S/\Pmass$,
which might possibly be measurable in a not too distant future: the
secular precession of the orbital plane of binary systems affects
the principal directions and waveforms of the emitted gravitational
radiation~\cite{Apostolatos:1994mx,LangHughes2006,Hannam:2013pra,Vecchio:2003tn,Stavridis:2009ys,Schmidt:2014iyl,Babak:2016tgq}.
In the absence of disk (thus of Magnus force), such precession reduces
to that caused by the spin-orbit and spin-spin couplings. Both are
expected to be detected in gravitational wave measurements in the
near future \cite{Hannam:2013pra,LangHughes2006,Vecchio:2003tn,Stavridis:2009ys,Schmidt:2014iyl}.
The former is is the leading one, and has magnitude of the form $\Omega_{{\rm SO}}\propto S/r^{3}$~\cite{Apostolatos:1994mx,Schmidt:2014iyl,Vecchio:2003tn};
in particular, for the precession caused by the force (\ref{eq:FBH}),
$\Omega_{{\rm SO}}\sim(M_{{\rm BH}}/\Pmass)S/r^{3}$, cf. Eqs. (\ref{eq:FGSCbobbings})-(\ref{eq:SecPrecess2}).
Comparing with the magnitude of the Magnus precession, 
\begin{equation}
\frac{\Omega}{\Omega_{{\rm SO}}}\sim\frac{\rho r^{3}}{M_{{\rm BH}}}=\frac{\rho}{\omega^{2}}\sim\frac{F_{{\rm Mag}}}{F_{{\rm SO}}}\ ,\label{eq:Omega_Omegas}
\end{equation}
cf. Eq. (\ref{eq:FmagFSO}). The the orbital precession originated
by the spin-spin couplings has approximate magnitude $\Omega_{{\rm SS}}\sim S_{{\rm BH}}S/(\Pmass vr^{4})$
(cf. e.g. Eq. (11.a) of \cite{Apostolatos:1994mx}); hence
\begin{equation}
\frac{\Omega}{\Omega_{{\rm SS}}}\sim\frac{\rho vr^{4}}{S_{{\rm BH}}}\sim\frac{F_{{\rm Mag}}}{F_{{\rm SS}}}\ ,\label{eq:Omega_SS}
\end{equation}
cf. Eq. (\ref{eq:FmagFSS}). So, the ratios amount to those of the
corresponding forces. This means that what is said in Sec. \ref{sub:Orders-of-magnitude}
and Fig. \ref{fig:fmax}, concerning the relative orders of magnitude
of the forces, applies here to the precessions. Namely, in SMBH binaries
emitting low frequency GWs, such as those within the pulsar timing
arrays band, the plots in Fig. \ref{fig:fmax} show that the Magnus
precession $\Omega$ can be the leading spin-induced orbital precession,
larger than both $\Omega_{{\rm SS}}$ and $\Omega_{{\rm SO}}$. As
an example, taking $M_{{\rm BH}}\sim10^{10}M_{{\rm \odot}}$, we have
that $\Omega/\Omega_{{\rm SO}}\gtrsim1$ for $f_{{\rm GW}}\lessapprox2\times10^{-7}{\rm Hz}({\rm \approx6yr^{-1}})$,
cf. Fig. \ref{fig:fmax}, the approximate equalities corresponding
to $\tilde{r}\sim10$, $\rho\sim6\times10^{-3}{\rm kg\,{\rm m^{-3}}}$
{[}cf. Eqs. (\ref{eq:FmagFSO}) and (\ref{eq:rhorealistic}){]}. Taking
a ``test'' companion of mass $\Pmass\sim0.1M_{{\rm BH}}=10^{9}M_{\odot}$
and spin $S=0.5\Pmass^{2}$, the Magnus precession for such setting,
$\Omega\sim0.5\rho\Pmass=10^{-9}{\rm Hz}$, amounts to an orbital
inclination of $\sim0.6$ degrees per cycle, reaching a peak angle
$\alpha_{{\rm peak}}\sim2\text{º}$ {[}cf. Eq. (\ref{eq:aproxamplitude}){]}
after a time interval $t_{{\rm peak}}=\pi/\Omega_{{\rm s}}=1{\rm yr}$
($\approx$ 3.4 cycles). If (as above) the central black hole has
spin $S_{{\rm BH}}=0.1M_{{\rm BH}}^{2}$, we have $\Omega\sim32\Omega_{{\rm SS}}$.
Considering instead, for the same binary, a frequency one order of
magnitude smaller, $f_{{\rm GW}}\sim2\times10^{-8}{\rm Hz}$, corresponding
to $\tilde{r}\sim47$, $\rho=6\times10^{-4}{\rm kg\,m^{-3}}$, the
Magnus precession $\Omega\sim10^{-10}{\rm Hz}$ becomes the dominant
spin-induced precession: $\Omega\sim10\Omega_{{\rm SO}}\sim7\times10^{2}\Omega_{{\rm SS}}$,
amounting to an orbital inclination of $\sim0.6\text{º}$ per cycle,
reaching a peak angle $\alpha_{{\rm peak}}\sim9\text{º}$ in a time
interval $t_{{\rm peak}}=\pi/\Omega_{{\rm s}}\approx50{\rm yr}$ ($\approx$
15.5 cycles).

Moreover, LISA's band is just above the frequency for which $\Omega$
meets the magnitude of $\Omega_{{\rm SS}}$ (according to the density
profile in Eq. (\ref{eq:rhorealistic}), for the conservative choice
of parameters made above); $\Omega$ being already within the same
order of magnitude as $\Omega_{{\rm SS}}$ in the lower part of LISA's
band ($f_{{\rm GW}}\sim10^{-5}{\rm Hz}$).

\section{Magnus effect in cosmology: the FLRW metric\label{sub:FLWR}}

A setup of especial interest to consider is the gravitational Magnus
effect for a spinning body moving through an homogeneous medium (or
``cloud'') representing the large scale matter distribution of our
universe. As discussed in Sec. \ref{sub:Infinite-clouds}, this is
problematic in the framework of a PN approximation, and of an analogy
with electromagnetism, in particular when the cloud is assumed infinite
(in all directions), due to the indeterminacy of $F_{{\rm Weyl}}^{\alpha}$.
General relativity however admits a well known exact solution corresponding
to an homogeneous isotropic universe (finite or not) --- the FLRW
spacetime, believed to represent the large scale structure of our
universe. The metric is 
\begin{equation}
ds^{2}=-dt^{2}+a^{2}(t)\left(\frac{dr^{2}}{1-kr^{2}}+r^{2}d\theta^{2}+r^{2}\sin^{2}\theta d\phi^{2}\right)\ .\label{eq:FLWRmetric}
\end{equation}
It is well known that this is a conformally flat metric, that is,
its Weyl tensor vanishes: $C_{\alpha\beta\gamma\delta}=0$. This makes
this metric remarkable in this context: the Weyl force vanishes, and
therefore, \emph{the total spin-curvature force exerted on a spinning
body} (if any) \emph{reduces to the Magnus force}, Eq. (\ref{eq:FMagGrav}),
\begin{equation}
F_{{\rm Weyl}}^{\alpha}=0\quad\Rightarrow\quad F^{\alpha}=F_{{\rm Mag}}^{\alpha}\ .\label{eq:FMagFLWR}
\end{equation}
Let $U^{\alpha}=U^{0}(1,v^{i})$, where $U^{0}=dt/d\tau$ and $v^{i}=U^{i}/U^{0}=dx^{i}/dt$,
be the 4-velocity of some arbitrary observer. The gravitomagnetic
tidal tensor it measures has, as only nonvanishing components, 
\begin{equation}
\mathbb{H}_{ij}=\mathbb{H}_{[ij]}=\epsilon_{ijk0}v^{k}A(t,r,\theta)\label{eq:HijFLRW}
\end{equation}
where 
\[
A(t,r,\theta)\equiv\frac{(U^{0})^{2}}{a^{2}(t)}\left[k+\dot{a}(t)^{2}-a(t)\ddot{a}(t)\right]\ .
\]
It thus reduces to the current term (responsible for the Magnus effect):
$\mathbb{H}_{\alpha\beta}=\mathbb{H}_{[\alpha\beta]}=-4\pi\epsilon_{\alpha\beta\sigma\gamma}U^{\gamma}J^{\sigma}$,
cf. Eq. (\ref{eq:H_ab_Decomp}).

For the observers at rest ($\vec{v}=0$) in the coordinate system
of (\ref{eq:FLWRmetric}), the gravitomagnetic tidal tensor vanishes,
$\mathbb{H}_{\alpha\beta}=0$. Therefore, the spin-curvature force
on a spinning body at rest vanishes: $F^{\alpha}=-\mathbb{H}^{\beta\alpha}S_{\beta}=0$.
This is the expected result: a body at rest in the coordinates of
(\ref{eq:FLWRmetric}) is well known to be comoving with the background
fluid, so relative to it the spatial mass/energy current $\vec{J}$
is zero, implying that the Magnus force (\ref{eq:FMagvec}) vanishes.

Consider now an observer of 4-velocity $U^{\alpha}=U^{0}(1,v^{i})$,
moving with respect to the coordinate system of (\ref{eq:FLWRmetric});
i.e, with a \emph{``peculiar''} velocity $\vec{v}\ne0$. Such an
observer measures a nonzero \emph{antisymmetric} gravitomagnetic tidal
tensor (\ref{eq:HijFLRW}); this means that a spinning body moving
with velocity $\vec{v}$ suffers a spin-curvature force $F^{\alpha}=-\mathbb{H}^{\beta\alpha}S_{\beta}$,
whose components read, in terms of the metric parameters, 
\begin{align}
F^{0}=0;\qquad & \vec{F}=A(t,r,\theta)\vec{S}\times\vec{v}\ ,\label{eq:ForceFLWR1}
\end{align}
where $(\vec{v}\times\vec{S})^{i}\equiv\epsilon_{\ jk0}^{i}v^{j}S^{k}$,
and the coordinate system is that of (\ref{eq:FLWRmetric}). On the
other hand, the energy-momentum tensor corresponding to the metric
(\ref{eq:FLWRmetric}) is that of a perfect fluid, $T^{\alpha\beta}=(\rho+p)u^{\alpha}u^{\beta}+pg^{\alpha\beta}$,
where $u^{\alpha}$ is the fluid's 4-velocity ; thus $J^{\alpha}=-T^{\alpha\beta}U_{\beta}=\gamma(\rho+p)u^{\alpha}-pU^{\alpha}$,
where $\gamma\equiv-u_{\alpha}U^{\alpha}$. From Eq. (\ref{eq:FMagFLWR}),
we have that 
\begin{equation}
F^{\alpha}=4\pi\gamma(\rho+p)\epsilon_{\ \beta\sigma\gamma}^{\alpha}u^{\beta}S^{\sigma}U^{\gamma}\ ;\label{eq:FLWRCov}
\end{equation}
since the fluid is at rest in the coordinate system of (\ref{eq:FLWRmetric}),
we have that $u^{\alpha}=\delta_{0}^{\alpha}$ and $\gamma=U^{0}$,
and therefore\footnote{One may check that Eq. (\ref{eq:ForceFLWR1}) indeed equals (\ref{eq:ForceFLWR2})
using the Friedman equations (e.g. Eqs. (5.11)-(5.12) of \cite{Hawking:1973uf})
\[
\frac{\dot{a}(t)^{2}+k}{a(t)^{2}}=\frac{8\pi\rho+\Lambda}{3};\qquad-\frac{\ddot{a}(t)}{a(t)}=\frac{4\pi}{3}(\rho+3p)-\frac{\Lambda}{3}
\]
} 
\begin{equation}
\vec{F}=-4\pi(\rho+p)(U^{0})^{2}\vec{v}\times\vec{S}\ .\label{eq:ForceFLWR2}
\end{equation}
This equation leads to an important conclusion: in the general case
that $\rho+p\ne0$, a spinning body \emph{arbitrarily moving} in the
FLRW metric \emph{suffers} \emph{a net force in the direction of the
Magnus effect}; such force is \emph{the only force} that acts on the
body, deviating it from geodesic motion. If the weak energy condition
holds, $\rho+p\ge0$ (cf. e.g. Eq. (9.2.19) of Ref. \cite{Wald:1984}),
we have that $\vec{F}$ ($=\vec{F}_{{\rm Mag}}$) is parallel to $\vec{S}\times\vec{v}$,
\emph{similarly to the Magnus effect of fluid dynamics}.\footnote{We note that this result is contrary to that estimated in \cite{Cashen:2016neh}.
Equation (\ref{eq:ForceFLWR2}) is, supposedly, the exact relativistic
solution for the problem addressed in \cite{Cashen:2016neh}: the
force exerted on a spinning body moving in a medium (``field of stars'')
of uniform density $\rho$, representing the large scale stellar distribution
of the universe. For the accuracy at hand in \cite{Cashen:2016neh},
Eq. (\ref{eq:ForceFLWR2}) yields $\vec{F}=-4\pi\rho\vec{v}\times\vec{S}$.
The force estimated in \cite{Cashen:2016neh} however does not agree
with this result, having even the opposite direction (i.e., it is
anti-Magnus).} This is the case for ordinary matter, radiation, or DM. In the case
$\rho=-p$, corresponding to cosmological constant/dark energy, the
Magnus force vanishes, $\vec{F}=0$. This can also be equivalently
seen from the fact that the Ricci tensor for a cosmological constant
is $R_{\alpha\beta}=\Lambda g_{\alpha\beta}$, which, via Eqs. (\ref{Riemann decomp-R})-(\ref{eq:H_ab_Decomp-1})
(with $C_{\alpha\beta\gamma\delta}=0$) implies $\mathbb{H}_{\alpha\beta}=0$
for all observers. Or from the fact that the effective energy-momentum
tensor of a cosmological constant is $T^{\alpha\beta}=pg^{\alpha\beta}$,
which does not lead to any \emph{spatial} mass-energy currents with
respect to any observer: $J^{\alpha}=-pU^{\alpha}$, so (see Eq. (\ref{eq:Spaceprojector}))
$h_{\ \beta}^{\alpha}J^{\beta}=0$ for all $U^{\alpha}$. Other dark
energy models have been proposed however, for which $\rho\ne-p$ (e.g.
\cite{Chavanis2014,ZlatevWangSteinhard_PRL1999,Gorini:2004by,Vikman:2004dc}),
and that, as such, would generate a Magnus force. Candidates even
exist for which $\rho<-p$ \cite{Vikman:2004dc} (violating the weak
energy condition), leading to an anti-Magnus force. 

We thus come to another interesting conclusion: the gravitational
Magnus force on spinning celestial bodies acts as a probe for the
matter/energy content of the universe, in particular, for the ratio
$\rho/p$, and for the different dark energy candidates. The bodies
that should be more affected are rotating galaxies (which one can
treat as extended bodies) with large peculiar velocities $\vec{v}$.
The effect is any case very small, given the constrains in place:
WMAP results \cite{WMAP} show our universe to be nearly flat, implying
an average density close to the critical value $\rho\approx10^{-26}{\rm kg}\,{\rm m}^{-3}$;
assuming an equation of state of the form $p=-w\rho$, the parameter
$w$ is constrained from observations to be within $-1.2\lesssim w<-1/3$
(e.g. \cite{EllisMaartensMacCallum,Vikman:2004dc}). Taking $w\sim-0.8$,
considering a galaxy with the same diameter $2R\approx55{\rm kpc}$
and Møller radius $S/m$ as the Milky way, and moving with a peculiar
velocity $v\sim10^{3}{\rm km}\,{\rm s}^{-1}$ (a reasonably high value
\cite{Springob:2007vb}), it would take about $10^{4}\times$ age
of the universe in order for the deflection caused by the Magnus force,
$\Delta x\sim(1/2)\Delta t^{2}F/m\sim2\pi\Delta t^{2}\rho vS/m$,
be of order the galaxy's size.

Finally, we note that a reciprocal force $\vec{F}_{{\rm {\rm body,cloud}}}$,
in the likes of that computed in Secs. \ref{sub:Toy-model-revisited}-\ref{sub:Dark-matter-halo},
cannot be computed here; such integrals assume that everywhere the
metric can be taken as nearly flat, so that vectors at different points
can be added. The metric (\ref{eq:FLWRmetric}), however, is not asymptotically
flat ($g_{rr}$ diverging at infinity), so any such integrals are
not valid mathematical operations here. This is the general situation
in the exact theory: $\vec{F}_{{\rm {\rm cloud,body}}}\equiv\vec{F}$
(the spin-curvature force) is a physical force always well defined,
whereas its reciprocal, $\vec{F}_{{\rm {\rm body,cloud}}}$, is not.

\section{Conclusions}

In the wake of earlier works \cite{Font:1998sc,Okawa:2014sxa} where
a gravitational analogue to the Magnus effect of fluid dynamics has
been suggested, we investigated its existence in the rigorous equations
of motion for spinning bodies in general relativity (Mathisson-Papapetrou
equations). We have seen that not only the effect takes place, as
it is a fundamental part of the spin-curvature force. Indeed, as made
manifest by writing it in tidal tensor form, such force can be exactly
split into two parts: one due to the magnetic part of the Weyl tensor
(the Weyl force $F_{{\rm Weyl}}^{\alpha}$), plus the Magnus force
($F_{{\rm Mag}}^{\alpha}$), which arises whenever, relative to the
body, there is a spatial mass/energy current nonparallel to its spin
axis, and has the same direction as the Magnus effect of fluid dynamics.
The effect was seen moreover to have a close analog in electromagnetism;
namely in the force exerted in a magnetic dipole inside a current
slab. Such setting, and its gravitational counterpart, provided useful
toy models for the understanding of the contrast between the two parts
of the spin-curvature force: the Magnus force, which depends only
on the body's angular momentum and on the mass-density current of
the medium relative to it, and the dependence of the Weyl force on
the details and boundary conditions of the system. This dependence
shows clearly in the astrophysical systems studied, and means also
that some problems tried to be addressed in earlier literature were
not well posed.

Gravitational Magnus effects could have interesting signatures in
cosmology or in astrophysics. They are shown to lead to secular orbital
precessions that might be detectable by future astrometric or gravitational-wave
observations. These effects are considered here in three astrophysical
settings: DM halos, BH accretion disks, and the FLRW metric. In DM
halos, due to their low density, the effects are typically small,
being more noticeable for bodies with large ``Møller radius'' $S/\Pmass$,
yielding a further possible test for the existence of DM and its density
profile (in addition e.g. to the dynamical friction effect proposed
in \cite{Pani:2015qhr}). In accretion disks, due to their high density,
the orbital precession caused by the Magnus force is more important;
it can be comparable, or larger, than the spin-spin and spin-orbit
precessions, and, in the future, possibly detectable in the gravitational
radiation emitted by binary systems with a disk. In the FLRW spacetime
(describing the standard cosmological model), is is shown that a Magnus
force acts on any spinning body moving with respect to the background
fluid, it is the only covariant force acting on the body (deviating
it from geodesic motion), and has the same direction of its fluid
dynamics counterpart. It should affect primarily galaxies with large
peculiar velocities $\vec{v}$. All forms of matter/energy give rise
to such Magnus force except for dark energy \emph{if} modeled with
a cosmological constant, so it acts as a probe for the nature of the
energy content of the universe. 
\begin{acknowledgments}
We thank J. Natário for enlightening discussions, and C. Moore and
D. Gerosa for correspondence and very useful suggestions. We also
thank the anonymous Referees for valuable remarks and suggestions
that helped us improve this work. V.C. acknowledges financial support
provided under the European Union's H2020 ERC Consolidator Grant ``Matter
and strong-field gravity: New frontiers in Einstein's theory'' Grant
Agreement No. MaGRaTh--646597. Research at Perimeter Institute is
supported by the Government of Canada through Industry Canada and
by the Province of Ontario through the Ministry of Economic Development
$\&$ Innovation. L.F.C. is funded by FCT/Portugal through Grant No.
SFRH/BDP/85664/2012. This project has received funding from the European
Union's Horizon 2020 research and innovation programme under the Marie
Sklodowska-Curie Grant Agreement No. 690904, and from FCT/Portugal
through the project UID/MAT/04459/2013. The authors would like to
acknowledge networking support by the COST Action CA16104. 
\end{acknowledgments}

\appendix

\section{Infinite clouds and Fubini's theorem\label{sec:Fubini_Theorem}}

In Sec.~\ref{sub:EM_Reciprocal-probem} we considered two different
ways of obtaining an infinite cloud in all directions (a slab orthogonal
to the $y$ axis delimited by $-h/2<y<h/2$, and a slab orthogonal
to the $z$ axis delimited by $-h/2<z<h/2$, in the limit $h\rightarrow\infty$),
and seen that the force $\vec{F}_{{\rm dip,cloud}}$ exerted on them
by the magnetic dipole is different in each case. We consider here
yet another route: an infinite cloud obtained by taking the limit
$R\rightarrow\infty$ of a sphere of radius $R$. From Eq. (\ref{eq:InteriorIntegral}),
which holds for any sphere containing the dipole (finite or infinite),
we have then $\vec{F}_{{\rm dip,cloud}}=-8\pi\mu j\vec{e}_{y}/3$;
which is yet another different result, comparing to (\ref{eq:ForcedipoleCloudy}),
and to the force exerted on the slab orthogonal to $z$ ($\vec{F}_{{\rm dip,cloud}}=0$).
These inconsistencies stem from a fundamental a mathematical principle,
embodied in Fubini's theorem \cite{Fubini,PeterWalker} (in turn related
with Riemann's series theorem, e.g. \cite{VICKERS2009}); namely,
that the multiple integral of a function which is not \emph{absolutely
convergent} (i.e., the integral of the absolute value of the integrand
does not converge) depends on the way the integration is performed.
This is the case of the integrals mentioned above. Take e.g. the ``spherical''
infinite cloud; we have 
\begin{align*}
\int_{r_{0}<r<R}|B_{{\rm dip}}^{z}|d^{3}x= & \mu\int_{0}^{\pi}\int_{0}^{2\pi}\int_{r_{0}}^{R}\frac{|1-3\cos^{2}\theta|}{r}\\
 & \sin\theta drd\phi d\theta\ =\ \frac{16\pi\mu}{3\sqrt{3}}\ln\left[\frac{R}{r_{0}}\right]
\end{align*}
where $r_{0}$ is the radius of some minimal sphere enclosing the
dipole. This integral \emph{diverges} for $R\rightarrow\infty$ (and/or
$r_{0}\rightarrow0$), whereas $\int_{r_{0}<r<R}B_{{\rm dip}}^{z}d^{3}\vec{x}=0$.

To make the connection with Fubini's theorem explicit, we go back
to the slabs of Sec. \ref{sub:EM_Reciprocal-probem}, and write the
integrals therein in terms of Cartesian coordinates. Since both slabs
are infinite in the $x$ direction, we have\footnote{The lower bound $R$ in the integrals actually amounts to leave a
cube of side $2R$ outside the integral, not a sphere of radius $R$;
that does not however have any effect on the outcome, in the limit
$h\rightarrow\infty$.}

\begin{align*}
 & \int_{r>R}\vec{B}_{{\rm dip}}d^{3}x=8\mu\vec{e}_{z}\int_{R}^{y_{{\rm m}}}\int_{R}^{z_{{\rm m}}}\int_{R}^{\infty}\frac{2z^{2}-x^{2}-y^{2}}{r^{5}}dxdzdy\\
 & =-8|_{y=R}^{y=y_{{\rm m}}}|_{z=R}^{z=z_{{\rm m}}}\left\{ |_{x=R}^{x=\infty}\arctan\frac{Ry}{zr}\right\} \mu\vec{e}_{z}\\
 & =8|_{y=R}^{y=y_{{\rm m}}}|_{z=R}^{z=z_{{\rm m}}}\left[\arctan\frac{Ry}{z\sqrt{y^{2}+z^{2}+R^{2}}}-\arctan\frac{y}{z}\right]\mu\vec{e}_{z}
\end{align*}
where $y_{{\rm m}}$ and $z_{{\rm m}}$ denote the upper integration
bounds for the respective coordinates. In the first equality we noticed
that the only surviving component is along $\vec{e}_{z}$, and that,
by symmetry, one needs only to integrate over the octant $x>R$, $y>R$,
$z>R$, multiplying then the result by a factor of 8. The slab orthogonal
to the $y$ axis corresponds to setting $z_{{\rm m}}=\infty$, $y_{m}=h/2$;
taking \emph{afterwards} the limit $h\rightarrow\infty$ amounts to
\begin{align}
 & \int_{r>R}\vec{B}_{{\rm dip}}d^{3}x=8|_{y=R}^{y=\infty}\left\{ |_{z=R}^{z=\infty}\left[-\arctan\frac{y}{z}\right.\right.\nonumber \\
 & \left.\left.+\arctan\frac{Ry}{z\sqrt{y^{2}+z^{2}+R^{2}}}\right]\right\} \mu\vec{e}_{z}\ =\ \frac{4}{3}\pi\mu\vec{e}_{z}\ .\label{eq:Intycart}
\end{align}
The slab orthogonal to the $z$ axis corresponds to $y_{{\rm m}}=\infty$,
$z_{{\rm m}}=h/2$; taking \emph{afterwards} the limit $h\rightarrow\infty$
amounts to 
\begin{align}
 & \int_{r>R}\vec{B}_{{\rm dip}}d^{3}x=8|_{z=R}^{z=\infty}\left\{ |_{y=R}^{y=\infty}\left[-\arctan\frac{y}{z}\right.\right.\nonumber \\
 & \left.\left.+\arctan\frac{Ry}{z\sqrt{y^{2}+z^{2}+R^{2}}}\right]\right\} \mu\vec{e}_{z}=-\frac{8\pi}{3}\mu\vec{e}_{z}\ .\label{eq:Intzcart}
\end{align}
The integrals (\ref{eq:Intycart}) and (\ref{eq:Intzcart}) differ
only in the order of the integrations (or of the infinite limits)
over the $y$ and $z$ coordinates; yet the outcome is very different.
This is a consequence of Fubini's theorem \cite{Fubini,PeterWalker}:
the double integral of a function which is not absolutely convergent
\emph{is not}, \emph{in general}, well defined; when written as a
iterative integral, the result may depend on the order of integration.
Since the problem of considering an infinite cloud by taking initially
a slab of width $h$ either along $y$ or along $z$ (and taking afterwards
the limit $h\rightarrow\infty$), boils down to the order of the iterations
in the multiple integrals (\ref{eq:Intycart}) and (\ref{eq:Intzcart}),
this just tells us that $\vec{F}_{{\rm dip,cloud}}$, as defined by
Eq. (\ref{eq:F_dip_cloud}), \emph{is not a well defined quantity
for an infinite (in all directions) cloud}.

\section{Action-reaction law and magnetic and gravitomagnetic interactions\label{sec:Action-reaction-law}}

In the previous literature \cite{Okawa:2014sxa,Cashen:2016neh} where
a gravitational analogue of the Magnus effect (or a ``gravitomagnetic
dynamical friction'') was implied, the force on the spinning body
was not directly computed from the concrete equations of motion for
the body as done herein, but instead indirectly inferred from estimating
the body's effect on the surrounding matter/other bodies, and then
naively applying a Newtonian-like action-reaction principle. This
is problematic however. Although for the toy model, \emph{stationary
settings} of Sec. \ref{sub:Toy-model-revisited}, we have shown that
the force $\vec{F}_{{\rm body,cloud}}$ exerted by the body on the
cloud indeed equals minus its reciprocal $\vec{F}\equiv\vec{F}_{{\rm cloud,body}}$
(the force exerted by the cloud on the body), this is not true in
general \emph{dynamics}: the gravitomagnetic interactions, similarly
to their magnetic counterparts, \emph{do not obey} an action-reaction
law of the type of $\vec{F}_{{\rm A,B}}=-\vec{F}_{{\rm B,A}}$. For
instance, the force exerted by the spinning body on an individual
particle of the cloud does not (contrary to the belief in some literature)
equal minus the force exerted by the particle on the spinning body.
This is a leading order effect, which is a consequence of the interchange
between field momentum and the ``mechanical'' momentum that the
bodies/matter possess. Below we discuss this issue in detail, starting
by the electromagnetic case.

\subsection{Magnetism\label{sub:3rdLawMagnetism}}

It is well known that electromagnetic forces do not obey the action
reaction law, and that the center of mass position of a system of
charged bodies is not a fixed point. Notice that this does not imply
any violation of the conservation equations for the total energy-momentum
tensor; in fact, it is a necessary consequence of the interchange
between mechanical momentum of the bodies and electromagnetic field
momentum . We analyze next some examples relevant to the problem at
hand. 
\begin{figure}
\includegraphics[width=1\columnwidth]{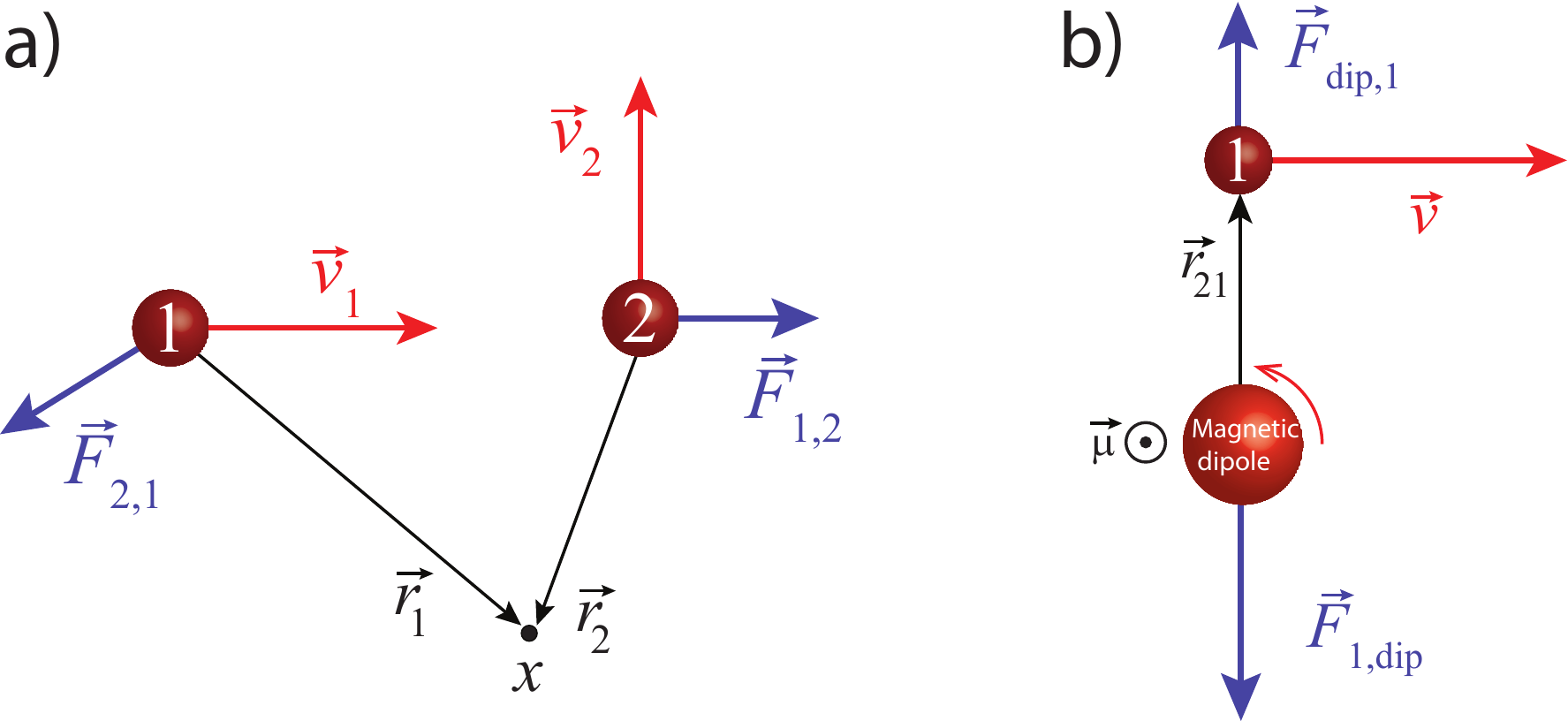} \protect\protect\protect\protect\protect\caption{\label{fig:ActionReactionEM}Two situations where action-reaction
law is not obeyed in electrodynamics: a) two orthogonality moving
point charges (Feynman paradox); b) interaction of a magnetic dipole
with a particle of the cloud. In a), the electric forces each particle
exerts on the other are nearly opposite, but particle 2 exerts a nonvanishing
magnetic force $q\vec{v}_{1}\times\vec{B}_{2}$ on particle 1, whereas
particle 1 does not exert any magnetic force on particle 2. In b),
the force exerted by the dipole (particle 2) on the moving particle
1 is only minus half the force that the latter exerts on the former,
$\vec{F}_{{\rm 1,dip}}=-2\vec{F}_{{\rm dip,1}}$.}
\end{figure}

\subsubsection{Simplest example: Two moving point charges (Feynman paradox)\label{sub:FeymanParadox}}

Consider the setup in Fig. \ref{fig:ActionReactionEM}a: a pair of
point particles with equal charge $q$ and mass $M$, and with orthogonal
velocities, one (particle 1) moving directly towards the other with
$\vec{v}_{1}=v_{1}\vec{e}_{x}$, and the other moving orthogonally
with $\vec{v}_{2}=v_{2}\vec{e}_{y}$. To first ``post-Coulombian''
\cite{Will:1993ns,Costa:2016iwu,Gralla:2010xg} order, the electric
and magnetic fields generated by a moving charge are \cite{Costa:2016iwu}
(cf. also \cite{Will:1993ns,Jackson:1998nia}) 
\begin{equation}
\vec{E}_{{\rm a}}=q(1+\varphi_{{\rm a}})\frac{\vec{r}_{{\rm a}}}{r_{{\rm a}}^{3}}-\frac{1}{2}q\frac{\vec{a}_{{\rm a}}}{r_{{\rm a}}};\qquad\vec{B}_{{\rm a}}=\frac{q}{r_{{\rm a}}^{3}}\vec{v}_{{\rm a}}\times\vec{r}_{{\rm a}}\label{eq:BPC}
\end{equation}
where $\vec{r}_{{\rm a}}\equiv\vec{x}-\vec{x}_{{\rm a}}$, $\vec{x}$
is the point of observation, $\vec{x}_{{\rm a}}$ is the instantaneous
position of particle ``${\rm a}$'', $\vec{a}_{{\rm a}}$ its acceleration,
and 
\[
\varphi_{{\rm a}}\equiv\frac{v_{{\rm a}}^{2}}{2}-\frac{1}{2}(\vec{r}_{{\rm a}}\cdot\vec{a}_{{\rm a}})-\frac{3}{2}\frac{(\vec{r}_{{\rm a}}\cdot\vec{v}_{{\rm a}})^{2}}{r_{{\rm a}}^{2}}\ .
\]
For a system of two bodies, to 1PC accuracy, $\vec{a}_{{\rm a}}$
\emph{in the equation above} is to be taken as the Coulomb force caused
by the other body, divided by the mass: $\vec{a}_{{\rm a}}=(q^{2}/M)\vec{r}_{12}/r_{12}^{3}$.
The electric force exerted by particle 1 on particle 2 is then 
\[
\vec{F}_{{\rm EL1,2}}=q\vec{E}_{1}=q^{2}(1+\varphi_{1}-\frac{q^{2}}{2Mr_{12}})\frac{\vec{r}_{12}}{r_{12}^{3}}
\]
and its reciprocal, the force $\vec{F}_{{\rm EL2,1}}=q\vec{E}_{2}$
exerted by particle 2 on particle 1, is 
\[
\vec{F}_{{\rm EL2,1}}=q^{2}(1+\varphi_{2}-\frac{q^{2}}{2Mr_{21}})\frac{\vec{r}_{21}}{r_{21}^{3}}=-q^{2}(1+\varphi_{2}-\frac{q^{2}}{2Mr_{12}})\frac{\vec{r}_{12}}{r_{12}^{3}}
\]
where we noted that $\vec{r}_{12}\equiv\vec{r}_{1}-\vec{r}_{2}=-\vec{r}_{21}$.
Thus, the electric forces are of opposite direction and of nearly
equal magnitude: $\vec{F}_{{\rm EL1,2}}\approx-\vec{F}_{{\rm EL2,1}}$
(they slightly differ because $\mbox{\ensuremath{\varphi_{1}\ne\varphi_{2}}}$).
The same however does not apply to the magnetic forces: particle 1
exerts no magnetic force on particle 2, $\vec{F}_{{\rm M1,2}}=0$,
since, at the site of particle 2, $\vec{B}_{1}=q\vec{v}_{1}\times\vec{r}_{12}/r_{12}^{3}=0$,
whereas particle 2 exerts a nonvanishing magnetic force on particle
1: 
\[
\vec{F}_{{\rm M2,1}}=q\vec{v}_{1}\times\vec{B}_{2}=-\frac{q^{2}v_{1}v_{2}}{r_{12}^{2}}\vec{e}_{y}\ .
\]
Therefore 
\[
\vec{F}_{1,2}=\vec{F}_{{\rm EL1,2}}\ne\vec{F}_{2,1}=\vec{F}_{{\rm EL2,1}}+\vec{F}_{{\rm M2,1}}
\]
showing that an action-reaction law (in a naive Newtonian sense) does
not apply here. This example, sometimes called the ``Feynman paradox,''
is due to Feynman, see Ref. \cite{Feynman:1963uxa} p. 26-5 and 27-11,
and Fig. 26-6 therein. Further discussions on this problem are given
in e.g Sec. 8.2.1 of Ref. \cite{GriffithsBook}, and, in more detail,
in Ref. \cite{PageAdams}.

\subsubsection{Interaction of a magnetic dipole with individual particles of the
cloud\label{sub:Magnetic dipole Cloud particle}}

Consider a system composed of a magnetic dipole $\vec{\mu}=\mu\vec{e}_{z}$
placed at the origin (call it particle 2), and a particle of the cloud
(particle 1, of charge $q$) in the equatorial plane, and at the instantaneous
position depicted in Fig. \ref{fig:ActionReactionEM}b. The magnetic
field created by the magnetic dipole is given by Eq. (\pageref{eq:Bdip})
; the force it exerts on particle 1 is thus 
\begin{equation}
\vec{F}_{{\rm dip,1}}=q\vec{v}\times\vec{B}_{{\rm dip}}=-q\frac{\vec{v}\times\vec{\mu}}{r_{21}^{3}}=q\frac{v\mu}{r_{21}^{3}}\vec{e}_{y}\ .\label{eq:Fdipcharge}
\end{equation}
Let us now compute the force that particle 1 exerts on the dipole.
The magnetic field created by a generically moving charge is, from
Eq. (\ref{eq:BPC}), $\vec{B}_{{\rm charge}}=q\vec{v}\times\vec{r}/r^{3}$;
the force it exerts on the dipole is $F_{{\rm charge,dip}}^{i}=B_{{\rm charge}}^{j,i}\mu_{j}\equiv\nabla^{i}(\vec{\mu}\cdot\vec{B}_{{\rm charge}})$,
cf. Eqs. (\ref{eq:FEM_TT})-(\ref{eq:FEMBook}); explicitly: 
\begin{equation}
\vec{F}_{{\rm charge,dip}}=q\frac{\vec{\mu}\times\vec{v}}{r^{3}}-3q\frac{(\vec{v}\times\vec{r})\cdot\vec{\mu}}{r^{5}}\vec{r}\ .\label{eq:Fchargedip}
\end{equation}
Hence, the force exerted by particle 1 on the dipole is 
\[
\vec{F}_{{\rm 1,dip}}=q\frac{\vec{\mu}\times\vec{v}}{r_{21}^{3}}-3q\frac{(\vec{v}\times\vec{r}_{12})\cdot\vec{\mu}}{r_{21}^{5}}\vec{r}_{12}=-\frac{2v\mu q}{r_{21}^{3}}\vec{e}_{y}\ .
\]
Comparing with Eq. (\ref{eq:Fdipcharge}), again we see that action
does not meet reaction: $\vec{F}_{{\rm 1,dip}}=-2\vec{F}_{{\rm dip,1}}$
(the sign is opposite as expected, but the magnitudes do not meet).

Let us now consider particle 3 lying at $\vec{x}_{3}=r_{3}\vec{e}_{z}$,
and moving (again) with velocity $\vec{v}=v\vec{e}_{x}$. The force
that the dipole exerts on it, $\vec{F}_{{\rm dip,3}}=q\vec{v}\times\vec{B}_{{\rm dip}}(x_{3})$,
is, from Eq. (\ref{eq:Bdip}), 
\[
\vec{F}_{{\rm dip,3}}=-q\frac{\vec{v}\times\vec{\mu}}{r_{23}^{3}}+q\frac{3(\vec{\mu}\cdot\vec{r}_{23})\vec{v}\times\vec{r}_{23}}{r_{23}^{5}}=-2\frac{v\mu q}{r_{23}^{3}}\vec{e}_{y}\ .
\]
Its reciprocal (the force that particle 3 exerts on the dipole) is,
from Eq. (\ref{eq:Fchargedip}), 
\[
\vec{F}_{{\rm 3,dip}}=q\frac{\vec{\mu}\times\vec{v}}{r_{23}^{3}}-3q\frac{(\vec{v}\times\vec{r}_{32})\cdot\vec{\mu}}{r_{23}^{5}}\vec{r}_{32}=\frac{v\mu q}{r_{23}^{3}}\vec{e}_{y}
\]
since the second term of the second expression vanishes. Thus, again,
action does not meet reaction, only now it is the magnitude of the
force on the particle that is twice that on the dipole: $\vec{F}_{{\rm dip,3}}=-2\vec{F}_{{\rm 3,dip}}$.

\subsubsection{A magnetic dipole and an infinite straight wire\label{sub:magnetic-dipole straight wire}}

Consider again a magnetic dipole $\vec{\mu}=\mu\vec{e}_{z}$ placed
at the origin, and an infinitely long wire placed along the straightline
(parallel to the $x$ axis) defined by $y=y_{0}$, $z=0$, with a
current $\vec{I}=\Sigma\vec{j}$ flowing through it in the positive
$x$ direction, $\vec{j}=j\vec{e}_{x}$. $\Sigma$ is the cross sectional
area of the wire. The force exerted by the magnetic dipole (placed
at the origin, and with $\vec{\mu}=\mu\vec{e}_{z}$) on the wire is,
from Eq. (\ref{eq:Bdip}), 
\[
\vec{F}_{{\rm dip,wire}}=\int_{{\rm wire}}\vec{j}\times\vec{B}_{{\rm dip}}=-\int_{{\rm wire}}\frac{\vec{j}\times\vec{\mu}}{r^{3}}=\frac{2\mu I}{y_{0}^{2}}\vec{e}_{y}.
\]
Let us now compute its reciprocal, i.e. the force that the wire exerts
on the dipole. The magnetic field generated by the wire is well known
to be (e.g. Sec. 5.3 of \cite{GriffithsBook}, or Eqs. (14.22)-(14.24)
of \cite{Feynman:1963uxa}) 
\[
\vec{B}_{{\rm wire}}=\frac{2I}{(z^{2}+y'^{2})}\left[y'\vec{e}_{z}-z\vec{e}_{y}\right]
\]
where $y'=y-y_{0}$. Thus $B_{{\rm wire}}^{i,j}$ has the only nonvanishing
components $B_{{\rm wire}}^{z,y}=B_{{\rm wire}}^{y,z}=2I(z^{2}-y'^{2})(z^{2}+y'^{2})^{-2}$.
Therefore, the force exerted on the dipole, $\vec{F}_{{\rm wire,dip}}=B_{{\rm wire}}^{j,i}\mu_{j}\vec{e}_{i}\equiv\nabla(\vec{\mu}\cdot\vec{B}_{{\rm wire}})$,
is 
\[
\vec{F}_{{\rm wire,dip}}=-\frac{2\mu I}{y_{0}^{2}}\vec{e}_{y}=-\vec{F}_{{\rm dip,wire}}\ .
\]
So, in this case, the action-reaction law \emph{is obeyed}, just like
for the current slabs in Sec. \ref{sub:A-current-slab}. This is the
expected result because one is dealing here with \emph{magnetostatics},
where there cannot be an exchange between mechanical and field momentum,
for the latter is constant and equal to zero (the Poynting vector
is zero, since the electric field is zero).

\subsection{Gravitomagnetism\label{sub:Gravitomagnetism_action_reaction}}

Analogous examples to the ones above can be given in gravity --- two
point masses momentarily in orthogonal motion, the interaction of
the spinning body with individual particles of the cloud (Sec. \ref{sub:PN_slab_Reciprocal_Fporce}),
and with the entire cloud --- with entirely analogous conclusions.
(As for the infinite wire of Sec. \ref{sub:magnetic-dipole straight wire},
it cannot be mirrored here since the metric of an infinitely long
cylindrical mass is not asymptotically flat). Below we discuss in
detail the especially important second example.

\subsubsection{Interaction of a spinning body with individual particles of the cloud\label{sub:Gravity interaction with individual particles of the cloud}}

\begin{figure}
\includegraphics[width=0.22\textwidth]{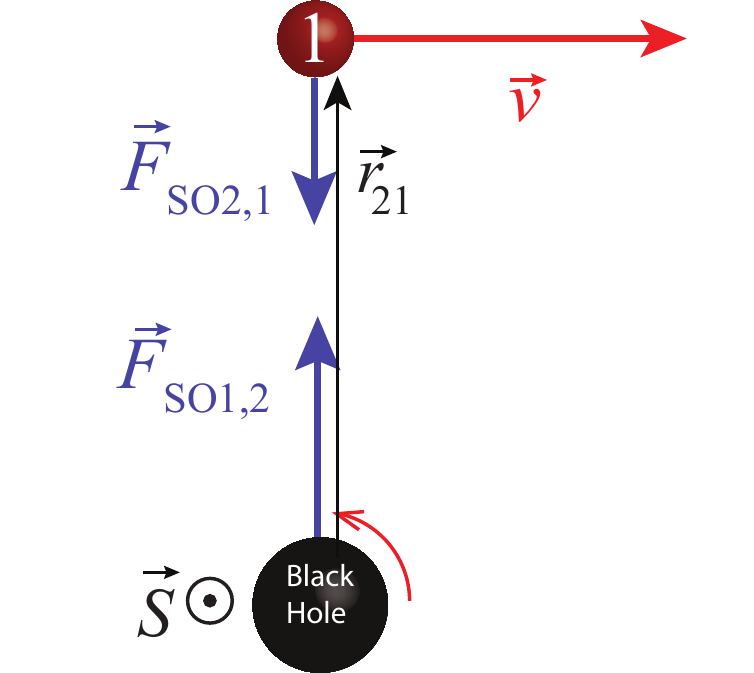} \protect\protect\protect\protect\protect\caption{\label{fig:ActionReactionGrav}A situation where the action reaction
law is not obeyed in PN gravity (analogue of Fig. \ref{fig:ActionReactionEM}b):
the spin-orbit force $\vec{F}_{{\rm SO}2,1}$ exerted by the spinning
body (e.g. a BH) on a moving particle is not the same as the spin-orbit
force $\vec{F}_{{\rm SO}1,2}$ that the latter exerts on the former:
$\vec{F}_{{\rm SO}1,2}=-3\vec{F}_{{\rm SO}2,1}/2$.}
\end{figure}

Consider a system composed of a spinning body \emph{momentarily} at
rest (body 2, of mass $M_{2}$, and angular momentum $\vec{S}$),
and a point mass (body 1, of mass $M_{1}$) moving with velocity $\vec{v}_{1}=\vec{v}$,
as illustrated in Fig. \ref{fig:ActionReactionGrav}. 

To 1.5PN order, the coordinate acceleration of a (spinning or nonspinning)
body in a gravitational field is generically given by Eqs. (\ref{eq:CoordAccelForce}),
(\ref{eq:FGPN}), (\ref{eq:FI}). The coordinate acceleration of the
spinning body (body 2), due to the gravitational field generated by
the moving point mass 1, is then (since $\vec{v}_{2}=0$) 
\begin{equation}
\frac{d^{2}\vec{x}_{2}}{dt^{2}}=(1-2U_{1})\vec{G}_{1}+\frac{1}{M_{2}}\vec{F}\label{eq:coorAcBody1}
\end{equation}
where $U_{1}=M_{1}/r_{1}$ is the Newtonian potential of body 1 and
$\vec{F}$ is the spin-curvature force on body 2. The latter amounts
to the whole spin-orbit force that acts on body 2, so we may write
$\vec{F}=\vec{F}_{{\rm SO}1,2}$. From Eq. (\ref{eq:FGPN}), 
\begin{equation}
F_{{\rm SO}1,2}^{j}=F^{j}=\frac{1}{2}H_{1}^{i,j}S_{i}-(\vec{S}\times\dot{\vec{G}}_{1})^{j}\ .\label{eq:FSCbody2}
\end{equation}
Here $\vec{H}_{1}$ is the gravitomagnetic field generated by the
translational motion of body 1; it is given by $\vec{H}_{1}=\nabla\times\vec{\mathcal{A}_{1}}$,
with $\vec{\mathcal{A}_{1}}=-4M_{1}\vec{v}/r_{1}$ (cf. e.g. \cite{Costa:2016iwu,Kaplan:2009}),
or, explicitly 
\begin{equation}
\vec{H}_{1}=-4\frac{M_{1}}{r_{1}^{3}}\vec{v}\times\vec{r}_{1}\label{eq:Htrans}
\end{equation}
The gravitoelectric field $\vec{G}_{1}$, to the accuracy at hand,
is to be taken, \emph{in this expression,} as the leading term $\vec{G}_{1}\simeq-M\vec{r}_{1}/r_{1}^{3}$.
In order to compute $\dot{\vec{G}}_{1}$, one notes that $\dot{\vec{r}}_{1}=-\vec{v}$,
and that $\dot{r}_{1}=-\vec{r}_{1}\cdot\vec{v}/r_{1}$. One gets\footnote{Transforming this expression to body 1's rest frame (by noticing that
the velocity $\vec{v}_{1}=\vec{v}$ of body 1 in the rest frame of
body 2 equals minus the velocity $\vec{v}_{2}$ of body 2 in body
1's rest frame: $\vec{v}=-\vec{v}_{2}$), yields Eq. (\ref{eq:FBH}).} 
\begin{align}
 & \vec{F}_{{\rm SO}1,2}=\nonumber \\
 & \frac{3M_{1}}{r_{12}^{3}}\left[\vec{v}\times\vec{S}+\frac{2\vec{r}_{12}[(\vec{v}\times\vec{r}_{12})\cdot\vec{S}]}{r_{12}^{2}}+\frac{(\vec{v}\cdot\vec{r}_{12})\vec{S}\times\vec{r}_{12}}{r_{12}^{2}}\right].\label{eq:FSO12_0}
\end{align}
Notice that this amounts to the whole spin-orbit force that acts on
body 2: $\vec{F}_{{\rm SO}1,2}=\vec{F}$.

The coordinate acceleration of body 1 (the point mass) is, from Eqs.
(\ref{eq:CoordAccelForce})-(\ref{eq:FI}), 
\begin{equation}
\frac{d^{2}\vec{x}_{1}}{dt^{2}}=(1+v^{2}-2U_{2})\vec{G}_{2}-4(\vec{G}_{2}\cdot\vec{v})\vec{v}+\vec{v}\times\vec{H}_{2}\label{eq:cooraccelBody2}
\end{equation}
where $U_{2}=M_{2}/r_{2}$ is the Newtonian potential of the spinning
body ($\vec{F}=0$ in this case, since body 1's spin is zero). The
gravitomagnetic field $\vec{H}_{2}$ generated by the spinning body
2 is given by Eq. (\ref{eq:GMfieldSpinning}) (replacing therein $\vec{r}\rightarrow\vec{r}_{2}$);
therefore the gravitomagnetic ``force'' $M_{1}\vec{v}\times\vec{H}_{2}$
it exerts on body 1, which amounts to the whole spin-orbit force $\vec{F}_{{\rm SO}2,1}$
acting on body 1, is given by 
\begin{equation}
\vec{v}\times\vec{H}_{2}=\frac{1}{r_{12}^{3}}\left[2\vec{v}\times\vec{S}-\frac{6}{r_{12}^{2}}(\vec{r}_{21}\cdot\vec{S})\vec{v}\times\vec{r}_{21}\right]=\frac{\vec{F}_{{\rm SO}2,1}}{M_{1}}\ .\label{eq:FSO21_0}
\end{equation}
Using the vector identity (5.2a) of \cite{Faye:2006gx}, we can re-write
this result as 
\begin{align}
 & \vec{F}_{{\rm SO}2,1}=\nonumber \\
 & -\frac{3M_{1}}{r_{12}^{3}}\left[\frac{4}{3}\vec{v}\times\vec{S}+\frac{2\vec{r}_{12}[(\vec{v}\times\vec{r}_{12})\cdot\vec{S}]}{r_{12}^{2}}+\frac{2(\vec{v}\cdot\vec{r}_{12})\vec{S}\times\vec{r}_{12}}{r_{12}^{2}}\right],\label{eq:FSO21_1}
\end{align}
cf. Eq. (5.3a) of \cite{Faye:2006gx}. Comparing with (\ref{eq:FSO12_0}),
we see that the spin-orbit interactions do not obey an action-reaction
law: $\vec{F}_{{\rm SO}1,2}\ne-\vec{F}_{{\rm SO}2,1}$. In other words,
the spin-curvature force (\ref{eq:FSO12_0}) exerted by body 1 on
body 2, \emph{is different} from the gravitomagnetic ``force'' (\ref{eq:FSO21_0})
exerted by body 2 on body 1. Notice that $\vec{F}_{{\rm SO}1,2}$
is the analogue of the electromagnetic force $\vec{F}_{1,{\rm dip}}$
of Sec.~\ref{sub:Magnetic dipole Cloud particle}, and $\vec{F}_{{\rm SO}2,1}$
the analogue of $\vec{F}_{{\rm dip,1}}$. Therefore, the overall coordinate
accelerations of the two bodies do not likewise obey an action-reaction
law: 
\[
M_{2}\frac{d^{2}\vec{x}_{2}}{dt^{2}}\ne M_{1}\frac{d^{2}\vec{x}_{1}}{dt^{2}}\ .
\]
In fact, comparing (\ref{eq:coorAcBody1}) to (\ref{eq:cooraccelBody2}),
we see that actually \emph{all the PN terms} (not only the spin-orbit
ones) differ; only the \emph{Newtonian parts} of $M_{1}\vec{G}_{2}$
and $M_{2}\vec{G}_{1}$ match up to sign.

For the setup in Fig. \ref{fig:ActionReactionGrav}, where particle
1 lies in the equatorial plane at the instantaneous position $\vec{x}_{1}=r_{1}\vec{e}_{z}$,
and has velocity $\vec{v}_{1}=\vec{v}=v\vec{e}_{x}$, we have 
\begin{equation}
\vec{F}_{{\rm SO}1,2}=\frac{3M_{1}}{r_{12}^{3}}\left[\vec{v}\times\vec{S}+\frac{2\vec{r}_{12}[(\vec{v}\times\vec{r}_{12})\cdot\vec{S}]}{r_{12}^{2}}\right]=\frac{3M_{1}Sv}{r_{12}^{3}}\vec{e}_{y}\label{eq:FSO12}
\end{equation}
and 
\begin{equation}
\vec{F}_{{\rm SO}2,1}=M_{1}\vec{v}\times\vec{H}_{2}=\frac{2}{r_{12}^{3}}M_{1}\vec{v}\times\vec{S}=-\frac{2M_{1}Sv}{r_{12}^{3}}\vec{e}_{y}\ .\label{eq:FSO21}
\end{equation}
Therefore, $\vec{F}_{{\rm SO}1,2}=-3\vec{F}_{{\rm SO}2,1}/2$.

Let us now consider particle 3 lying at $\vec{x}_{3}=r_{3}\vec{e}_{z}$
(i.e., on top of the spinning body, above the equatorial plane), and,
again, with velocity $\vec{v}_{3}=\vec{v}=v\vec{e}_{x}$, see Fig.
\ref{fig:Magnus_GMDeflection}b. The spin-orbit force that it exerts
on the spinning body, $\vec{F}_{{\rm SO}3,2}=\vec{F}$, is obtained
from Eq. (\ref{eq:FSO12_0}), replacing therein $\vec{r}_{12}\rightarrow\vec{r}_{32}$,
with $\vec{r}_{32}\equiv\vec{r}_{3}-\vec{r}_{2}$, 
\begin{equation}
\vec{F}_{{\rm SO}3,2}=\frac{3M_{3}}{r_{32}^{3}}\left[\vec{v}\times\vec{S}+\frac{2\vec{r}_{32}[(\vec{v}\times\vec{r}_{32})\cdot\vec{S}]}{r_{32}^{2}}\right]=-\frac{3M_{3}Sv}{r_{23}^{3}}\vec{e}_{y}\ ,\label{eq:FSO32}
\end{equation}
where we noted that $(\vec{v}\times\vec{r}_{32})\cdot\vec{S}=0$.
The spin-orbit force exerted by the spinning body on particle 3, $\vec{F}_{{\rm SO}2,3}=M_{3}\vec{v}\times\vec{H}_{{\rm 2}}$,
is 
\begin{equation}
\vec{F}_{{\rm SO}2,3}=\frac{2M_{3}}{r_{23}^{3}}\left[\vec{v}\times\vec{S}-\frac{3}{r_{23}^{2}}(\vec{r}_{23}\cdot\vec{S})\vec{v}\times\vec{r}_{23}\right]=\frac{4M_{3}Sv}{r_{23}^{3}}\vec{e}_{y}.\label{eq:FSO23}
\end{equation}
Thus, again, action does not equal minus reaction: $\vec{F}_{{\rm SO}2,3}=-4\vec{F}_{{\rm SO}3,2}/3$.

Analogously to the electromagnetic case, this mismatch between action
and reaction does not mean a violation of any conservation principle;
it can be cast as an interchange between mechanical momentum of the
bodies and field momentum (in the sense of the Landau-Lifshitz pseudotensor,
see \cite{Kaplan:2009}). It is the same principle that is behind
the famous bobbings in binary systems \cite{Keppel:2009tc}, where
the center of mass \emph{of the whole binary} bobs up and down.

The examples above illustrate an important aspect depicted in Fig.
\ref{fig:Magnus_GMDeflection}b: cloud particles in the equatorial
plane deflect in the negative $y$ direction (i.e., to the left in
Fig. \ref{fig:Magnus_GMDeflection}b), cf. Eq. (\ref{eq:FSO21}),
and push the spinning body to the right, as Eq. (\ref{eq:FSO12})
shows; however, cloud particles in regions outside the equatorial
plane do the opposite: they deflect to the right {[}cf. Eq. (\ref{eq:FSO23}){]},
and push the spinning body to the left, with a force whose magnitude
is twice that of the force exerted by the particles at the equatorial
plane, as shown by Eq. (\ref{eq:FSO32}). It is the effect of the
latter that eventually prevails in the cloud slab (orthogonal to $y$)
of Fig. \ref{fig:MagnusGrav}, where the net force on the cloud points
in the positive $y$ direction (and the force on the spinning body
points in the negative $y$ direction), whereas in the \emph{special}
\emph{case} of a slab orthogonal to $z$ (i.e., to the body's spin
axis) the two effects exactly cancel out.

The above explains also why in the earlier work \cite{Okawa:2014sxa}
the authors were misled into concluding that the force on the spinning
body was opposite to the Magnus effect (``anti-Magnus'', upwards
in Fig. \ref{fig:MagnusGrav}): the argument therein is based on the
asymmetric accretion that occurs in a spinning BH, i.e., the absorption
cross section being larger for counterrotating particles than for
corotating ones. This is a gravitomagnetic phenomenon, due to the
fact that the gravitomagnetic force $\vec{F}_{{\rm GM}}=M\vec{v}\times\vec{H}$
is attractive for counterrotating particles, and repulsive in the
corotating case. That can easily be seen considering again particles
with velocity $\vec{v}=v\vec{e}_{x}$, as in Fig. \ref{fig:Magnus_GMDeflection}b;
from Eq. (\ref{eq:GMfieldSpinning}), it follows that the radial component
of the gravitomagnetic force is $\vec{F}_{{\rm GM}}\cdot\vec{r}/r=-2MvSy/r^{4}$
(attractive for positive $y$, repulsive for negative $y$). However,
it is crucial here to distinguish between the radial component of
$\vec{F}_{{\rm GM}}$ (which determines its attractive/repulsive nature),
from the force itself, and the overall deflection it causes. If one
looks only at the equatorial plane, the reasoning in \cite{Okawa:2014sxa}
is qualitatively correct, since, as depicted in Fig. \ref{fig:Magnus_GMDeflection}b,
particles in the equatorial plane suffer a deflection opposite to
that corresponding to a Magnus effect. That however overlooks the
key fact that, as exemplified by the particles along the axis in Fig.
\ref{fig:Magnus_GMDeflection}b, there are regions outside the equatorial
plane where particles are deflected in the opposite direction, i.e.,
in the direction expected from a Magnus effect.

 \bibliographystyle{utphys}
\bibliography{Ref}

\providecommand{\href}[2]{#2}\begingroup\raggedright\begin{thebibliography}{10}

\bibitem{rubinow_keller_1961}
S.~I. Rubinow and J.~B. Keller, ``The transverse force on a spinning sphere
  moving in a viscous fluid,''
  \href{http://dx.doi.org/10.1017/S0022112061000640}{{\em Journal of Fluid
  Mechanics} {\bfseries 11} no.~3, (1961) 447}.

\bibitem{TsujiMorikawaMizuno1985}
Y.~Tsuji, Y.~Morikawa, and O.~Mizuno, ``{Experimental measurement of the Magnus
  force on a rotating sphere at low Reynolds numbers},''
  \href{http://dx.doi.org/10.1115/1.3242517}{{\em J. Fluids Eng.} {\bfseries
  107} (1985) 484}.

\bibitem{WattsFerrer1987}
R.~G. Watts and R.~Ferrer, ``The lateral force on a spinning sphere:
  Aerodynamics of a curveball,'' \href{http://dx.doi.org/10.1119/1.14969}{{\em
  American Journal of Physics} {\bfseries 55} no.~1, (1987) 40}.

\bibitem{MunsunYoungOkiishi}
B.~Munson, D.~Young, and T.~Okiishi, {\em {Fundamentals of fluid mechanics}}.
\newblock Wiley \& Sons, 1998.

\bibitem{CiufoliniWheeler}
I.~{Ciufolini} and J.~A. {Wheeler}, {\em {Gravitation and Inertia}}.
\newblock Princeton Series in Physics, Princeton, NJ, 1995.

\bibitem{Font:1998sc}
J.~A. Font, J.~M. Ibanez, and P.~Papadopoulos, ``{Nonaxisymmetric relativistic
  Bondi-Hoyle accretion onto a Kerr black hole},''
  \href{http://dx.doi.org/10.1046/j.1365-8711.1999.02459.x}{{\em Mon. Not. Roy.
  Astron. Soc.} {\bfseries 305} (1999) 920},
\href{http://arxiv.org/abs/astro-ph/9810344}{{\ttfamily arXiv:astro-ph/9810344
  [astro-ph]}}.

\bibitem{Okawa:2014sxa}
H.~Okawa and V.~Cardoso, ``{Black holes and fundamental fields: Hair, kicks,
  and a gravitational Magnus effect},''
  \href{http://dx.doi.org/10.1103/PhysRevD.90.104040}{{\em Phys. Rev.}
  {\bfseries D90} no.~10, (2014) 104040},
\href{http://arxiv.org/abs/1405.4861}{{\ttfamily arXiv:1405.4861 [gr-qc]}}.

\bibitem{Cashen:2016neh}
B.~Cashen, A.~Aker, and M.~Kesden, ``{Gravitomagnetic dynamical friction},''
  \href{http://dx.doi.org/10.1103/PhysRevD.95.064014}{{\em Phys. Rev.}
  {\bfseries D95} no.~6, (2017) 064014},
\href{http://arxiv.org/abs/1610.01590}{{\ttfamily arXiv:1610.01590 [gr-qc]}}.

\bibitem{Mathisson:1937zz}
M.~Mathisson, ``{Neue mechanik materieller systemes},''
\href{http://dx.doi.org//10.1007/s10714-010-0939-y}{{\em Acta Phys. Polon.}
  {\bfseries 6} (1937) 163}.

\bibitem{Papapetrou:1951pa}
A.~Papapetrou, ``{Spinning test particles in general relativity. 1.},''
\href{http://dx.doi.org/10.1098/rspa.1951.0200}{{\em Proc. Roy. Soc. Lond.}
  {\bfseries A209} (1951) 248--258}.

\bibitem{Dixon1964}
W.~G. Dixon, ``{A covariant multipole formalism for extended test bodies in
  general relativity},'' {\em Il Nuovo Cimento} {\bfseries 34} (1964) 317.

\bibitem{Dixon:1970zza}
W.~G. Dixon, ``{Dynamics of extended bodies in general relativity. I. Momentum
  and angular momentum},''
\href{http://dx.doi.org/10.1098/rspa.1970.0020}{{\em Proc. Roy. Soc. Lond.}
  {\bfseries A314} (1970) 499--527}.

\bibitem{Gralla:2010xg}
S.~E. Gralla, A.~I. Harte, and R.~M. Wald, ``{Bobbing and Kicks in
  Electromagnetism and Gravity},''
  \href{http://dx.doi.org/10.1103/PhysRevD.81.104012}{{\em Phys. Rev.}
  {\bfseries D81} (2010) 104012},
\href{http://arxiv.org/abs/1004.0679}{{\ttfamily arXiv:1004.0679 [gr-qc]}}.

\bibitem{Tulczyjew}
W.~Tulczyjew, ``{Motion of multipole particles in general relativity theory},''
  {\em Acta Phys. Polon.} {\bfseries 18} (1959) 393.

\bibitem{Costa:2012cy}
L.~F.~O. Costa, J.~Nat\'ario, and M.~Zilh\~{a}o, ``{Spacetime dynamics of
  spinning particles: Exact electromagnetic analogies},''
  \href{http://dx.doi.org/10.1103/PhysRevD.93.104006}{{\em Phys. Rev.}
  {\bfseries D93} no.~10, (2016) 104006},
\href{http://arxiv.org/abs/1207.0470}{{\ttfamily arXiv:1207.0470 [gr-qc]}}.

\bibitem{FilipeCosta:2006fz}
L.~Filipe~Costa and C.~A.~R. Herdeiro, ``{Gravitoelectromagnetic analogy based
  on tidal tensors},'' \href{http://dx.doi.org/10.1103/PhysRevD.78.024021}{{\em
  Phys. Rev.} {\bfseries D78} (2008) 024021}.

\bibitem{Costa:2016iwu}
L.~F.~O. Costa, L.~Wylleman, and J.~Nat\'ario, ``{Gravitomagnetism and the
  significance of the curvature scalar invariants},''
\href{http://arxiv.org/abs/1603.03143}{{\ttfamily arXiv:1603.03143 [gr-qc]}}.

\bibitem{GriffithsBook}
D.~J. Griffiths, {\em {Introduction to electrodynamics}}.
\newblock Pearson Benjamim Cummings, San Francisco, 2008.

\bibitem{Jackson:1998nia}
J.~D. Jackson, {\em {Classical Electrodynamics}}.
\newblock Wiley,
1998.
\newblock

\bibitem{Fubini}
\url{{https://en.wikipedia.org/wiki/Fubini_theorem}}.

\bibitem{PeterWalker}
P.~Walker, \href{http://dx.doi.org/10.1007/978-0-85729-380-0}{{\em Examples and
  Theorems in Analysis}}.
\newblock Springer-Verlag, London, 2004.

\bibitem{Costa:2014nta}
L.~F.~O. Costa and J.~Nat{\'a}rio,
  \href{http://dx.doi.org/10.1007/978-3-319-18335-0_6}{``Center of mass, spin
  supplementary conditions, and the momentum of spinning particles,''} in {\em
  Equations of Motion in Relativistic Gravity}, D.~Puetzfeld,
  C.~L{\"a}mmerzahl, and B.~Schutz, eds., pp.~215--258.
\newblock Springer, Cham, 2015 [Fund. Theor. Phys. \textbf{179}, 215].
\newblock
\href{http://arxiv.org/abs/1410.6443}{{\ttfamily arXiv:1410.6443}}.
\newblock

\bibitem{Pirani:1956tn}
F.~A.~E. Pirani, ``{On the physical significance of the Riemann tensor},''
  \href{http://dx.doi.org/10.1007/s10714-009-0787-9}{{\em Acta Phys. Polon.}
  {\bfseries 15} (1956) 389--405}.
[Gen. Relativ. Gravit. \textbf{41},1215 (2009)].

\bibitem{Dadhich:1999eh}
N.~Dadhich, ``{Electromagnetic duality in general relativity},''
  \href{http://dx.doi.org/10.1023/A:1001913409254}{{\em Gen. Rel. Grav.}
  {\bfseries 32} (2000) 1009--1023},
\href{http://arxiv.org/abs/gr-qc/9909067}{{\ttfamily arXiv:gr-qc/9909067
  [gr-qc]}}.

\bibitem{EllisMaartensMacCallum}
G.~F.~R. Ellis, R.~Maartens, and M.~A.~H. MacCallum, {\em Relativistic
  Cosmology}.
\newblock Cambridge University Press, Cambridge, UK, 2012.

\bibitem{MaartensBasset1997}
R.~Maartens and B.~A. Bassett, ``Gravito-electromagnetism,'' {\em Classical and
  Quantum Gravity} {\bfseries 15} no.~3, (1998) 705.

\bibitem{Damour:1990pi}
T.~Damour, M.~Soffel, and C.-m. Xu, ``{General relativistic celestial
  mechanics. 1. Method and definition of reference systems},''
\href{http://dx.doi.org/10.1103/PhysRevD.43.3272}{{\em Phys. Rev.} {\bfseries
  D43} (1991) 3272--3307}.

\bibitem{WillPoissonBook}
E.~{Poisson} and C.~M. {Will}, {\em {Gravity: Newtonian, Post-Newtonian,
  Relativistic}}.
\newblock Cambridge University Press, Cambridge, UK, 2014.

\bibitem{Misner:1974qy}
C.~W. Misner, K.~S. Thorne, and J.~A. Wheeler, {\em {Gravitation}}.
\newblock W. H. Freeman, San Francisco,
1973.
\newblock

\bibitem{Will:1993ns}
C.~M. Will, {\em {Theory and experiment in gravitational physics}}.
\newblock Cambridge University Press, Cambridge, UK, 1993.

\bibitem{Kaplan:2009}
J.~D. Kaplan, D.~A. Nichols, and K.~S. Thorne, ``{Post-Newtonian approximation
  in Maxwell-like form},''
  \href{http://dx.doi.org/10.1103/PhysRevD.80.124014}{{\em Phys. Rev.}
  {\bfseries D80} (2009) 124014},
\href{http://arxiv.org/abs/0808.2510}{{\ttfamily arXiv:0808.2510 [gr-qc]}}.

\bibitem{Binietal1994}
D.~Bini, P.~Carini, R.~T. Jantzen, and D.~Wilkins, ``{Thomas precession in
  post-Newtonian gravitoelectromagnetism},''
  \href{http://dx.doi.org/10.1103/PhysRevD.49.2820}{{\em Phys. Rev. D}
  {\bfseries 49} (1994) 2820--2827}.

\bibitem{Costa:2017kdr}
L.~F.~O. Costa, G.~Lukes-Gerakopoulos, and O.~Semer\'{a}k, ``{Spinning
  particles in general relativity: Momentum-velocity relation for the
  Mathisson-Pirani spin condition},''
  \href{http://dx.doi.org/10.1103/PhysRevD.97.084023}{{\em Phys. Rev.}
  {\bfseries D97} (2018) 084023},
\href{http://arxiv.org/abs/1712.07281}{{\ttfamily arXiv:1712.07281 [gr-qc]}}.

\bibitem{Wald:1972sz}
R.~M. Wald, ``{Gravitational spin interaction},''
\href{http://dx.doi.org/10.1103/PhysRevD.6.406}{{\em Phys. Rev.} {\bfseries D6}
  (1972) 406--413}.

\bibitem{Costa:2015hlh}
L.~F. Costa and J.~Nat\'ario, ``{The Coriolis field},''
  \href{http://dx.doi.org/10.1119/1.4938056}{{\em Am. J. Phys.} {\bfseries 84}
  (2016) 388}, \href{http://arxiv.org/abs/1511.02458}{{\ttfamily
  arXiv:1511.02458}}.

\bibitem{VICKERS2009}
P.~Vickers, ``{Was Newtonian cosmology really inconsistent?},''
  \href{http://dx.doi.org/http://dx.doi.org/10.1016/j.shpsb.2009.05.001}{{\em
  Studies in History and Philosophy of Science Part B: Studies in History and
  Philosophy of Modern Physics} {\bfseries 40} (2009) 197 -- 208}.

\bibitem{McCrea1955}
W.~H. {McCrea}, ``{On the significance of Newtonian cosmology},''
  \href{http://dx.doi.org/10.1086/107226}{{\em Astronom. Journal} {\bfseries
  60} (1955) 271}.

\bibitem{Raifeartaigh2017}
C.~O'Raifeartaigh, M.~O'Keeffe, W.~Nahm, and S.~Mitton, ``Einstein's 1917
  static model of the universe: a centennial review,''
  \href{http://dx.doi.org/10.1140/epjh/e2017-80002-5}{{\em European Physical
  Journal H} {\bfseries 42} (2017) 431--474},
  \href{http://arxiv.org/abs/1701.07261}{{\ttfamily arXiv:1701.07261}}.

\bibitem{EinsteinPrinciple}
H.~Lorentz, A.~Einstein, H.~Minkowski, and H.~Weyl, {\em {The priciple
  Relativity: A Collection of Original Memoirs on the Special and General
  Theory of Relativity}}.
\newblock Dover Publications, Inc, New York, 1952.

\bibitem{EinsteinRelativityl}
A.~{Einstein}, {\em {Relativity: The special and the general theory, R.W.
  Lawson(trans.)}}.
\newblock Methuen, London, 1954.

\bibitem{BinneyTremaine}
J.~{Binney} and S.~{Tremaine}, {\em {Galactic Dynamics: Second Edition}}.
\newblock Princeton University Press, 2008.

\bibitem{Bertone:2005hw}
G.~Bertone and D.~Merritt, ``{Time-dependent models for dark matter at the
  Galactic Center},'' \href{http://dx.doi.org/10.1103/PhysRevD.72.103502}{{\em
  Phys. Rev.} {\bfseries D72} (2005) 103502},
\href{http://arxiv.org/abs/astro-ph/0501555}{{\ttfamily arXiv:astro-ph/0501555
  [astro-ph]}}.

\bibitem{Burkert:1995yz}
A.~Burkert, ``{The Structure of dark matter halos in dwarf galaxies},''
  \href{http://dx.doi.org/10.1086/309560}{{\em Astrophys. J.} {\bfseries 447}
  (1995) L25},
\href{http://arxiv.org/abs/astro-ph/9504041}{{\ttfamily arXiv:astro-ph/9504041
  [astro-ph]}}.

\bibitem{Karpov:2003bn}
O.~B. Karpov, ``{Gyroscope deviation from geodesic motion: Quasiresonant
  oscillations on a circular orbit},''
  \href{http://dx.doi.org/10.1134/1.1574531}{{\em J. Exp. Theor. Phys.}
  {\bfseries 96} (2003) 581--586},
  \href{http://arxiv.org/abs/gr-qc/0301056}{{\ttfamily arXiv:gr-qc/0301056
  [gr-qc]}}.
[Zh. Eksp. Teor. Fiz. \textbf{123}, 659 (2003)].

\bibitem{Barausse:2014tra}
E.~Barausse, V.~Cardoso, and P.~Pani, ``{Can environmental effects spoil
  precision gravitational-wave astrophysics?},''
  \href{http://dx.doi.org/10.1103/PhysRevD.89.104059}{{\em Phys. Rev.}
  {\bfseries D89} no.~10, (2014) 104059},
\href{http://arxiv.org/abs/1404.7149}{{\ttfamily arXiv:1404.7149}}.

\bibitem{Moller1949}
C.~M{\o}ller, ``Sur la dynamique des syst\`{e}mes ayant un moment angulaire
  interne,'' {\em Annales de l'institut Henri Poincar\'{e}} {\bfseries 11}
  no.~5, (1949) 251--278.

\bibitem{Pani:2015qhr}
P.~Pani, ``{Binary pulsars as dark-matter probes},''
  \href{http://dx.doi.org/10.1103/PhysRevD.92.123530}{{\em Phys. Rev.}
  {\bfseries D92} no.~12, (2015) 123530},
\href{http://arxiv.org/abs/1512.01236}{{\ttfamily arXiv:1512.01236}}.

\bibitem{Read:2014qva}
J.~I. Read, ``{The Local Dark Matter Density},''
  \href{http://dx.doi.org/10.1088/0954-3899/41/6/063101}{{\em J. Phys.}
  {\bfseries G41} (2014) 063101},
\href{http://arxiv.org/abs/1404.1938}{{\ttfamily arXiv:1404.1938
  [astro-ph.GA]}}.

\bibitem{Gilmore:1996pr}
G.~Gilmore, ``{The distribution of dark matter in the Milky Way galaxy},'' in
  {\em {Proceedings, 1st International Workshop on The identification of dark
  matter (IDM 1996): Sheffield, UK, September 8-12, 1996}}, pp.~73--82.
\newblock World Scientific, Singapore, 1997.
\newblock
\href{http://arxiv.org/abs/astro-ph/9702081}{{\ttfamily arXiv:astro-ph/9702081
  [astro-ph]}}.
\newblock

\bibitem{KommetAl}
R.~{Komm}, R.~{Howe}, B.~R. {Durney}, and F.~{Hill}, ``{Temporal Variation of
  Angular Momentum in the Solar Convection Zone},''
  \href{http://dx.doi.org/10.1086/367608}{{\em \apj} {\bfseries 586} (2003)
  650--662}.

\bibitem{vanderMarel:2013jza}
R.~P. van~der Marel and N.~Kallivayalil, ``{Third-Epoch Magellanic Cloud Proper
  Motions II: The Large Magellanic Cloud Rotation Field in Three Dimensions},''
  \href{http://dx.doi.org/10.1088/0004-637X/781/2/121}{{\em Astrophys. J.}
  {\bfseries 781} no.~2, (2014) 121},
\href{http://arxiv.org/abs/1305.4641}{{\ttfamily arXiv:1305.4641
  [astro-ph.CO]}}.

\bibitem{Gaia}
\url{http://sci.esa.int/gaia/}.

\bibitem{Genzel2003}
R.~Genzel {\em et~al.}, ``The stellar cusp around the supermassive black hole
  in the galactic center,'' \href{http://dx.doi.org/10.1086/377127}{{\em
  Astrophys. J.} {\bfseries 594} no.~2, (2003) 812},
  \href{http://arxiv.org/abs/astro-ph/0305423}{{\ttfamily
  arXiv:astro-ph/0305423}}.

\bibitem{NovikovThorne1973}
I.~D. {Novikov} and K.~S. {Thorne}, ``{Astrophysics of black holes.},'' in {\em
  Black Holes (Les Astres Occlus)}, C.~{Dewitt} and B.~S. {Dewitt}, eds.,
  pp.~343--450.
\newblock Gordon and Breach, New York, 1973.

\bibitem{Shakura:1972te}
N.~I. Shakura and R.~A. Sunyaev, ``{Black holes in binary systems.
  Observational appearance},''
{\em Astron. Astrophys.} {\bfseries 24} (1973) 337--355.

\bibitem{Nordtvedt:1988vt}
K.~Nordtvedt, ``{Existence of the Gravitomagnetic Interaction},''
\href{http://dx.doi.org/10.1007/BF00671317}{{\em Int. J. Theor. Phys.}
  {\bfseries 27} (1988) 1395--1404}.

\bibitem{Hannam:2013pra}
M.~Hannam, ``{Modelling gravitational waves from precessing black-hole
  binaries: Progress, challenges and prospects},''
  \href{http://dx.doi.org/10.1007/s10714-014-1767-2}{{\em Gen. Rel. Grav.}
  {\bfseries 46} (2014) 1767},
\href{http://arxiv.org/abs/1312.3641}{{\ttfamily arXiv:1312.3641 [gr-qc]}}.

\bibitem{LangHughes2006}
R.~N. Lang and S.~A. Hughes, ``{Measuring coalescing massive binary black holes
  with gravitational waves: The impact of spin-induced precession},''
  \href{http://dx.doi.org/10.1103/PhysRevD.75.089902,
  10.1103/PhysRevD.74.122001, 10.1103/PhysRevD.77.109901}{{\em Phys. Rev.}
  {\bfseries D74} (2006) 122001},
  \href{http://arxiv.org/abs/gr-qc/0608062}{{\ttfamily arXiv:gr-qc/0608062}}.
[Erratum: Phys. Rev. D\textbf{77},109901 (2008)].

\bibitem{Vecchio:2003tn}
A.~Vecchio, ``{LISA observations of rapidly spinning massive black hole binary
  systems},'' \href{http://dx.doi.org/10.1103/PhysRevD.70.042001}{{\em Phys.
  Rev.} {\bfseries D70} (2004) 042001},
\href{http://arxiv.org/abs/astro-ph/0304051}{{\ttfamily
  arXiv:astro-ph/0304051}}.

\bibitem{Lang:2007ge}
R.~N. Lang and S.~A. Hughes, ``{Localizing coalescing massive black hole
  binaries with gravitational waves},''
  \href{http://dx.doi.org/10.1086/528953}{{\em Astrophys. J.} {\bfseries 677}
  (2008) 1184},
\href{http://arxiv.org/abs/0710.3795}{{\ttfamily arXiv:0710.3795 [astro-ph]}}.

\bibitem{Stavridis:2009ys}
A.~Stavridis, K.~G. Arun, and C.~M. Will, ``{Precessing supermassive black hole
  binaries and dark energy measurements with LISA},''
  \href{http://dx.doi.org/10.1103/PhysRevD.80.067501}{{\em Phys. Rev.}
  {\bfseries D80} (2009) 067501},
\href{http://arxiv.org/abs/0907.4686}{{\ttfamily arXiv:0907.4686 [gr-qc]}}.

\bibitem{Schmidt:2014iyl}
P.~Schmidt, F.~Ohme, and M.~Hannam, ``{Towards models of gravitational
  waveforms from generic binaries II: Modelling precession effects with a
  single effective precession parameter},''
  \href{http://dx.doi.org/10.1103/PhysRevD.91.024043}{{\em Phys. Rev.}
  {\bfseries D91} no.~2, (2015) 024043},
\href{http://arxiv.org/abs/1408.1810}{{\ttfamily arXiv:1408.1810 [gr-qc]}}.

\bibitem{Berti:2004bd}
E.~Berti, A.~Buonanno, and C.~M. Will, ``{Estimating spinning binary parameters
  and testing alternative theories of gravity with LISA},''
  \href{http://dx.doi.org/10.1103/PhysRevD.71.084025}{{\em Phys. Rev.}
  {\bfseries D71} (2005) 084025},
\href{http://arxiv.org/abs/gr-qc/0411129}{{\ttfamily arXiv:gr-qc/0411129
  [gr-qc]}}.

\bibitem{Moore:2014lga}
C.~J. Moore, R.~H. Cole, and C.~P.~L. Berry, ``{Gravitational-wave sensitivity
  curves},'' \href{http://dx.doi.org/10.1088/0264-9381/32/1/015014}{{\em Class.
  Quant. Grav.} {\bfseries 32} no.~1, (2015) 015014},
\href{http://arxiv.org/abs/1408.0740}{{\ttfamily arXiv:1408.0740}}.

\bibitem{Audley:2017drz}
H.~{Audley} {\em et~al.}, ``{Laser Interferometer Space Antenna},''
  \href{http://arxiv.org/abs/1702.00786}{{\ttfamily arXiv:1702.00786
  [astro-ph.IM]}}.

\bibitem{Hobbs2009}
G.~Hobbs {\em et~al.}, ``The international pulsar timing array project: using
  pulsars as a gravitational wave detector,''
  \href{http://dx.doi.org/10.1088/0264-9381/27/8/084013}{{\em Classical and
  Quantum Gravity} {\bfseries 27} no.~8, (2010) 084013}.

\bibitem{Detweiler1979}
S.~Detweiler, ``Pulsar timing measurements and the search for gravitational
  waves,'' \href{http://dx.doi.org/10.1086/157593}{{\em Astrophys. J.}
  {\bfseries 234} (1979) 1100}.

\bibitem{Sazhin}
M.~V. Sazhin, ``Opportunities for detecting ultralong gravitational waves,''
  {\em Soviet Astronomy} {\bfseries 22} (1978) 36.

\bibitem{Moore:2014eua}
C.~J. Moore, S.~R. Taylor, and J.~R. Gair, ``{Estimating the sensitivity of
  pulsar timing arrays},''
  \href{http://dx.doi.org/10.1088/0264-9381/32/5/055004}{{\em Class. Quant.
  Grav.} {\bfseries 32} no.~5, (2015) 055004},
\href{http://arxiv.org/abs/1406.5199}{{\ttfamily arXiv:1406.5199}}.

\bibitem{Arzoumanian2016}
Z.~Arzoumanian {\em et~al.}, ``The nanograv nine-year data set: Limits on the
  isotropic stochastic gravitational wave background,''
  \href{http://dx.doi.org/10.3847/0004-637X/821/1/13}{{\em Astrophys. J.}
  {\bfseries 821} no.~1, (2016) 13},
  \href{http://arxiv.org/abs/1508.03024}{{\ttfamily arXiv:1508.03024}}.

\bibitem{Semerak:2012dw}
O.~Semer\'{a}k and P.~Sukov\'{a}, ``{Free motion around black holes with discs
  or rings: between integrability and chaos - I},''
  \href{http://dx.doi.org/10.1111/j.1365-2966.2009.16003.x}{{\em Mon. Not. Roy.
  Astron. Soc.} {\bfseries 404} (2010) 545--574},
\href{http://arxiv.org/abs/1211.4106}{{\ttfamily arXiv:1211.4106 [gr-qc]}}.

\bibitem{Morgan:1969jr}
T.~Morgan and L.~Morgan, ``{The Gravitational Field of a Disk},''
  \href{http://dx.doi.org/10.1103/PhysRevD.1.3522, 10.1103/PhysRev.188.2544,
  10.1103/PhysRev.183.1097}{{\em Phys. Rev.} {\bfseries 183} (1969)
  1097--1101}.
[Erratum: Phys. Rev. D \textbf{1}, 3522 (1970)].

\bibitem{Lemos:1988vf}
J.~P.~S. Lemos, ``{Self-similar relativistic discs with pressure},''
\href{http://dx.doi.org/10.1088/0264-9381/6/9/007}{{\em Class. Quant. Grav.}
  {\bfseries 6} (1989) 1219--1320}.

\bibitem{Lemos:1993uy}
J.~P.~S. Lemos and P.~S. Letelier, ``{Superposition of Morgan and Morgan discs
  with a Schwarzschild black hole},''
\href{http://dx.doi.org/10.1088/0264-9381/10/6/003}{{\em Class. Quant. Grav.}
  {\bfseries 10} (1993) L75--L78}.

\bibitem{Bicak:1993xat}
J.~Bicak, D.~Lynden-Bell, and J.~Katz, ``{Relativistic disks as sources of
  static vacuum spacetimes},''
\href{http://dx.doi.org/10.1103/PhysRevD.47.4334}{{\em Phys. Rev.} {\bfseries
  D47} no.~10, (1993) 4334}.

\bibitem{Espitia:2001cj}
O.~A. Espitia and G.~A. Gonzalez, ``{Relativistic static thin disks: The
  counterrotating model},''
  \href{http://dx.doi.org/10.1103/PhysRevD.68.104028}{{\em Phys. Rev.}
  {\bfseries D68} (2003) 104028},
\href{http://arxiv.org/abs/gr-qc/0107044}{{\ttfamily arXiv:gr-qc/0107044
  [gr-qc]}}.

\bibitem{Bicak:1993zz}
J.~Bicak and T.~Ledvinka, ``{Relativistic disks as sources of the Kerr
  metric},''
\href{http://dx.doi.org/10.1103/PhysRevLett.71.1669}{{\em Phys. Rev. Lett.}
  {\bfseries 71} (1993) 1669--1672}.

\bibitem{NishidaEriguchiLanza}
S.~{Nishida}, Y.~{Eriguchi}, and A.~{Lanza}, ``{General Relativistic Structure
  of Star-Toroid Systems},'' \href{http://dx.doi.org/10.1086/172090}{{\em \apj}
  {\bfseries 401} (1992) 618}.

\bibitem{Cizek:2017wzr}
P.~\v{C}\'{i}\v{z}ek and O.~Semer\'{a}k, ``{Perturbation of a Schwarzschild
  Black Hole Due to a Rotating Thin Disk},''
  \href{http://dx.doi.org/10.3847/1538-4365/aa876b}{{\em Astrophys. J. Suppl.
  Ser.} {\bfseries 232} (2017) 14},
\href{http://arxiv.org/abs/1710.07109}{{\ttfamily arXiv:1710.07109 [gr-qc]}}.

\bibitem{Kotlarik:2018nbd}
P.~Kotla\v{r}\'{i}k, O.~Semer\'{a}k, and P.~\v{C}\'{i}\v{z}ek, ``{Schwarzschild
  black hole encircled by a rotating thin disc: Properties of perturbative
  solution},'' \href{http://dx.doi.org/10.1103/PhysRevD.97.084006}{{\em Phys.
  Rev.} {\bfseries D97} (2018) 084006},
\href{http://arxiv.org/abs/1804.02010}{{\ttfamily arXiv:1804.02010 [gr-qc]}}.

\bibitem{Jaranowski:2014yva}
P.~Jaranowski, P.~Mach, E.~Malec, and M.~Pirog, ``{General-relativistic versus
  Newtonian: Geometric dragging and dynamic antidragging in stationary
  self-gravitating disks in the first post-Newtonian approximation},''
  \href{http://dx.doi.org/10.1103/PhysRevD.91.024039}{{\em Phys. Rev.}
  {\bfseries D91} no.~2, (2015) 024039},
\href{http://arxiv.org/abs/1410.8527}{{\ttfamily arXiv:1410.8527 [gr-qc]}}.

\bibitem{Miyamoto:1975zz}
M.~Miyamoto and R.~Nagai, ``{Three-dimensional models for the distribution of
  mass in galaxies},''
{\em Publ. Astron. Soc. Jap.} {\bfseries 27} (1975) 533--543.

\bibitem{Baes:2008td}
M.~Baes, ``{Exact potential-density pairs for flattened dark haloes},''
  \href{http://dx.doi.org/10.1111/j.1365-2966.2008.14174.x}{{\em Mon. Not. Roy.
  Astron. Soc.} {\bfseries 392} (2009) 1503},
\href{http://arxiv.org/abs/0810.5483}{{\ttfamily arXiv:0810.5483 [astro-ph]}}.

\bibitem{Vogt:2009sy}
D.~Vogt and P.~S. Letelier, ``{Analytical Potential-Density Pairs for Flat
  Rings and Toroidal Structures},''
  \href{http://dx.doi.org/10.1111/j.1365-2966.2009.14803.x}{{\em Mon. Not. Roy.
  Astron. Soc.} {\bfseries 396} (2009) 1487},
\href{http://arxiv.org/abs/0906.0919}{{\ttfamily arXiv:0906.0919}}.

\bibitem{Witzany:2015yqa}
V.~Witzany, O.~Semer\'{a}k, and P.~Sukov\'{a}, ``{Free motion around black
  holes with discs or rings: between integrability and chaos -- IV},''
  \href{http://dx.doi.org/10.1093/mnras/stv1148}{{\em Mon. Not. Roy. Astron.
  Soc.} {\bfseries 451} (2015) 1770},
\href{http://arxiv.org/abs/1503.09077}{{\ttfamily arXiv:1503.09077
  [astro-ph.HE]}}.

\bibitem{Apostolatos:1994mx}
T.~A. Apostolatos, C.~Cutler, G.~J. Sussman, and K.~S. Thorne, ``{Spin induced
  orbital precession and its modulation of the gravitational wave forms from
  merging binaries},''
\href{http://dx.doi.org/10.1103/PhysRevD.49.6274}{{\em Phys. Rev.} {\bfseries
  D49} (1994) 6274--6297}.

\bibitem{Babak:2016tgq}
S.~Babak, A.~Taracchini, and A.~Buonanno, ``{Validating the effective-one-body
  model of spinning, precessing binary black holes against numerical
  relativity},'' \href{http://dx.doi.org/10.1103/PhysRevD.95.024010}{{\em Phys.
  Rev.} {\bfseries D95} no.~2, (2017) 024010},
\href{http://arxiv.org/abs/1607.05661}{{\ttfamily arXiv:1607.05661 [gr-qc]}}.

\bibitem{Hawking:1973uf}
S.~W. Hawking and G.~F.~R. Ellis,
  \href{http://dx.doi.org/10.1017/CBO9780511524646}{{\em {The Large Scale
  Structure of Space-Time}}}.
\newblock Cambridge Monographs on Mathematical Physics. Cambridge University
  Press,
1973.
\newblock

\bibitem{Wald:1984}
R.~M. Wald, {\em {General Relativity}}.
\newblock The University of Chicago Press, Chicago, 1984.

\bibitem{Chavanis2014}
P.-H. Chavanis, ``{Models of universe with a polytropic equation of state: II.
  The late universe},''
  \href{http://dx.doi.org/10.1140/epjp/i2014-14222-0}{{\em EPJ Plus} {\bfseries
  129} no.~10, (2014) 222}, \href{http://arxiv.org/abs/1208.0801}{{\ttfamily
  arXiv:1208.0801}}.

\bibitem{ZlatevWangSteinhard_PRL1999}
I.~Zlatev, L.~Wang, and P.~J. Steinhardt, ``{Quintessence, Cosmic Coincidence,
  and the Cosmological Constant},''
  \href{http://dx.doi.org/10.1103/PhysRevLett.82.896}{{\em Phys. Rev. Lett.}
  {\bfseries 82} (1999) 896--899}.

\bibitem{Gorini:2004by}
V.~Gorini, A.~Kamenshchik, U.~Moschella, and V.~Pasquier,
  \href{http://dx.doi.org/10.1142/9789812704030_0050}{``{The Chaplygin gas as a
  model for dark energy},''} in {\em {On recent developments in theoretical and
  experimental general relativity, gravitation, and relativistic field
  theories. Proceedings, 10th Marcel Grossmann Meeting, MG10, Rio de Janeiro,
  Brazil, July 20-26, 2003. Pt. A-C}}, pp.~840--859.
\newblock World Scientific, Singapore, 2006.
\newblock
\href{http://arxiv.org/abs/gr-qc/0403062}{{\ttfamily arXiv:gr-qc/0403062}}.
\newblock

\bibitem{Vikman:2004dc}
A.~Vikman, ``{Can dark energy evolve to the phantom?},''
  \href{http://dx.doi.org/10.1103/PhysRevD.71.023515}{{\em Phys. Rev.}
  {\bfseries D71} (2005) 023515},
  \href{http://arxiv.org/abs/astro-ph/0407107}{{\ttfamily
  arXiv:astro-ph/0407107}}.

\bibitem{WMAP}
C.~L. Bennett {\em et~al.}, ``{Nine-year Wilkinson Microwave Anisotropy Probe
  (WMAP) Observations: Final Maps and Results},''
  \href{http://dx.doi.org/10.1088/0067-0049/208/2/20}{{\em Astrophys. J. Suppl.
  Ser.} {\bfseries 208} no.~2, (2013) 20}.

\bibitem{Springob:2007vb}
C.~M. Springob {\em et~al.}, ``{SFI++ II: A New I-band Tully-Fisher Catalog,
  Derivation of Peculiar Velocities and Dataset Properties},''
  \href{http://dx.doi.org/10.1088/0067-0049/182/1/474}{{\em Astrophys. J.
  Suppl. Ser.} {\bfseries 172} (2007) 599--614},
  \href{http://arxiv.org/abs/0705.0647}{{\ttfamily arXiv:0705.0647
  [astro-ph]}}.
[Erratum: Astrophys. J. Suppl. Ser. \textbf{182}, 474 (2009)].

\bibitem{Feynman:1963uxa}
R.~P. Feynman, R.~B. Leighton, and M.~Sands, {\em {The Feynman Lectures on
  Physics, Vol. II}}.
\newblock Addison Wesley, 1964.
\newblock
\url{http://www.feynmanlectures.info/}.
\newblock

\bibitem{PageAdams}
L.~{Page} and N.~I. {Adams}, ``{Action and Reaction Between Moving Charges},''
  \href{http://dx.doi.org/10.1119/1.1990689}{{\em Am. J. Phys.} {\bfseries 13}
  (1945) 141--147}.

\bibitem{Faye:2006gx}
G.~Faye, L.~Blanchet, and A.~Buonanno, ``{Higher-order spin effects in the
  dynamics of compact binaries. I. Equations of motion},''
  \href{http://dx.doi.org/10.1103/PhysRevD.74.104033}{{\em Phys. Rev.}
  {\bfseries D74} (2006) 104033},
\href{http://arxiv.org/abs/gr-qc/0605139}{{\ttfamily arXiv:gr-qc/0605139
  [gr-qc]}}.

\bibitem{Keppel:2009tc}
D.~Keppel, D.~A. Nichols, Y.~Chen, and K.~S. Thorne, ``{Momentum Flow in Black
  Hole Binaries. I. Post-Newtonian Analysis of the Inspiral and Spin-Induced
  Bobbing},'' \href{http://dx.doi.org/10.1103/PhysRevD.80.124015}{{\em Phys.
  Rev.} {\bfseries D80} (2009) 124015},
\href{http://arxiv.org/abs/0902.4077}{{\ttfamily arXiv:0902.4077 [gr-qc]}}.

\end{thebibliography}\endgroup

\end{document}